\documentclass[]{interact}
\pdfoutput=1
\usepackage{epstopdf}
\usepackage[caption=false]{subfig}

\usepackage[numbers,sort&compress]{natbib}
\bibpunct[, ]{[}{]}{,}{n}{,}{,}
\renewcommand\bibfont{\fontsize{10}{12}\selectfont}
\makeatletter
\def\NAT@def@citea{\def\@citea{\NAT@separator}}
\makeatother

\theoremstyle{plain}

\theoremstyle{definition}

\theoremstyle{remark}

\usepackage[hyphens]{url}
\usepackage{color}
\usepackage{graphicx}
\usepackage{hyperref}
\usepackage{wrapfig} 
\usepackage{tikz}
\usetikzlibrary{positioning}
\usepackage{floatrow}
\usepackage{framed}

\begin{document}


\title{Scientific Discovery with the James Webb Space Telescope}

\author{
\name{Jason Kalirai\textsuperscript{a,b}\thanks{Accepted for Publication in Contemporary Physics (2018)}\thanks{CONTACT Jason Kalirai. Email: jkalirai@stsci.edu}}
\affil{\textsuperscript{a}Space Telescope Science Institute (STScI), 3700 San Martin Drive, Baltimore MD, USA, 21218}
\affil{\textsuperscript{b}Center for Astrophysical Sciences, 3400 N.\ Charles Street, Johns Hopkins University, Baltimore MD, USA, 21218}}
\maketitle

\begin{abstract}
For the past 400 years, astronomers have sought to observe and interpret the Universe by building more powerful telescopes.  These incredible instruments extend the capabilities of one of our most important senses, sight, towards new limits such as increased sensitivity and resolution, new dimensions such as exploration of wavelengths across the full electromagnetic spectrum, new information content such as analysis through spectroscopy, and new cadences such as rapid time-series views of the variable sky.  The results from these investments, from small to large telescopes on the ground and in space, have completely transformed our understanding of the Universe; including the discovery that Earth is not the center of the Universe, that the Milky Way is one among many galaxies in the Universe, that relic cosmic background radiation fills all space in the early Universe, that that the expansion rate of the Universe is accelerating, that exoplanets are common around stars, that gravitational waves exist, and much more.  For modern astronomical research, the next wave of breakthroughs in fields ranging over planetary, stellar, galactic, and extragalactic science motivate a general-purpose observatory that is optimized at near- and mid-infrared wavelengths, and that has much greater sensitivity, resolution, and spectroscopic multiplexing than all previous telescopes.  This scientific vision, from measuring the composition of rocky worlds in the nearby Milky Way galaxy to finding the first sources of light in the Universe to other topics at the forefront of modern astrophysics, motivates the state-of-the-art James Webb Space Telescope ({\it Webb}).  In this review paper, I summarize the design and technical capabilities of {\it Webb} and the scientific opportunities that it enables.
\end{abstract}



\section{Telescopes}\label{telescopes}

In 1610, a picture seen through a telescope changed forever our most basic understanding of the Universe and what our place within it was.  The idea that the Earth was not at the center of the Universe existed at the time of Galileo's observation, but verifying this Copernican model required the human eye to see something in support of its predictions.  In this case, Galileo's 37 mm telescope collected about $\sim$50$\times$ more light than the human eye, and revealed that there were faint moons orbiting the planet Jupiter, and, therefore, not everything was orbiting Earth.  This scientific discovery sparked rapid advances in optics and telescope design to see deeper and sharper into space, and, since then, progress in astronomy has been tied directly to advances in telescope engineering.  With each successive increase in aperture size, new instrumental capability, and other technology, we are able to see the Universe in a different light and reshape our understanding of what our place within it is.  This progression began with larger refractors on the ground, then the first reflectors, and, in the modern age, with very large ground-based telescopes and space-based observatories.

\subsection{Ground-Based Telescopes}

Refracting telescopes with focal lengths of many tens of meters were being constructed by the mid 1600s and the first reflectors also started coming online.  These telescopes expanded the census of moons in our solar system and stars in the night sky, and also provided important lessons to fine tune the engineering strategy for building a robust telescope.  By the 1700s, astronomers were making multi-year voyages to the Southern hemisphere to catalog thousands of stars and discover strange nebulous objects \citep{delacaille1763,messier1764,messier1781}.  In the late 1700s, Herschel built several reflectors including a 1.26~meter telescope that was the world's largest.  He used his telescopes to discover several moons of Saturn, the planet Uranus, and also cataloged nebulae and clusters that formed the foundation of the New General Catalog (NGC).  To study these objects in more detail, astronomers constructed even larger telescopes in the coming decades such as Lord Rosse's 1.83~m Leviathan of Parsonstown.  Mid-1800s drawings of objects observed through this telescope, such as the Whirlpool Galaxy M51, reveal clear spiral structure and the companion NGC 5195 galaxy.

The era of modern large reflecting telescopes, with silvered mirrors and other new technologies, began a century ago with the 1.5~m Hale (1908) and then the 2.5~m Hooker (1917) telescopes at Mt. Wilson Observatory.  Observations by Edwin Hubble with the latter telescope resolved ``The Great Debate'', and showed that nebulae such as Andromeda in the night sky had distances much greater than the size of the Milky Way galaxy, and, therefore, were entire galaxies themselves \citep{hubble1926}.  By the 1930s and 40s, astronomers were motivating much larger telescopes such as the 200-inch at Palomar, which would be completed in 1948.  Now considered a ``moon shot'' of its generation, this telescope was a marvel of engineering.  It was used to measure the expansion rate of the Universe \citep{sandage1958}, to discover quasars and active galactic nuclei in the distant reaches of the Universe \citep{schmidt1968}, and to time-tag stellar populations in the Milky Way to better understand the formation process of our Galaxy.  By the 1970s and 80s, multiple 4-meter class telescopes were online throughout the world, and by the 1990s, the first of the 10-meter class telescopes were built using a new technology involving a segmented mirror design (i.e., the Keck Observatory).  Today, astronomers are building multiple 20 -- 40~meter segmented telescopes.

\subsection{Space-Based Telescopes}

During the same time that the 200-inch at Palomar was about to begin operations, astronomer Lyman Spitzer published a paper in 1946 titled, ``Astronomical Advantages of an Extra-Terrestrial Observatory'' \citep[reprinted as][]{spitzer1990}.  Spitzer identified several limitations of ground-based astronomy, including the extinction of gamma rays, x-rays, ultraviolet, far-infrared, and long radio wavelengths by the Earth's atmosphere, the blurring of visible light that passes through the atmosphere, and engineering limitations of the then large telescopes due to flexure by gravity.  To push astrophysics beyond these limitations, Spitzer presented a bold vision to place a large telescope in space.\footnote{The original idea of launching a telescope into space via a rocket was outlined by Herman Oberth in 1923.}  This observatory, ``The Large Space Telescope'', now {\it Hubble}, became a national priority for the USA and was completed in 1985, launched in 1990, and first serviced in 1993.  Due to its stable environment, low background, and advanced technologies from multiple servicing missions, {\it Hubble} remains today, 28 years later, our most powerful telescope to study the nearby and distant Universe.  Similar to Galileo's and Edwin Hubble's discoveries, {\it Hubble} revealed to us how much more complex the Universe is than we perceived.  {\it Hubble} showed us that our Milky Way galaxy is just one out of hundreds of billions of galaxies in the Universe \citep{williams1996,ferguson2000,beckwith2006}.

Since the time of {\it Hubble}, the astronomical community has prioritized technologies for space astronomy through a scientific process known as ``The Decadal Survey''.  Once every decade, hundreds of astronomers write papers to describe the leading research problems in their respective fields.  The synthesis of this input motivates new telescopes.  Moving forward with the {\it Hubble Space Telescope} was the top priority in the 1972 Decadal Survey.  The {\it Chandra X-Ray Observatory} was the top priority in the 1982 Decadal Survey and the {\it Spitzer Space Telescope} was the top priority in the 1991 Decadal Survey.  Combined with the {\it Compton Gamma Ray Observatory}, this array of multi-wavelength facilities became known as astronomy's ``Great Observatories''.  Most areas of modern-day astrophysics have been transformed by these missions, as well as by the many dozens of smaller space telescopes launched during each of these eras.

\section{The James Webb Space Telescope}

The James Webb Space Telescope ({\it Webb}) is one the largest and most complex science programs ever undertaken, and is designed to answer forefront questions about the origins of planets, stars, and galaxies in the Universe \citep{seery2003,sabelhaus2004,gardner2006}.  Representing the next step in the evolution of space telescopes, {\it Webb} is built with innovative technologies to explore the Universe beyond {\it Hubble} and {\it Spitzer's} reach,

\begin{itemize}
\item An 18-segmented 6.5~meter primary mirror coated in gold, with adjustable optics and built-in integrated wavefront sensing,
\item A 21 $\times$ 14~meter sunshield with 5-layers of 25 -- 50 micrometer thick Kapton to passively cool the observatory to a cryogenic temperature of 40 Kelvin,
\item An advanced array of infrared-optimized instruments with dozens of high-resolution imaging, multiplexed spectroscopy, and coronagraphy observing modes.  {\it Webb's} sensitivity will extend from the red part of the visible spectrum at 0.6~microns to mid-infrared wavelengths of 28.8~microns,
\item A deployable design that enables launch from a rocket with a 5-meter fairing,
\item A 10-year science mission goal from a thermally stable solar orbit 1 million miles from Earth at the second Lagrange point.
\end{itemize}

The {\it Webb} project is led by the USA's National Aeronautics and Space Administration (NASA), working in conjunction with the European Space Agency (ESA) and the Canadian Space Agency (CSA).  The formal partnership on {\it Webb} is 80\% NASA, 15\% ESA, and 5\% CSA.  The mission lead is NASA's Goddard Space Flight Center (GSFC) and the observatory contractor is Northrop Grumman Aerospace Systems.  The science and flight operations center is the Space Telescope Science Institute (STScI).  Dozens of other industry and academic partners are integral to the development of {\it Webb}.

\begin{figure}
\centering
\includegraphics[width=13cm]{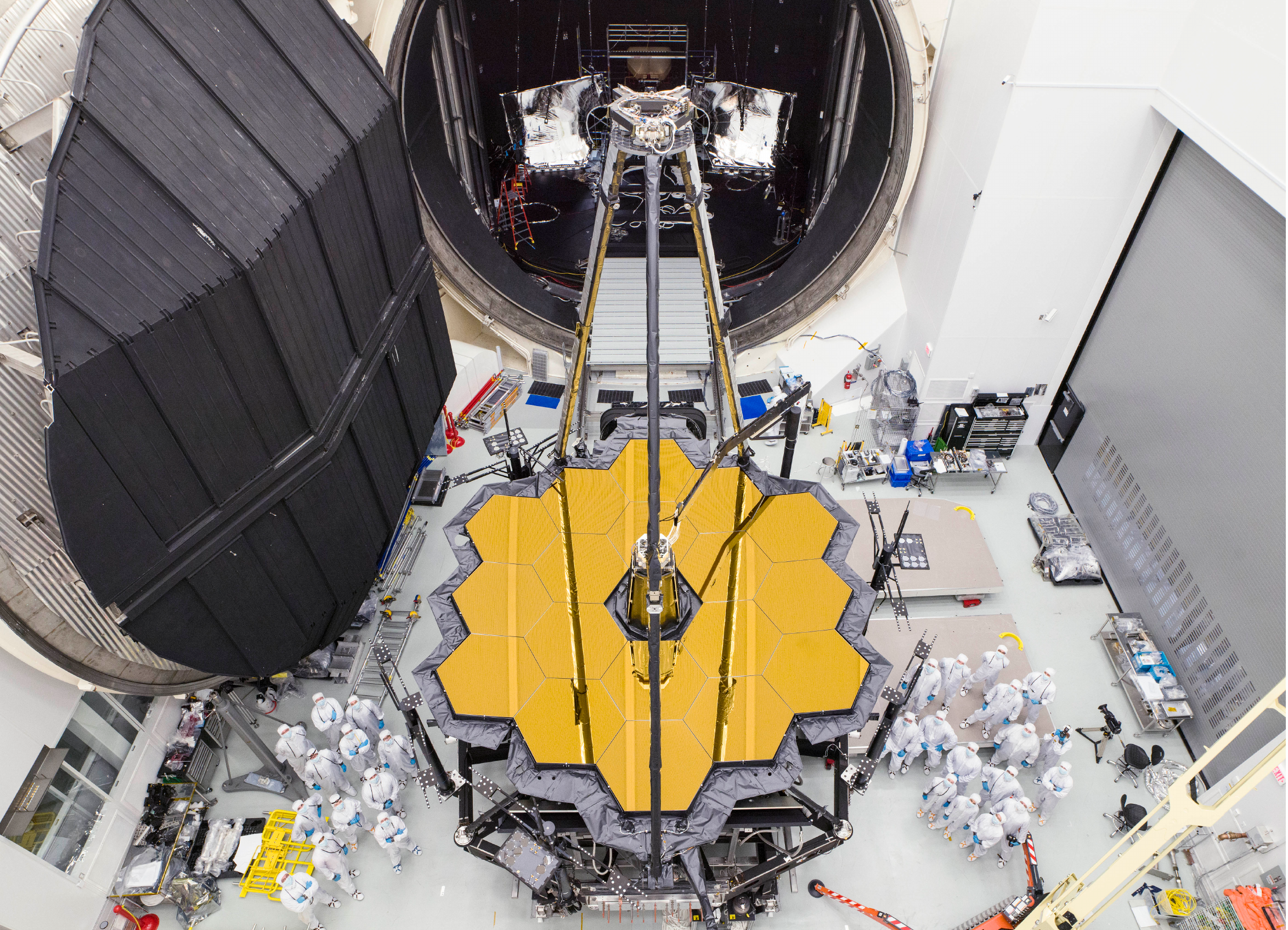} 
\caption{A birds-eye view of the 6.5~meter 18-segmented primary mirror and attached secondary mirror of {\it Webb}. This picture is taken just after the telescope and instruments (mated to the back of the mirror) were removed from the NASA Johnson Space Center's Chamber A test facility, after successfully completing a $\sim$100~day cryovacuum test campaign in 2017 \citep{feinberg2006,kimble2012}.  A number of the {\it Webb} engineers involved in the test surround the mirror. {\it Image Credit: NASA/C.\ Gunn}} \label{Fig:Mirrors}
\end{figure}

Figure~\ref{Fig:Mirrors} shows a beautiful picture of {\it Webb's} optics following a 2017 cryovacuum test campaign at NASA.  This 6.5~meter ``Golden Eye'' is the most powerful telescope ever built.


\subsection{Observatory Design}
{\it Webb} has a novel ``open'' design that flows directly from its science requirements \citep{nella2004,gardner2006,clampin2008,menzel2010}.  The top and bottom sides of the observatory are separated by a 21 $\times$ 14~meter 5-layer sunshield.  As {\it Webb} orbits the Sun once every year, the bottom-side of the sunshield will always face the Sun, Earth, and Moon and will directly receive Solar radiation and reach temperatures of 358~Kelvin.  Each layer of the sunshield blocks and re-directs the heat to protect the top-side of the observatory and maintain it at a cryogenic temperature of $\sim$40~Kelvin.  The optics and instruments of {\it Webb} are on the cold (top) side of the observatory and the spacecraft is on the bottom (hot) side.

The optics of the telescope are a three-mirror anastigmat with focal ratio f/20 and an effective focal length of 131~meters \citep[e.g., see][]{lightsey2012,glassman2016}.  The primary mirror of {\it Webb} contains 18 thin and light-weight hexagonal beryllium mirror segments, each measuring 1.32~m across (flat-to-flat) and weighing just 20~kg without the support infrastructure.  The surface aberrations on each mirror segment are $<$30~nm rms \citep{feinberg2012}.  After bouncing off the primary mirror, photons will reflect off the 0.74~m secondary and enter the Aft Optics System located at the center of the primary \citep[e.g.,][]{martinez2007,atkinson2012}.  This contains a 0.73 $\times$ 0.52~m concave aspheric tertiary mirror that receives the incoming light beam, cancels out aberrations, and sends it to a flat fine steering mirror for image stabilization \citep{arneson2010}.  The fine steering mirror has a surface figure of $<$25~nanometer rms and adjusts continuously along two axes to suppress jitter and deliver diffraction limited performance (i.e., the optical system will produce images with a resolution as good as the theoretical limit for the telescope)\citep{ostaszewski2007}. The light beam then goes to the scientific instruments (see \S\,\ref{instruments}), which are mounted to the back of the primary mirror.  All of the {\it Webb} mirrors are coated in a 1000 angstrom thin layer of gold to achieve $>$98\% reflection of infrared light \citep{keski-kuha2012}.  

The top side of the telescope is connected to the bottom (hot) side by a deployable tower that passes through the sunshield.  The bottom side of {\it Webb} contains the spacecraft and several other components that are essential to {\it Webb's} functionality, as described in \cite{arenberg2016}.  This includes the attitude control system that maintains {\it Webb's} pointing accuracy and stability (see \S\,\ref{pointing}).  {\it Webb} also has thrusters that will be used initially to insert the observatory into its orbit, and then during operations to dump angular momentum that is built up through the execution of its observing program and to maintain the telescope's orbit (see \S\,\ref{commissioning}).  The thrusters operate using fuel, which is the only consumable on {\it Webb} (intended to last $\gtrsim$10~years).  To help conserve fuel, a momentum trim tab is located on the back of the observatory to balance out solar pressure and manage the effects of the reaction wheels.  The bottom-side of {\it Webb}, which will receive constant illumination, also contains solar panels to power {\it Webb's} computer and the warm electronics in the spacecraft.  Communications with {\it Webb} will use NASA's Deep Space Network. A high-gain antenna on the bottom of the observatory will receive commands and also transmit scientific data and telemetry from {\it Webb's} computer to Earth \citep{johns2008}.  

\subsection{Launch, Deployment, and Orbit}\label{commissioning}

\begin{figure*}
\centering
\includegraphics[width=14cm]{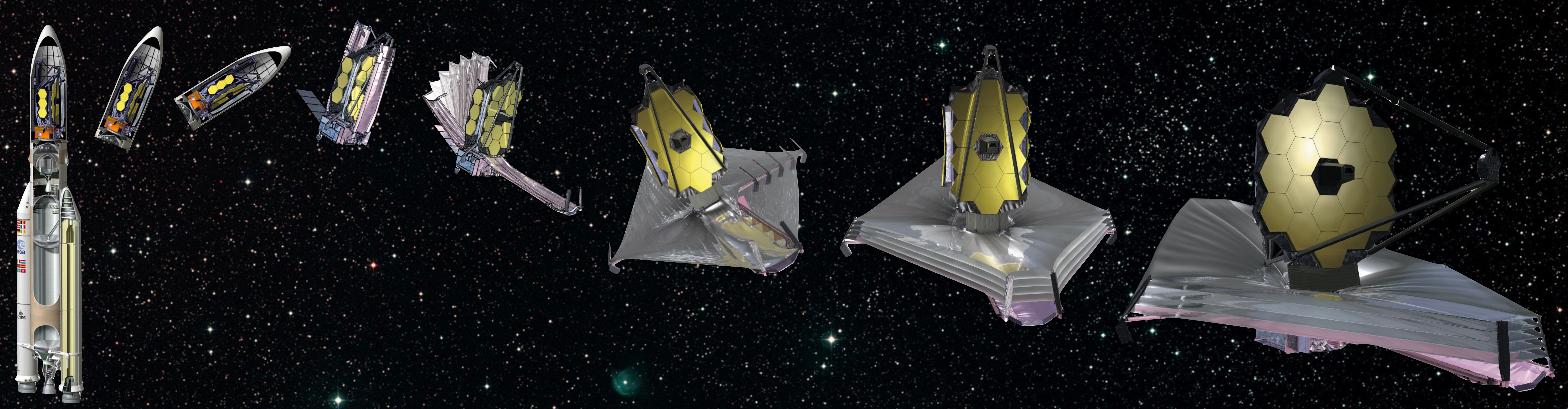} 
\caption{A graphical illustration of the {\it Webb} launch and deployment.  The images on the left show the telescope folded up and stowed at the top of the Ariane 5 launch rocket, and subsequent images to the right show the folding down of the long ``palettes'' that house the sunshield, telescoping booms on the side of the observatory to extend the folded sunshield, tensioning of the individual sunshield layers, and the deployment of the secondary mirror and primary mirror segments. {\it Image Credit: Pictures of Webb stages obtained from Northrop Grumman}} \label{Fig:Deployment}
\end{figure*}

The launch and deployment of {\it Webb} is a major technical challenge.  The size of the observatory in the fully deployed configuration is $>$10~meters in two dimensions and $>$20~meters in the third, whereas the fairing of the ESA's Ariane 5 launch vehicle has a diameter of 4.6~m and a length of 16.2~m.  To fit inside the rocket, {\it Webb's} components have been built to carefully fold together such that the effective volume of the observatory is significantly reduced.  Just minutes after the launch when {\it Webb} enters space, the first deployments (e.g., the solar panel) will be undertaken and the unfolding of the observatory will begin (see Figure~\ref{Fig:Deployment}).

{\it Webb's} deployments will take approximately one month to complete, and include many small components such as the antenna, momentum trim tab, radiator shield, primary mirror actuators, etc.  This first phase of the commissioning of the mission also includes several mid-course corrections, subsystem check outs, and appropriate modifications to electric heaters in the instrument module that prevent condensation from forming as residual water in {\it Webb's} components escapes into space.  At the end of the first month, {\it Webb} will be $\sim$1.4~million km from Earth and will begin its L2 halo orbit.  It is during this time that {\it Webb} enters phase 2 of its commissioning timeline, which will last three months.  First, the Near Infrared Camera (NIRCam) and Fine Guidance Sensor will be activated and used to obtain images of a bright star in order to locate and identify the positions all of {\it Webb's} mirrors.  Initially, the mirrors will be misaligned by several millimeters. Through an iterative process involving segment-level wavefront control \citep{acton2004}, {\it Webb's} mirrors will be aligned to tens of nanometers through adjustments in six degrees of freedom and in radius of curvature by way of high-precision actuators on the back of each segment \citep[e.g.,][]{barto2008}.  During this phase of commissioning, the other science instruments will also be activated and checked out, and the Mid Infrared Instrument (MIRI) will begin its active cooldown via a cryocooler (see \S\,\ref{miri}).  Co-phasing of {\it Webb's} mirrors will proceed through an iterative process involving multiple science instruments and multi-field sensing and control \citep{acton2012,perrin2016}.  The final two months of commissioning comprises phase 3, where each science instrument is independently focused, calibrated, and characterized and used in observatory level tests such as thermal slew, stray light, mechanism disturbance, and moving target tests.  Once {\it Webb} has passed all of its commissioning milestones, approximately six months after launch, the scientific program will begin.  Throughout operations, fine adjustments to {\it Webb's} alignment will be made periodically to ensure that the telescope performs as a single 25.4~m$^{2}$ primary mirror \citep{contos2006}.

{\it Webb} will orbit the L2 point approximately once every six months at a distance ranging from 250,000 to 832,000~km.


\subsection{Sky Visibility}\label{visibility}

Unlike {\it Hubble} and other low Earth orbit satellites, {\it Webb's} sightline to astronomical objects will not be periodically occulted by the Earth or Moon.  However, to fully block out light lines from the Sun, {\it Webb}'s boresight (the path from the center of the primary through the secondary) must be between 85 -- 135~degrees of the Sun line at all times.  Over this range, the optics of the observatory will always be in the shadow of the sunshield.  The torus formed around the sky from this field of regard includes about 40\% of the sky, which represents the instantaneous visibility region for {\it Webb}.  As {\it Webb} moves around the Sun, the torus moves across the sky to sweep out all right ascension and declinations.  Given the design and orbit of {\it Webb}, the sky visibility includes all of the sky for 51 days of continuous observing each year, 30\% of the sky for 197 days per year, and a small part of the sky for 365 days per year (i.e., ``continuous viewing zones'' that are within five degrees of each ecliptic pole) \citep{kinzel2012}.

\subsection{Pointing Stability}\label{pointing}

{\it Webb's} pointing accuracy and stability is achieved first through coarse pointing by its attitude control system.  This is a combination of three star trackers (small telescopes that use star patterns for coarse alignment) that point the observatory, six reaction wheels that rotate the observatory, and eight gyroscopes that track and control how fast the observatory is turning \citep[][A.\ Cohen, private communication]{arenberg2016}.  Finer pointing for astronomical observations requires {\it Webb's} Fine Guidance Sensor, an infrared camera with two 2.15 $\times$ 2.15~arcmin usable fields and $\lambda$ = 0.6 -- 5.0~micron sensitivity \citep{rowlands2004}.  The Fine Guidance Sensor will identify and guide on stars as faint as $J$ = 19.5 (AB mag), which gives {\it Webb} a 95\% probability of finding a guide star anywhere on the sky.  The final pointing stability for relative offsets is 6~milliarcsec rms and the absolute accuracy is limited to the 1~arcsec rms accuracy of the guide star catalog \citep{rowlands2016}.

\subsection{Thermal Background and Optical Performance}\label{thermal}

As an infrared observatory, {\it Webb's} performance is critically dependent on achieving a stable thermal environment with excellent control of stray light (i.e., anything that doesn't come from the science field of view).  Taking full advantage of {\it Webb's} high-resolution instrumentation also requires superb wavefront sensing and alignment to deliver excellent image quality from the optics to the science instruments \citep{knight2012,rohrbach2016}.  {\it Webb's} giant sunshield is the observatory's first line of protection from the radiation of the Sun, Earth and Moon, as well as other stray light on the back side of the observatory.  The sunshield will reduce the incident radiation on {\it Webb's} top-side by a factor of 10~million, down to milliwatt levels \citep{arenberg2016}.  

{\it Webb} includes several additional design elements to eliminate potential stray light lines from other celestial objects, zodiacal light from the solar system and Milky Way, and the observatory's own self-emission \citep{lightsey2014b}.  These include a ``frill'' around the primary mirror, a thermal shield (``bib'') under the primary mirror, and a baffle, pupil mask, and field stop in other locations of the optical system \citep{wei2006,elliott2013}.  At most wavelengths with $\lambda$ $<$ 15~microns, the background is dominated by in-field zodiacal emission whereas observations at $\lambda$ $>$ 15~microns will be dominated by thermal self-emission from {\it Webb} itself.

The optics are designed to deliver diffraction limited performance at $\lambda$ $>$ 2~microns, and therefore {\it Webb} will have excellent image quality over most of its wavelength range \citep{lightsey2012,lightsey2014a}.  More information on the image quality across all {\it Webb} instruments is described in \cite{perrin2014}.

\section{The Scientific Instruments}\label{instruments}

Behind {\it Webb}'s 18-segmented 6.5~meter primary mirror are four complex instruments and their supporting subsystems \citep{greenhouse2011}.  The science instruments enable an array of imaging, spectroscopy, and coronagraphy modes covering the wavelength range $\lambda$ = 0.6 to 28.8 microns \citep{davila2004,greenhouse2004,greenhouse2016}.  Unlike {\it Hubble}, {\it Webb's} instruments cannot be upgraded so the initial suite has been designed with technologies to maximize competitiveness over the lifetime of the mission.  The capabilities include multiple ultra-sensitive and high-resolution imagers operating at the diffraction limit, the first space-based multiobject spectrograph, integral field spectroscopy over five separate wavelength regions, single object and wide-field slitless spectroscopy, coronagraphy at multiple near- and mid-infrared wavelengths, and more.  The instrument modes on {\it Webb} offer many specific configurations (e.g., read out patterns, subarrays, time series) to allow for broad science applications targeting a wide range of astronomical targets (e.g., moving bodies in the solar system, bright stars hosting planets, rapidly varying objects, etc.).

The four {\it Webb} instruments are,

\subsection{The Near Infrared Camera (NIRCam)}\label{nircam}

NIRCam is {\it Webb's} workhorse imaging instrument at near-infrared wavelengths, and offers significant gains over previous high-resolution cameras on space-based telescopes \citep{rieke2005,beichman2010,beichman2012,rieke2011,greene2017}.  The instrument has dual modules that are independent and redundant, and each points to adjacent fields on the sky separated by 44~arcsec.  Each of the NIRCam detectors has sensitivity from $\lambda$ = 0.6 -- 5.0 microns, but a dichroic mirror splits the beam into a short wavelength ($\lambda$ = 0.6 -- 2.3 microns) and long-wavelength (2.4 -- 5.0 microns) channel.  Within a module, the two channels observe the same field.  The typical imaging observation can therefore achieve the full 2.2 $\times$ 4.4~arcmin field of view in both a short wavelength filter and a long wavelength filter (i.e., the simultaneous field of view of a NIRCam image will be 2.2 $\times$ 4.4~arcmin).  The filter complement on NIRCam includes 10 broadband filters ($R$ = $\lambda$/{$\Delta$}{$\lambda$} $\sim$ 4), 11 medium band filters ($R \sim$ 10), and 3 narrow band filters ($R \sim$ 100).

NIRCam's detectors have a pixel scale of 0.032~arcsec on the short wavelength channel and 0.064~arcsec on the long wavelength channel, so both detectors achieve Nyquist sampling of the diffraction limit at 2.0$+$ and 4.0$+$ microns.  In comparison {\it Hubble}/WFC3-IR does not achieve Nyquist sampling of the diffraction limit and {\it Spitzer} only reaches it at $\lambda$ $>$ 24~microns.  To enable observations of brighter astrophysical sources (e.g., Jupiter) up to a factor of 200 and/or to collect higher-cadence data, NIRCam detectors can be read out in smaller subarrays in imaging and time-series observing modes.  NIRCam's imaging mode is expected to be used for a very wide range of astrophysical investigations in full frame, subarray, and multi-field mosaic configurations.


\begin{figure}
\centering
\includegraphics[width=12cm]{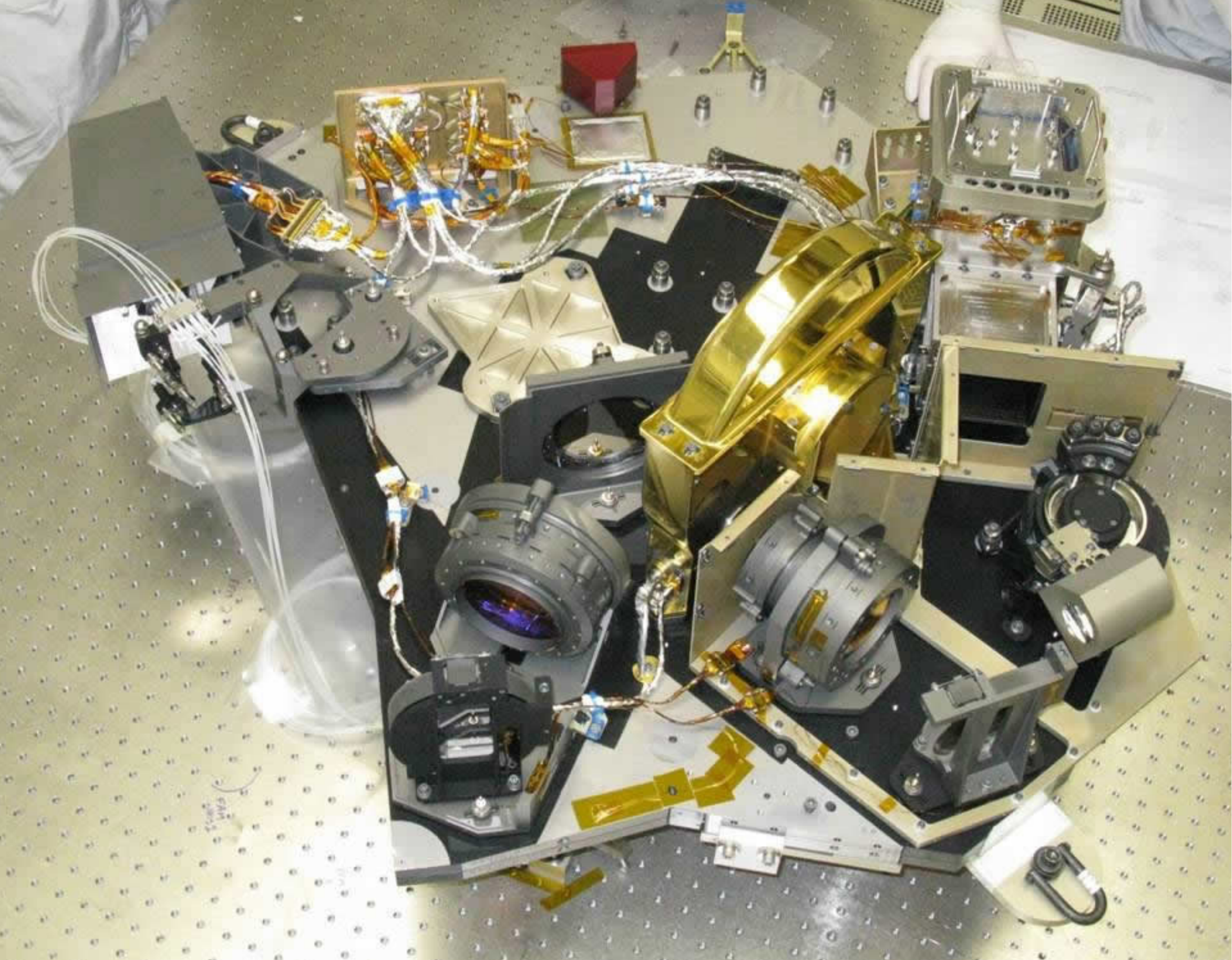} 
  \caption{The Near-Infrared Camera (NIRCam) on {\it Webb} can be used for high-resolution imaging, wide-field and time-series single object slitless grism spectroscopy, and coronagraphy at near-infrared wavelengths (see \S\,\ref{nircam}).  NIRCam was built by a team at the University of Arizona and Lockheed-Martin's Advanced Technology Center. The Principle Investigator of NIRCam is Marcia Rieke.  {\it Image Credit: Lockheed Martin}} \label{Fig:NIRCam}
\end{figure}


In addition to high-resolution imaging, NIRCam has a mode for wide-field slitless grism spectroscopy on its long-wavelength channel, over $\lambda$ = 2.4 -- 5.0~microns \citep{greene2007,greene2016a,greene2017}.  This mode can obtain a spectrum of every object in the 2.2 $\times$ 2.2~arcmin field of view, with spectral resolving power $R$ = 1600 at $\lambda$ = 4~microns decreasing to $R$ = 1130 at $\lambda$ = 2.4~microns.  Time-series observations of single sources with the grism can also utilize rectangular subarrays for shorter frame times.  NIRCam's spectroscopic mode is expected to be used heavily for planetary transit detections, but also for line-emitting galaxy and other targets.

NIRCam also contains a mode for Lyot coronagraphy at both short and long wavelengths \citep{krist2007,krist2010}.  Coronagraphy enables high-contrast imaging by suppressing the light of a bright object to reveal fainter companions.  The inner working angle ranges from 0.13 to 0.88 arcsec half-width half-maximum (HWHM) and the contrast is $\sim$10$^{-5}$ at 0.5~arcsec separation and $\sim$10$^{-6}$ at 1.5~arcsec separation.  NIRCam coronagraphy is expected to be used to discover exoplanets, to image the circumstellar environments of massive stars, and for other studies of bright central region sources as such active galactic nuclei host galaxies.

A picture of NIRCam is shown in Figure~\ref{Fig:NIRCam}.

\subsection{The Near Infrared Spectrograph (NIRSpec)}\label{nirspec}

NIRSpec is {\it Webb's} workhorse spectroscopic instrument at near-infrared wavelengths \citep{ferruit2012,birkmann2016,dorner2016,teplate2016}.  The instrument has sensitivity over a broad wavelength range of $\lambda$ = 0.6 -- 5.3 microns, designed to complement NIRCam (and NIRISS) imaging.  NIRSpec offers three distinct spectroscopic modes; single slit spectroscopy, integral field spectroscopy, and multi-object spectroscopy, each with resolving powers of $R$ $\sim$ 100, 1000, and 2700 depending on the disperser choice.  The full NIRSpec field of view is 3.6 $\times$ 3.4~arcmin and the detectors have a pixel scale of 0.1~arcsec \citep{rauscher2014}.

Spectroscopy with NIRSpec in its lowest resolution mode ($R$ $\sim$ 100) utilizes a prism with sensitivity over the full $\lambda$ = 0.6 -- 5.3~micron wavelength range.  For higher resolution spectroscopy in any of the modes, NIRSpec offers three medium resolution gratings with $R$ $\sim$ 500 -- 1300 and three high resolution gratings with $R$ $\sim$ 1500 -- 3500.

Slit spectroscopy with NIRSpec can achieve higher sensitivity and contrast than other modes as the slits are cut into an opaque mounting plate.  There are five slits available with three different sizes; 0.2 $\times$ 3.2~arcsec, 0.4 $\times$ 3.65~arcsec, and 1.6 $\times$ 1.6~arcsec.  This mode is expected to be used to target single objects over a range of brightnesses, from faint high-redshift galaxies to young Milky Way stars in star forming regions with variable backgrounds.  A specific bright object time-series mode (with the 1.6 $\times$ 1.6~arcsec slit) is optimized for exoplanet transits observations that require high-precision time-series spectroscopy over long durations \citep{ferruit2014}.


\begin{figure}
\centering
\includegraphics[width=13cm]{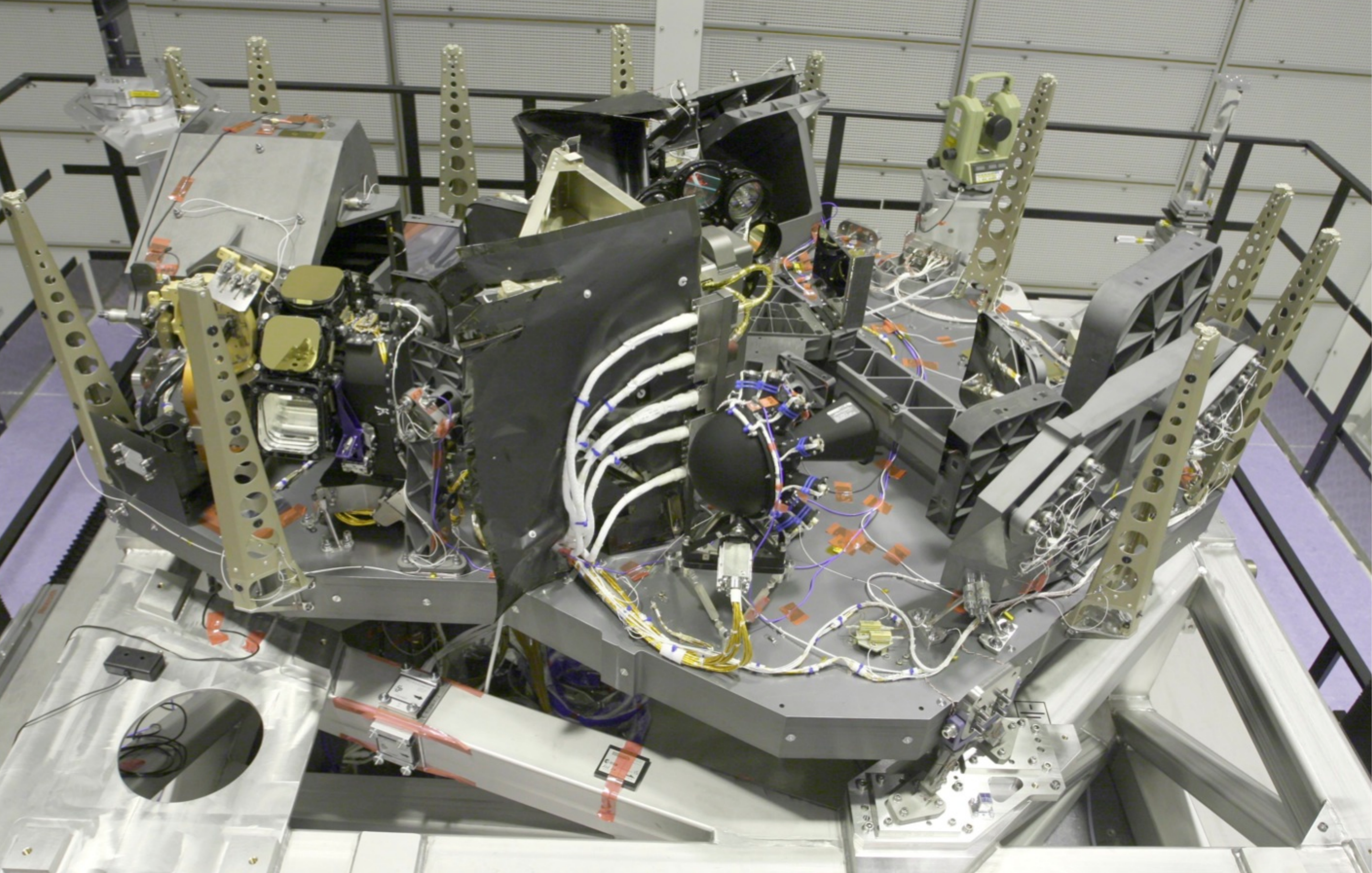} 
  \caption{The Near-Infrared Spectrograph (NIRSpec) on {\it Webb} can be used for low and medium resolution slit, integral field unit, and multiobject spectroscopy at near-infrared wavelenths (see \S\,\ref{nirspec}).  NIRSpec was built for ESA by Airbus Industries.  The microshutter array and detectors were built by NASA.  The Principle Investigator of NIRSpec prior to Dec 2011 was Peter Jakobsen and the instrument's development since that time has been guided by Pierre Ferruit. {\it Image Credit: Astrium: NIRSpec}} \label{Fig:NIRSpec}
\end{figure}


NIRSpec's integral field unit is a powerful tool to enable 3D spectral imaging over a contiguous area of the sky \citep{closs2008,purll2010}.  The aperture for the spatially resolved spectroscopy is 3.0 $\times$ 3.0~arcsec.  Each of the spatial elements comprising the data cube are 0.1 $\times$ 0.1~arcsec, so there are 30 image slices per observation (each 0.1 $\times$ 3.0~arcsec).  {\it Webb} + NIRSpec therefore makes it possible to obtain spatial and kinematic characterization of bright and faint spatially extended sources on many scales, from measuring variations in the atmospheric and surface composition of moons of solar system planets to mapping the structure and environments of distant emission line galaxies.

NIRSpec offers the first space-based multi-object spectroscopy through a four quadrant microshutter array spanning the full 3.6 $\times$ 3.4~arcmin field of view \citep{moseley2004,kutyrev2008,li2010}.  In total, the instrument contains $\sim$250,000 of these microshutters, each one spanning an open area of 0.20 $\times$ 0.46~arcsec.  The microshutters can be configured from high-precision astrometry and opened as individual, small groups, or columns to form multiple apertures.  This mode allows {\it Webb} to simultaneously target many sources (up to $\gtrsim$100) at high-spatial resolution for near-infrared spectroscopic applications within rich fields of stars and galaxies and extended targets.

{\it Webb's} tremendous sensitivity combined with NIRSpec's wavelength coverage, high spectral resolution, and diversity of modes will bring forward breakthroughs in spectral characterization of astrophysical sources.  Large spectral windows from $\sim$1.3 -- 1.9~microns, $\sim$2.2 -- 3.2~microns, and $\sim$3.6 -- 5.3~microns are all accessible at $R$ $>$ 2500 with NIRSpec.  Any metric combining sensitivity and resolution for NIRSpec is orders of magnitude more powerful than previous space-based observations.

A picture of NIRSpec is shown in Figure~\ref{Fig:NIRSpec}.

\subsection{The Near Infrared Imager and Slitless Spectrograph (NIRISS)}\label{niriss}

NIRISS is a near-infrared imager, spectrograph, and interferometer with sensitivity over $\lambda$ = 0.6 -- 5~microns \citep{doyon2012}.  The imaging capabilities are similar to the long wavelength channel of NIRCam and can be used as a parallel camera to increase {\it Webb's} near-infrared survey efficiency.  The field of view is 2.2 $\times$ 2.2~arcmin, the pixel scale is 0.065~arcsec.

The spectroscopic capabilities of NIRISS include both single-object and wide-field slitless modes.  The single-object mode provides slitless grism spectroscopy from $\lambda$ = 0.6 -- 2.8~microns, at a spectral resolving power of $R$ $\sim$ 700.  This mode offers two rectangular subarrays (96 $\times$ 2048~pixels and 256 $\times$ 2048~pixels) that are optimized for time-resolved spectroscopy of bright objects (up to $J$ = 6.75 in Vega mags).  NIRISS slitless spectroscopy is expected to be used for observations that require high-precision and spectrophotometric stability, such as transiting exoplanet studies \citep[e.g.,][]{louie2017}.


\begin{figure}
\centering
\includegraphics[width=11cm]{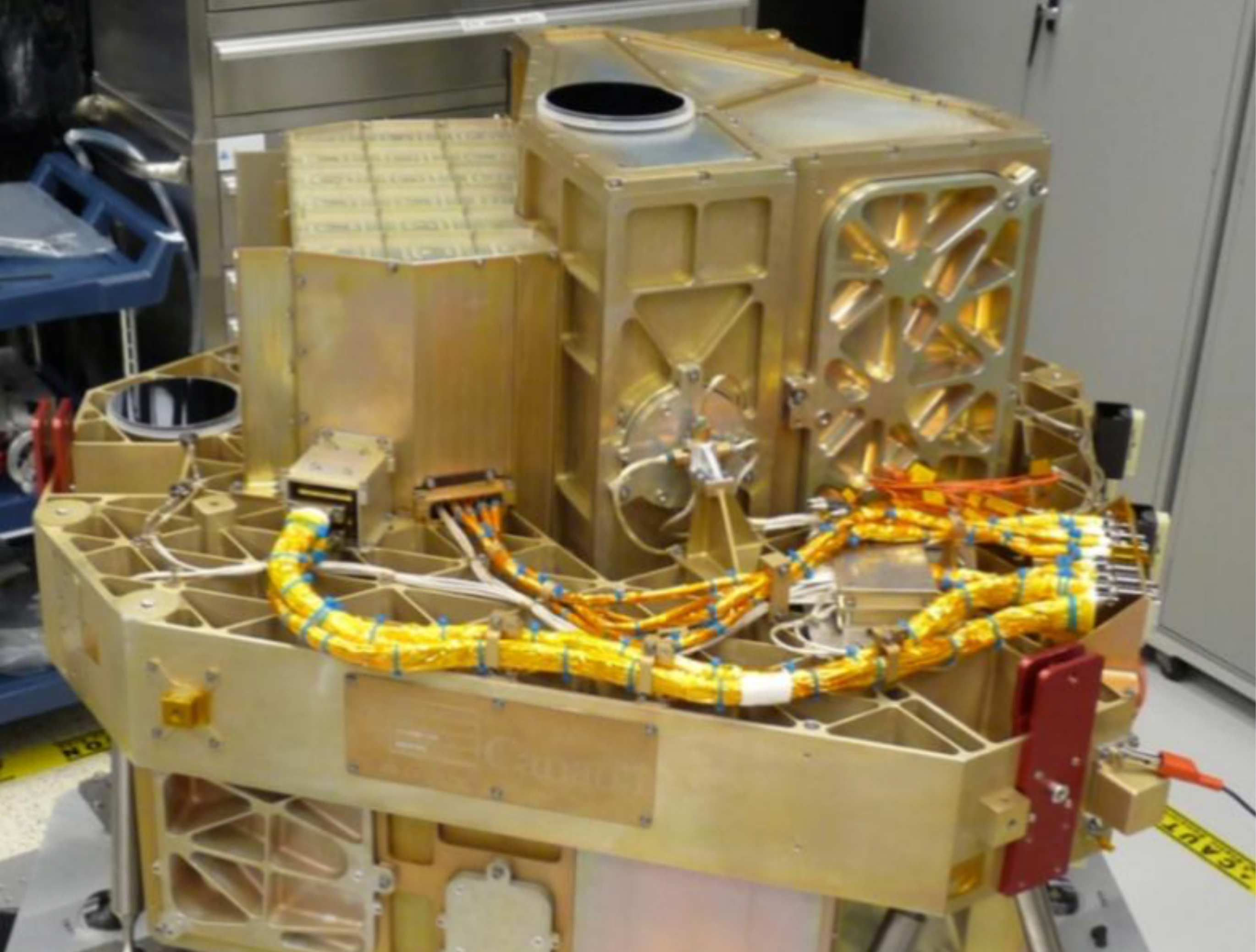} 
\caption{The Near-Infrared Imager and Slitless Spectrograph (NIRISS) on {\it Webb} can be used for parallel imaging, single-object and wide-field slitless spectroscopy, and interferometry at near-infrared wavelengths (see \S\,\ref{niriss}).  NIRISS was built by the CSA.  The prime contractor is Honeywell (formerly COM DEV).  The Principle Investigator of NIRISS is Ren\'e Doyon.   {\it Image Credit: Honeywell / COM DEV}} \label{Fig:NIRISS}
\end{figure}


The wide-field slitless mode uses two orthogonally-oriented grisms to mitigate blending and enable low resolution spectroscopy with $R \sim$ 150 over a wavelength range of $\lambda$ = 0.8 -- 2.2~microns.  Most objects within the 2.2 $\times$ 2.2~arcmin field of view will have a spectrum (e.g., applications targeting faint emission-line galaxies at high redshift explored through Ly$\alpha$).
  
NIRISS also offers a first-time space-based opportunity with aperture mask interferometry \citep{monnier2003} through a seven aperture non-redundant mask and three medium-band or one wide-band (red) filters \citep{monnier2003,sivaramakrishnan2012,artigau2014,greenbaum2015}.  This mode enables high-contrast imaging by reducing the full JWST aperture to seven much smaller apertures through a pupil mask.  The interferogram produced by these apertures consists of a narrow central diffraction core surrounded by an extended pattern of faint fringes that sample a unique (i.e., ``non-redundant'') set of spatial frequencies.  This mode can reach 10$^{-4}$ contrast ratio at separations of 70 -- 400 milliarcsec, at wavelengths of 3.8, 4.3, and 4.8~microns.  Observations at multiple wavelengths can constrain the spectral properties of the companion target.  NIRISS aperture mask interferometry can observe bright targets in an 80 $\times$ 80~pixel subarray (for short frame times and quick read outs), though options to perform aperture synthesis imaging are also available.  Aperture mask interferometry can be used to detect brown dwarfs and exoplanets that are up to 9 magnitudes fainter than their host stars \citep{sivaramakrishnan2010}, to study winds around massive stars, to measure the structure of nearby active galactic nuclei \citep{ford2014}, and other similar applications.

The NIRISS instrument is packaged together with {\it Webb's} Fine Guidance Sensor, but the two are independent.

A picture of NIRISS is shown in Figure~\ref{Fig:NIRISS}.

\subsection{The Mid Infrared Instrument (MIRI)}\label{miri}

MIRI is a versatile instrument that extends {\it Webb's} sensitivity into mid-infrared wavelengths with four core capabilities; photometric imaging, low-resolution slit or slitless spectroscopy, medium-resolution integral field spectroscopy, and coronagraphy.  Unlike {\it Webb's} other instruments, MIRI will be actively cooled to 6.7 deg Kelvin with an electric cryocooler.  The sensitivity range of MIRI is designed to start where the near-infrared cameras end (i.e., $\lambda$ = 4.9 microns) and extend well out into the mid-infrared spectrum with a red cutoff of $\lambda$ = 28.8~microns \citep[see][for more information]{wright2010,rieke2015a,wright2015}.

MIRI offers high-resolution imaging over wavelengths of $\lambda$ = 5.0 -- 27.5 microns.  The field of view is 1.23 $\times$ 1.88~arcmin on a single 1k $\times$ 1k detector with pixel scale of 0.11~arcsec \citep{bouchet2015,rieke2015b}.  This pixel scale is 12$\times$ finer than that of the InfraRed Array Camera (IRAC) on {\it Spitzer}, 25$\times$ finer than that of the Multi-Band Imaging Photometer (MIPS) on {\it Spitzer} and is also finer that of the Wide Field Camera 3 (WFC3/IR) on {\it Hubble} that operates in the near infrared ($\lambda$ = 0.9 -- 1.67~micron).  MIRI observations achieve Nyquist sampling of the diffraction limit at all wavelengths longer than 6.25~microns.  MIRI includes 9 broadband filters with central wavelengths ranging from $\lambda$ = 5.0 to 27.5~microns.  Similar to NIRCam, MIRI imaging can also be executed with short integrations through several subarrays when observing brighter objects (up to $>$30$\times$) or high sky background regions.  MIRI imaging is expected to be used for a wide variety of astrophysical studies, ranging from solar system objects to the analysis of dust and gas in Milky Way star forming regions to the discovery of first galaxies in the Universe.

Spectroscopy on MIRI includes a low-resolution slit (0.51 $\times$ 4.7~arcsec) or slitless configuration \citep{kendrew2015} and a medium resolution mode through one of four integral field units \citep{wells2015}.  The low-resolution spectroscopy covers a wavelength range from $\lambda$ = 5 -- 12~microns with resolving power $R$ = 40 at $\lambda$ = 5~microns increasing to $R$ = 160 at $\lambda$ = 10~microns.  Slitless observations can be especially important for high-precision time-series spectrophotometry observations of bright stars with transiting exoplanets.  Spectroscopic time-series observations of exoplanets with MIRI can utilize a subarray mode with fast readout times (0.159~seconds), enabling observations of stars as bright as $K$ = 5.65 (vs $K$ = 8.3 with the slit).


\begin{figure}
\centering
\includegraphics[width=13cm]{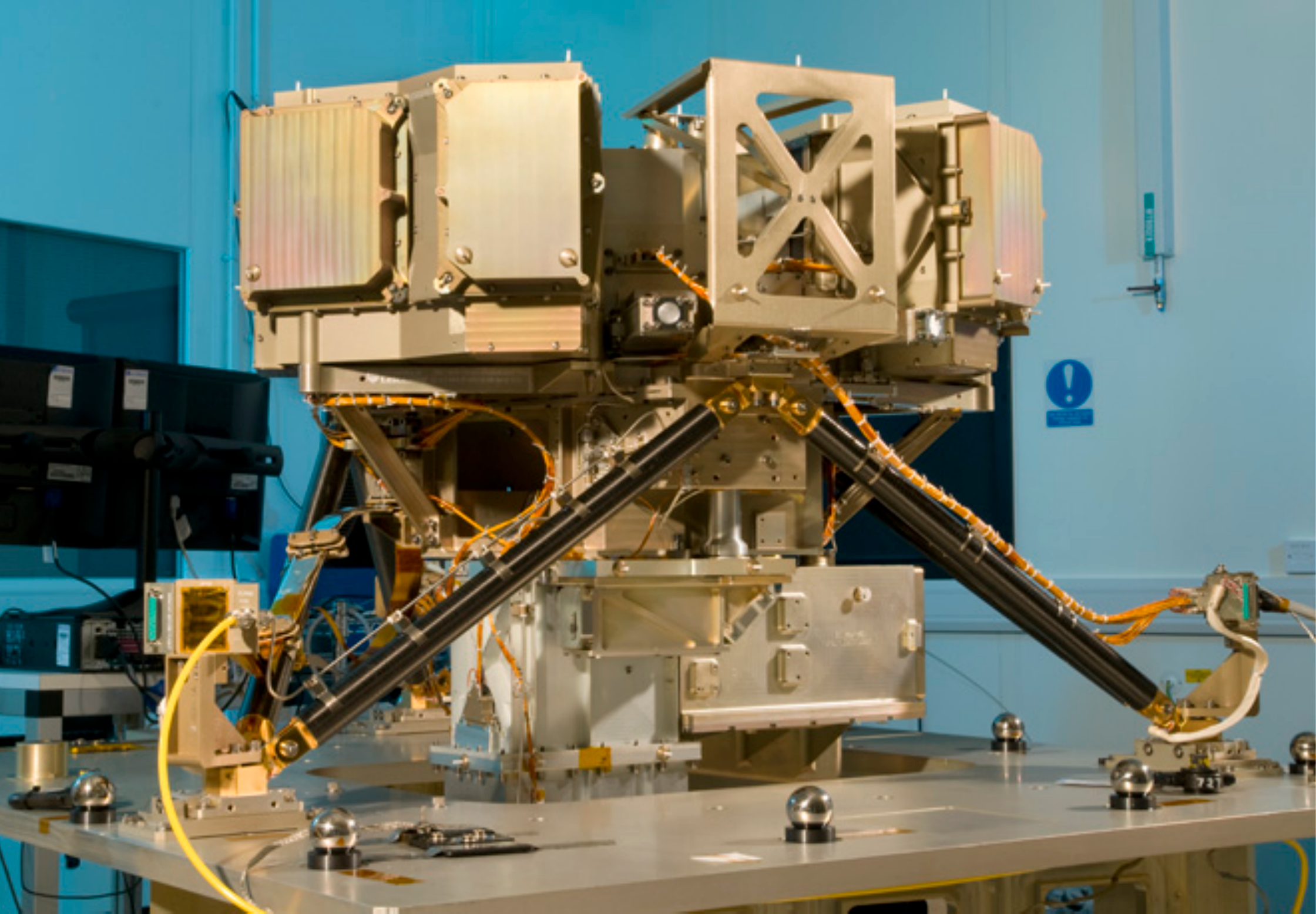} 
  \caption{The Mid Infrared Instrument (MIRI) on {\it Webb} can be used for high resolution imaging, low-resolution slit or slitless spectroscopy, medium-resolution integral field spectroscopy, and coronagraphy at mid-infrared wavelengths (see \S\,\ref{miri}).  MIRI was built by a European consortium led by Principle Investigator Gillian Wright.  The detectors were developed by Raytheon. NASA's Jet Propulsion Laboratory integrated and tested the core instrument flight software and detector systems.  The science team lead is George Rieke.  {\it Image Credit: Science and Technology Facilities Council (STFC)}} \label{Fig:MIRI}
\end{figure}


Compared to the low-resolution modes, MIRI's medium resolution spectrometer uses four integral field units (channels) to enable significantly higher resolving power and spatially-resolved spectroscopy with $R$ $\sim$ 1330 -- 3750 over $\lambda$ = 4.9 -- 28.5 microns.  The central wavelengths of each channel are 6.4, 9.2, 14.5, and 22.75~microns and the fields of view range from 3.5 to 7.5 arcsec going from the shortest to longest wavelength channels.  The integral field units split the field of view into spatial slices, each of which produces a separate dispersed "long-slit" spectrum.   The integral field units have 21 slices with widths of 0.176 arcsec in the shortest wavelength mode (4.89 -- 7.66 microns) and 12 slices with widths of 0.645 arcsec in the longest wavelength mode (17.66 -- 28.45 microns).  Post-processing produces a composite 3-dimensional (2 spatial and one spectral dimension) data cube combining the information from each of these spatial slices.  The sensitivity and fields of view of MIRI's integral field units are well suited to target everything from solar system ice giants to young disks in Galactic star forming regions to structure in galaxy lensing maps. This spectroscopy is therefore a key tool to study the early stages of star and planet formation and galaxy evolution.  MIRI integral field spectroscopy can be used simultaneously with the adjacent MIRI imager.

MIRI includes two different types of coronagraphy for mid-infrared high-contrast imaging at four wavelengths \citep{boccaletti2015}.  There are three four-quadrant phase masks at $\lambda$ = 10.65, 11.4, and 15.5~microns (field of view of 24 $\times$ 24~arcsec) and one Lyot coronagraph at $\lambda$ = 23~microns (field of view of 30 $\times$ 30~arcsec).  Each of the four-quadrant phase masks has coronagraph filters to automatically select a narrow spectral bandpass centered the wavelengths above.  The inner working angle of MIRI's four-quadrant phase masks ranges from 0.33 to 0.49~arcsec, and for the Lyot coronagraph it is 2.16~arcsec (at 22.75~microns).  MIRI's coronagraphs can be used extensively for direct imaging of young and self-luminous exoplanets with masses down to 0.1 -- 0.2~$M_{\it Jup}$.  \cite{beichman2010} demonstrate that the success rate of directly imaging more massive (Jupiter sized) planets at larger separations of $\sim$60~AU will be high.  The coronagraphs can also be used to study debris disks and quasar/AGN host galaxies.

A picture of MIRI is shown in Figure~\ref{Fig:MIRI}.

\subsection{Sensitivity of Instrument Modes}\label{sensitivity}

\begin{figure*}
\centering
\includegraphics[width=14cm]{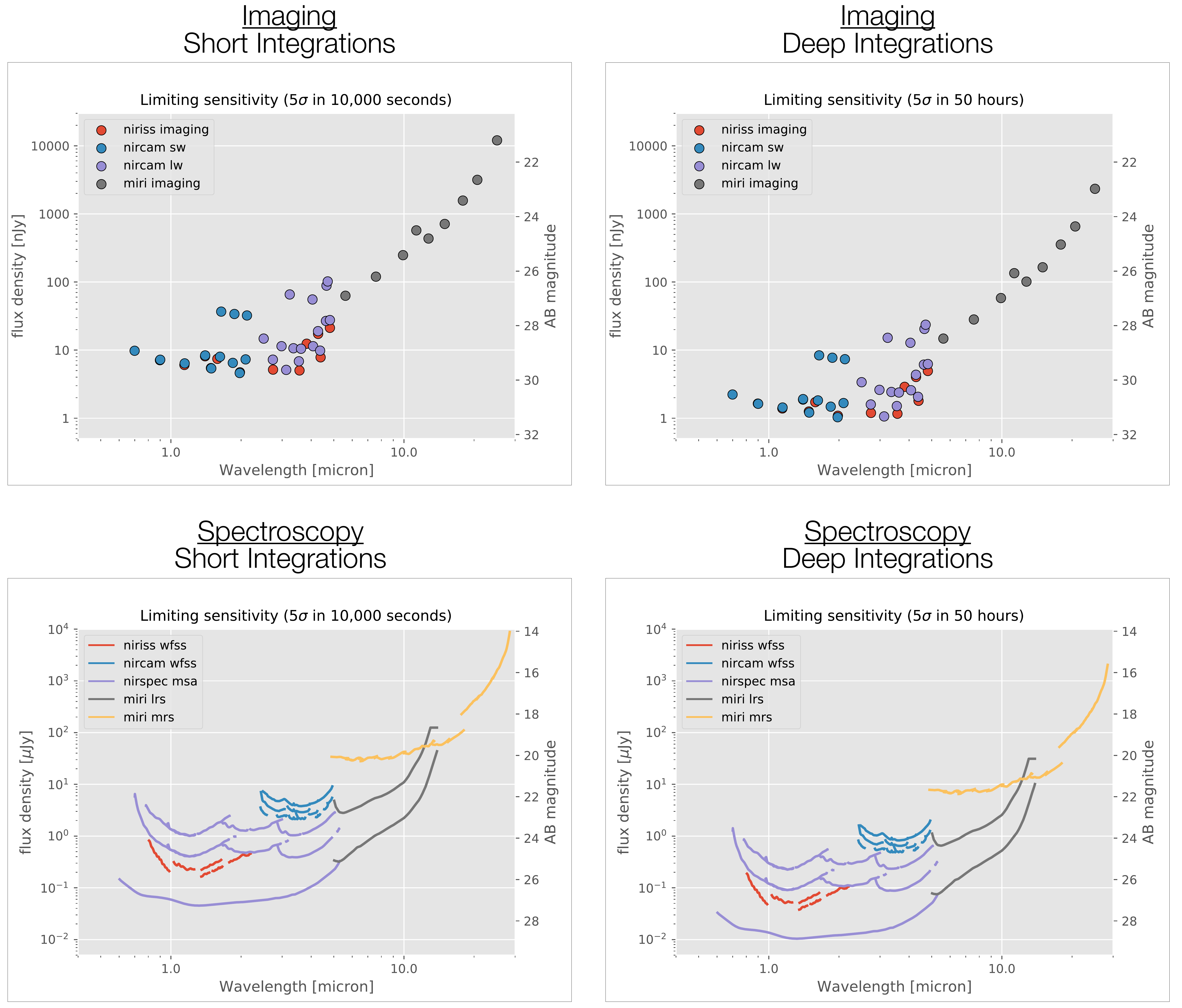} 
\caption{The 5$\sigma$ sensitivity of {\it Webb} for imaging observations is shown for integrations of 10,000~s (top-left) and 50~hour (top-right).  Similar diagrams for spectroscopy are shown in the bottom row.  The data points, curves, and colors represent distinct observing modes as indicated in the legend.  The limits that {\it Webb} reaches in most of its modes are substantially deeper than current capabilities. {\it Image Credit: K.\ Pontoppidan}} \label{Fig:Sensitivity}
\end{figure*}

The sensitivity of {\it Webb} across all instruments and imaging and spectroscopic modes is presented in Figure~\ref{Fig:Sensitivity} \citep{pontoppidan2016}.  For broad-band near-infrared imaging, {\it Webb} can reach near $>$29.5 AB mag (at 5$\sigma$) at 2 and 4~microns in just 10,000~s (top-left panel) and $>$31 AB mag in deeper integrations (e.g., 50 hours; top-right panel).  At 1.5~microns, these limits are 1.5~mag deeper than {\it Hubble} capabilities.  At mid-infrared wavelengths, the sensitivity at 10~microns is 25.5~AB~mag (5$\sigma$, 10,000~s).  These limits are more than 2.5~mag deeper than {\it Spitzer} capabilities.


For spectroscopy, {\it Webb} can reach down to $>$26.5 AB mag at near-infrared wavelengths using its low-resolution NIRSpec prism (available with all modes) in 10,000~s (bottom-left panel).  Higher-resolution modes have brighter limits.  The NIRISS wide-field slitless spectroscopy mode can reach $\sim$25.5 AB mag in this integration time.  For mid-infrared spectroscopy, MIRI can reach 25 AB~mag and 23 AB~mag at 5 and 10~microns in low resolution slit mode and 20~AB~mag at 5 -- 10~microns in medium resolution mode in 10,000s integrations.  At comparable resolutions and wavelengths, {\it Webb} spectroscopy is several orders of magnitude more sensitive than previous capabilities.

More information on the capabilities of all four {\it Webb} instruments is available online at several websites\footnote{\url https://jwst.stsci.edu/instrumentation} as well as in the {\it Webb} online user documentation system.\footnote{\url https://jwst-docs.stsci.edu}  A table summarizing all of {\it Webb's} instruments modes is provided in Appendix~\ref{instrumentmodes}.


\section{Science}\label{science}

The design and capabilities of {\it Webb} follow from four scientific pillars that were outlined in the 2000 Decadal Survey, ``Astronomy and Astrophysics in the New Millennium'' \citep{mckee2001}, and discussed extensively in \cite{gardner2006}.  These themes -- {\it Planetary Systems and the Origins of Life}, {\it Birth of Stars and Protoplanetary Systems}, {\it Assembly of Galaxies}, and {\it First Light and Reionization} -- capture leading research topics in modern astrophysics and many {\it Webb} projects related to the solar system, exoplanets, star formation, stellar populations, galaxy formation and evolution, cosmology, and other topics will form the foundation of understanding these global goals.


\subsection{Solar System}\label{solarsystem}

The most detailed information regarding objects in our solar system has always come from sending interplanetary missions to visit these worlds.  Over the last decade, this has included analysis of Mars from the {\it Mars Science Lab} and {\it Curiosity Rover} \citep{grotzinger2012}, characterization of the nucleus and composition of a comet from {\it Rosetta} and {\it Philae} \citep{glassmeier2007}, exquisite imaging of the surface composition, geology, and atmosphere of Pluto and Charon from {\it New Horizons} \citep{stern2015}, detailed investigation of the atmosphere structure, composition, magnetic field, and core properties of the planet Jupiter from {\it Juno} \citep{bolton2017}, and much more.

Complementing these high-resolution and in-situ studies, flagship observatories have yielded countless new discoveries in the solar system and have enhanced the scientific return from Planetary Science missions. For example, the discovery of small moons of Neptune and Pluto, water plumes being expelled from Europa, the largest ring on Saturn, and new Kuiper Belt Objects along the trajectory of the New Horizons mission were made by {\it Hubble} and/or {\it Spitzer} \citep[]{weaver2006,showalter2013,verbiscer2009,roth2014,sparks2016,showalter2013,schlichting2009}.

To continue this legacy, {\it Webb} is designed with specific capabilities to allow new breakthroughs in a wide range of solar system research themes,  This includes the ability to track moving targets with apparent rates of motion up to 30~milliarcsec/s and with a pointing stability of 0.05~arcsec, to observe naked-eye bright objects using subarray modes with very short integration times and rapid readout capability, and to better schedule and perform data analysis of moving target observations.  While imaging will still be challenging for the brightest objects, such as Mars, {\it Webb} will allow for spectroscopic observations of every solar system target observable.  Figure~\ref{Fig:Molecules} shows a simulated {\it Webb} near-and mid-infrared spectrum from $\lambda$ = 0.7 -- 28.8~microns, with a rich forest of molecular and ice features identified \citep{villanueva2012,villanueva2017}.  Observations with {\it Webb's} many modes can characterize the abundance and ratios of different molecules on asteroids, moons, rings, comets, and planets to understand composition and dynamics, clouds and weather patterns, surface activity and outgassing, and thermal and temperature properties. 

\subsubsection{Near-Earth Objects and Asteroids}

Among all objects in the solar system, small asteroids and near-Earth objects are two of the least well studied.  They represent some of the smallest bodies in the solar system (only $\sim$10\% are $>$1~km in size), and are the most difficult to visit with direct interplanetary missions.  Knowledge of the properties and orbits of these numerous small bodies, which have traveled from other parts of the solar system to our proximity, translate to crucial insights of the past dynamical history of the solar system.


\begin{figure}[t]
\centering
\includegraphics[width=10cm,trim={0.2cm 2cm 0 0}, clip]{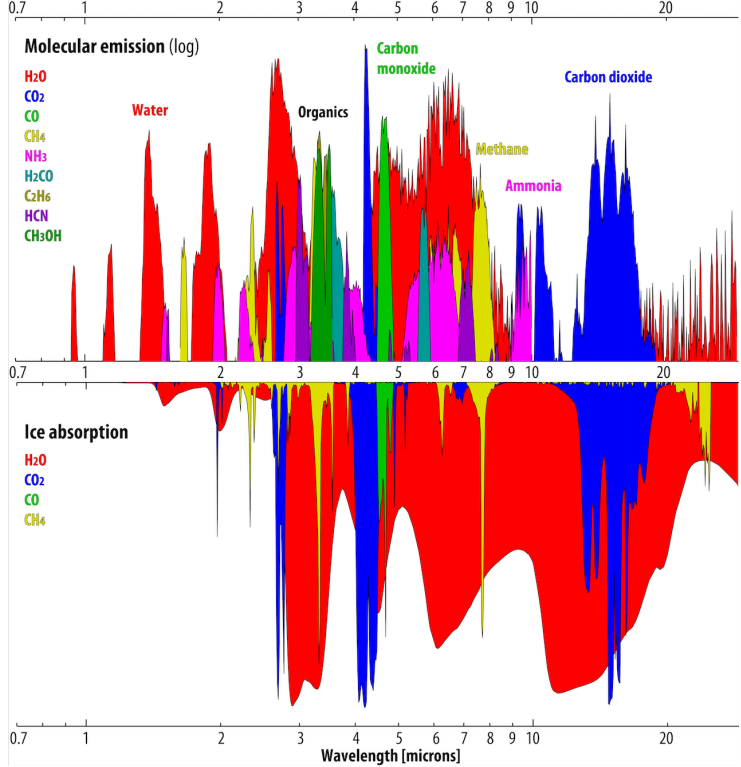} 
\caption{{\it Webb} can sample a rich forest of molecular and ice signatures in the $\lambda$ = 0.7 -- 28.8 micron spectral region for solar system objects and exoplanets.  The upper panel shows solar fluorescence fluxes \citep{villanueva2012} considering primordial / cometary abundances (H$_2$O/CO$_2$/CO/CH$_4$/NH$_3$/H$_2$CO/C$_2$H$_6$/HCN/CH$_3$OH = 100/10/10/1/1/1/1/0.2/2).  The lower panel shows ice absorptions computed with a Hapke model \citep{protopapa2017} assuming thin monolayers of 10/100/10/10~microns (H$_2$O/CO$_2$/CO/CH$_4$). The spectroscopic databases and models employed to synthesize these spectra are part of the online Planetary Spectrum Generator \citep[psg.gsfc.nasa.gov][]{villanueva2017}. {\it Image Credit: G.\ Villanueva}} \label{Fig:Molecules}
\end{figure}


{\it Webb} can observe $\sim$75\% of all known near-Earth objects with powerful imaging and spectroscopic capabilities to establish the diversity of small bodies.  Imaging observations with {\it Webb} can constrain shapes and potential outgassing, near-infrared spectroscopy can constrain surface composition (e.g., see $\lambda$ = 2.5 -- 5.0~micron region in Figure~\ref{Fig:Molecules}), and mid-infrared spectroscopy can characterize albedos, diameters, and shapes for objects much smaller (i.e., meter-size) than current studies \citep{thomas2016}.  {\it Webb's} high-spatial resolution imaging can also resolve asteroids as small as 80~km at distances of 2.8~AU.  For larger asteroids, near-infrared imaging and integral field spectroscopy can provide compositional mapping across the surfaces of the bodies \citep{rivkin2016}.  {\it Webb} can spectroscopically characterize nearly every selected object in the main asteroid belt with signal-to-noise ratio $>$10 in just 1000~s.  This capability is especially exciting in the upcoming era of wide-field and transient imaging with telescopes such as the Large Synoptic Survey Telescope ({\it LSST}), which is expected to increase the census of small bodies by factors of 10 to 100 over current numbers \citep{jones2016}.

\subsubsection{Mars}

One of the driving research questions in all of space science is to understand the past habitability of Mars.  {\it Webb} can advance this theme by building global spectroscopic maps of the atmosphere over the entire disk of the planet at high resolution in a very short amount of time.  Current missions to Mars provide only local measurements of composition, while {\it Webb} will provde a measure of the total atmospheric column across the planet \citep{villanueva2016}.  Specifically, secular monitoring on the timescales of months, seasons and years can explore variations of molecules in the atmosphere that are caused by changes in surface activity on the planet.  {\it Webb's} near infrared sensitivity is ideally suited to measure key species in the atmosphere.  Figure~\ref{Fig:Molecules} illustrates spectral features of particular interest for topics such as outbursts of sulfur species such as methane, ethane, and nitrous oxide that result from volcanic activity, hydrogen chloride in the atmosphere that vents from acidic (sub)surface liquid water, and variations in isotopic and abundance ratios of hydrogen that are related to the polar caps \citep[e.g.,][and references therein]{villanueva2013}.  These atmospheric data are important clues to understanding the processes that altered the chemical stability of the Martian atmosphere.  Near-infrared spectroscopy at $R \sim$ 2700 can achieve $\sim$100~km resolution on Mars and also be used to search for organic compounds. In addition to the spectroscopic characterization, near-infrared imaging of Mars from {\it Webb} can help characterize global dust storms and cloud systems on the planet, and carbon dioxide emission from Mars' nightside (Mars is so bright that saturation is a significant problem for dayside observations in the $\lambda$ = 0.9 -- 1.5~micron range). {\it Webb's} remote characterization of Mars can help plan future surface exploration of key regions by planetary science missions.

\subsubsection{Gas and Ice Giants}

Jupiter, Saturn, Uranus, and Neptune are analogs of the giant planets that are being discovered around nearby stars, and therefore offer a unique opportunity to test and refine models of planet formation and evolution.  The structure and dynamics of giant planet atmospheres can be complex, and many questions related to circulation, heat transport, chemistry, and clouds remain unanswered.  The spatial resolution of {\it Webb} is comparable to the largest ground-based telescopes, but enable observations at infrared wavelengths that are only accessible from space.

{\it Webb} will explore atmospheric physics in Jupiter and Saturn by studying the stratification of gases such as methane, ammonia, and carbon monoxide due to their varying strength of their absorption features.  The near-infrared part of the spectrum from $\lambda$ = 1 -- 5~microns is especially rich with absorption signatures from many molecules \citep{norwood2016b}.  Medium and narrow-band imaging with NIRCam in small subarrays, as well as near-infrared slit and integral field spectroscopy from NIRSpec, can isolate different physical process such as the absorption of gas by specific molecules, thermal radiation, and cloud opacity \citep[e.g., see]{sanchez-lavega2011}.  Such observations can constrain wind speeds as a function of altitude, latitude, and within specific storm systems such as the great red spot, and the vertical and horizontal cloud structures in the atmospheres of gas giants.

Mid-infrared subarray observations with MIRI can provide complementary studies of the thermal radiation from specific parts of Jupiter and Saturn (imaging), as well as absorption and emission features from molecules such as acetylene at 14~microns in spectroscopic modes (Figure~\ref{Fig:Molecules}).

Observations of the ice giants Uranus and Neptune with {\it Webb} will be transformative over current capabilities both in terms of spatial resolution and sensitivity.  At near-infrared wavelengths, {\it Webb} imaging can reveal cloud structures and spectroscopy can be used to correlate the depths of methane lines across different wavelengths to distinct physical processes occurring at different vertical layers (e.g., scattering from haze vs clouds).  Features in the redder part of the $\lambda$ = 1 -- 5~microns spectral window allow penetration of clouds to explore deeper into the planetary atmospheres. Specific NIRCam medium-band filters also overlap tritium, carbon monoxide, and other features that give complementary information on the vertical structure of the planets.  At mid-infrared wavelengths, {\it Webb} observations can explore the thermal emission from ice giants through a large number of emission features from hydrocarbons that cannot be observed from the ground. These observations can constrain photochemical processes, temporal variations in temperature, and help disentangle the causes of rotational modulation \citep[e.g.,][]{moses2005,dobrijevic2010}

One of the many exciting opportunities to study ice giants with {\it Webb} involves spatially resolved spectroscopy with integral field units to measure how the cloud properties and atmospheric structure changes with latitude and longitude.  These observations can constrain atmospheric circulation through dynamical models \citep{norwood2016b}, especially with short and long-term time-series observations.  The angular diameter of Uranus and Neptune are 3.6 and 2.3~arcsec, well-suited to the size of {\it Webb's} near and mid-infrared integral field units.

In addition to these core studies, {\it Webb} is a powerful telescope for target of opportunity investigations in giant planets such as large storm systems \citep[e.g.,][]{sayanagi2013}, auroral processes \citep[e.g.,][]{clarke1996}, and bombardment events and their subsequent impact on atmospheric dynamics \citep[e.g.,][]{hammel1995}.  {\it Webb} observations throughout the 2020s can establish the global context of weather patterns and other atmospheric events to complement high-resolution studies from current and future planetary science missions (e.g., {\it Juno}).

\subsubsection{Active Moons}

The moons of the giant planets of our solar system are believed to exhibit a wide range of active geology and potential biology.  Recent observations by {\it Hubble} and {\it Cassini} have revealed plumes of water vapor escaping into space from the icy worlds Europa and Enceladus \citep{sparks2017}, indicating the presence of subsurface oceans.  Similarly, Ganymede is believed to host a large underground saltwater ocean \citep{kivelson2002} and Titan is believed to have rainfalls and methane lakes \citep{stofan2007}.  Triton is an active moon with geysers driven by internal heat and Io is believed to be active volcanic worlds with geysers and lava flows \citep{soderblom1990,morabito1979}.  {\it Webb's} powerful infrared instruments are the next step in understanding the nature of these worlds \citep{keszthelyi2016}, and can help guide future missions to these exotic locations.

For Jupiter's moon Europa, {\it Webb's} mid-infrared spectroscopy with $\lambda$ = 5 -- 15~microns is an entirely new diagnostic to characterize the composition of the moon's surface and search for chemistry (e.g., hydrated minerals) expected from the ocean.  {\it Webb} data can test diagnostics of carbonic acid and other organic features across Europa's surface.  At near-infrared wavelengths, {\it Webb} can build high-resolution maps of the surface to search for ``hot spots'' that are indicative of active plumes and then rapidly follow up successful discoveries with spectroscopy to measure the composition and temperature.

For Saturn's moon Enceladus, {\it Webb} will complement {\it Cassini's} maps of the moon by obtaining spectroscopic observations for unique analysis of the surface features. Time-series observations during a plume outburst can infer the molecular composition of the subsurface ocean.  {\it Webb} can search for organic signatures, such as methane, ethane, and methanol, in the plumes using infrared spectroscopy (see Figure~\ref{Fig:Molecules}).

\begin{figure}
\centering
\includegraphics[width=9cm]{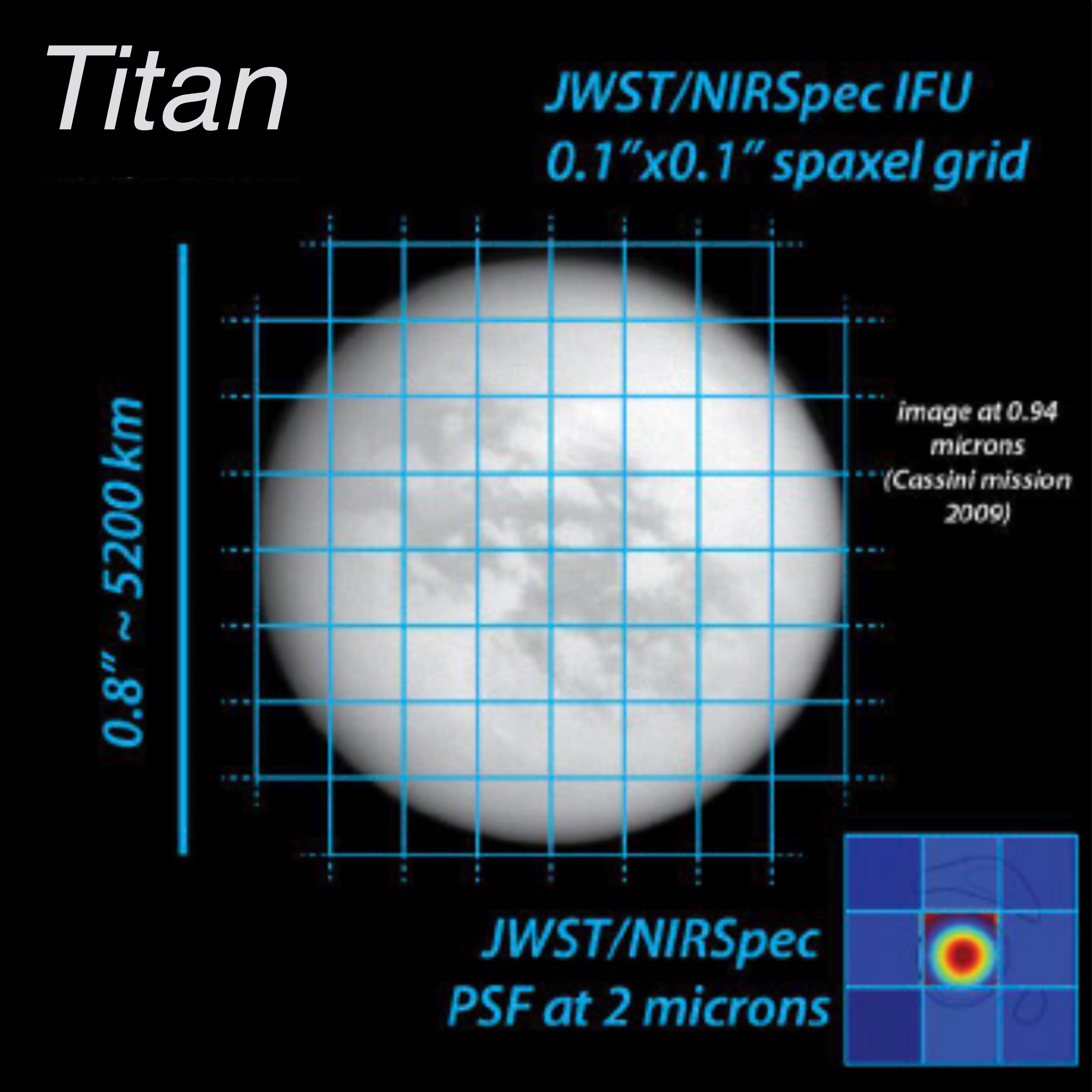} 
\caption{The resolved disk of Titan as viewed through a small part of the NIRSpec integral field unit and its 0.1 $\times$ 0.1~arcsec spatial scale.  The disk of Titan can be mapped with $>$50 pixels to obtain spatially-resolved spectroscopy in the near-infrared.  Long-term spectroscopic monitoring can establish the spatial distribution of molecules such as methane and the movement of clouds and hazes over seasonal time scales.  {\it Image Credit: Cassini-Huygens Mission 2009}} \label{Fig:IFUTitan}
\end{figure}

For Saturn's largest moon Titan, {\it Webb's} near-infrared spectroscopy will be much higher resolution than {\it Cassini} data, and, therefore, lead to a finer detection and analysis of surface organic molecules.  Specifically, the NIRSpec integral field unit can simultaneously give 3D spectroscopy over 50 spatial elements across the disk of the moon (see Figure~\ref{Fig:IFUTitan}).  Observations throughout {\it Webb's} 10-year mission goal can establish seasonal shifts and time-variable phenomena \citep{nixon2016}.  Longer wavelength imaging observations can also map the distribution of haze and clouds in the atmosphere across different seasons, and measure the temperatures and abundances of a number of gases.  These key data sets serve as input to test and refine dynamical models of Titan's atmosphere. {\it Webb} is very complementary to {\it Cassini}.  It provides more global coverage, better spectral resolution capable of identifying distinct molecular features compared to the VIMS instrument, and extends the timeline of data.


Jupiter's moon Io is active and exhibits periodic plume deposits due to volcanism.  The lava flows have temperatures ranging from 500 -- 2000~K, have diameters of hundreds of kilometers \citep{mcewen1998}, and have a cadence of several months.  {\it Webb's} aperture mask interferometry mode on NIRISS can spatially resolve the volcanoes \citep{thatte2016}, and multi-filter observations can constrain the specific temperatures of the lava flows.  These observations can yield the composition of the lavas and new insights on the state of Io's mantle \citep{keszthelyi2007}.

\subsubsection{Comets}

{\it Webb} can greatly expand our understanding of comets in the solar system by measuring their composition (e.g., gases and dust), coma and nuclei structure, and evolution over long temporal baselines.  As \cite{kelley2016} demonstrate, {\it Webb} can easily measure changes in water, carbon dioxide, and carbon monoxide in comets over different heliocentric distances ranging from 7~AU to 2~AU.  These differences relate to intrinsic cometary activity and the impact of solar energy in sublimating the ices for each species.  At mid-infrared wavelengths, {\it Webb} can study the heterogeneity of comet nuclei through high-resolution images and integral field unit spectral cubes.  For close approaches of $\lesssim$1~AU, {\it Webb} can observe features at the $\lesssim$200~km scale needed to probe distinct active regions.  These observations constrain theories of comet formation, for example, whether they are accreted planetesimals or collisional fragments \citep[e.g.,][]{weidenschilling1997}.  Mid-infrared observations with {\it Webb} can also establish the composition, temperature, grain-size distribution, and structure of the dusty comae and dust trails of comets \citep{rivkin2016}. {\it Webb} can also characterize water ice on comet surfaces and comae through low-resolution near-infrared spectroscopy with NIRSpec, detect activity driven by volatiles carbon monoxide and carbon dioxide at large heliocentric distances through sensitive near-infrared imaging with NIRCam, and measure the overall sizes, elongation, and albedo of comets \citep{kelley2016}.  Taken together, this research can assess the role that comets likely play in bringing water and other organics from the outskirts of the solar system to its inner regions (e.g., to Earth in our solar system).

\subsubsection{Kuiper Belt Objects and Dwarf Planets}

Large objects beyond the orbit of Neptune, such as Pluto, Eris, Makemake, Haumea, and Sedna, offer unique insights on the initial conditions and early evolution of the solar system.  Most of these objects retain almost all of the material from which they accreted, as evidenced by absorption features of volatiles such as methane and, on Pluto, Sedna and Eris, nitrogen \citep[e.g.,][]{emery2007,tegler2010}.  Much of this material is in the form of volatile ices that combine, via radiolysis, to form organic molecules.  Near-infrared spectroscopy at $\lambda$ = 3 -- 4~microns with {\it Webb} can characterize the molecular composition (nitrogen, carbon monoxide, carbon dioxide ices) of each of these objects in much greater detail than has been possible so far \citep[e.g.,][]{parker2016}, and connect the distribution of volatile ices to expected seasonal variations.

Mid-infrared thermal observations with {\it Webb} on objects such as Eris and Sedna can enhance the return from earlier {\it Spitzer} and {\it Herschel} data \citep[e.g.,][]{stansberry2008,lellouch2013} and contemporaneous {\it Atacama Large Millimeter Array} ({\it ALMA}) data \citep[e.g.,][]{lellouch2017}.  {\it Webb} can also characterize the thermal state of these Kuiper Belt dwarf planets as they go through seasonal changes.  For smaller Kuiper Belt Objects, Webb's mid-infrared photometry can provide the first thermal emission data for colder bodies that were too faint for {\it Spitzer} and {\it Herschel}.  Thermal data at two wavelengths directly measure the temperature, and therefore can be used to constrain diameters and albedos.  {\it Webb} can also, for the first time, resolve the mid-IR thermal emission from primaries and moons in the Kuiper belt. This can address many questions regarding these dwarf-planet systems, and is particularly synergistic with the even higher resolution capabilities of {\it ALMA}.

{\it Webb's} redder sensitivity and higher spatial resolution can improve upon {\it Hubble's} legacy of discovering rare binary (and even trinary) objects \citep{noll2008,parker2017} and measuring masses by characterizing the binary orbits. Combined with measurements of the diameters, determined from thermal observations or stellar occultations, the masses can be converted into bulk densities, which reflect the silicate to ice ratio, internal structure, and accretion conditions of these objects.

\subsubsection{Other {\it Webb} Opportunities for Solar System Research}

\cite{showalter2013} recently discovered the faintest moon of Neptune using {\it Hubble} imaging down to 26.5~mag.  This object is believed to have a diameter of 16--20~km.  {\it Webb} can extend this down to objects that are a few km in size at these distances, likely discovering large populations of small (faint) moons around giant planets.  A particular strength of {\it Webb} is the ability to image at wavelengths where strong methane absorption features in the atmospheres of the giant planets reduce their brightness by orders of magnitude, thereby increasing sensitivity to faint nearby objects.

{\it Webb's} resolution can provide exquisite separation and characterization of the rings of Uranus and Neptune as compared to current ground-based and {\it Hubble} studies, and {\it Webb} can also search for predicted rings around Mars, Pluto, and trans-Neptunian dwarf planets \citep{tiscareno2016}.  By measuring the composition of small moons and inner vs outer rings with molecular spectroscopy, {\it Webb} can help constrain models on the formation process and evolution of these structures.

For Saturn, {\it Webb's} contiguous near- and mid-infrared spectral coverage can fill in gaps in {\it Cassini} observations (e.g., at the 1.65~micron water absorption feature and between $\lambda$ = 5 -- 8~microns) and achieve higher sensitivity at similar spatial resolution.  {\it Webb} data can lead to the best maps of the rings in the thermal infrared to constrain temperature variations and also potentially discover silicate absorption features to understand aspects of the composition of the rings that are not water-ice based \citep{tiscareno2016}.

An exciting opportunity with {\it Webb} will be to enhance characterization of small bodies in the solar system by observing them during times of predicted occultation of background stars.  The technique offers unique data to reduce uncertainties in the derived sizes, shapes, and albedos of small bodies \citep{sicardy2011}, by taking advantage of the refraction, absorption, and diffraction of bright starlight as it interacts with the foreground object.  \cite{santos-sanz2016} predict 13 stellar occultation opportunities of minor bodies in the outer solar system during {\it Webb's} first five years, including Eris, Makemake, Quaoar, Sedna, Haumea, and others.  The technique can also be used to perform high-precision studies of the geometry and structure of planetary rings.

An excellent summary of {\it Webb's} broad applications for solar system research is available in a series of white papers \citep[see][and references therein]{milam2016}.

\subsection{Exoplanets}\label{exoplanets}

Since the discovery of the first Jupiter-mass object orbiting a Sun-like star in 1995 \citep{mayor1995}, the field of exoplanet research has exploded.  The census of confirmed exoplanets grew from tens in the early 2000s to hundreds by 2010 to several thousands today.  The {\it Kepler Space Telescope} has told us that essentially all stars have planets, and that the size distribution of planets is such that most of them are small objects between the radius of Earth and Neptune \citep{batalha2013}.

\begin{figure*}[t]
\centering
\includegraphics[width=14cm]{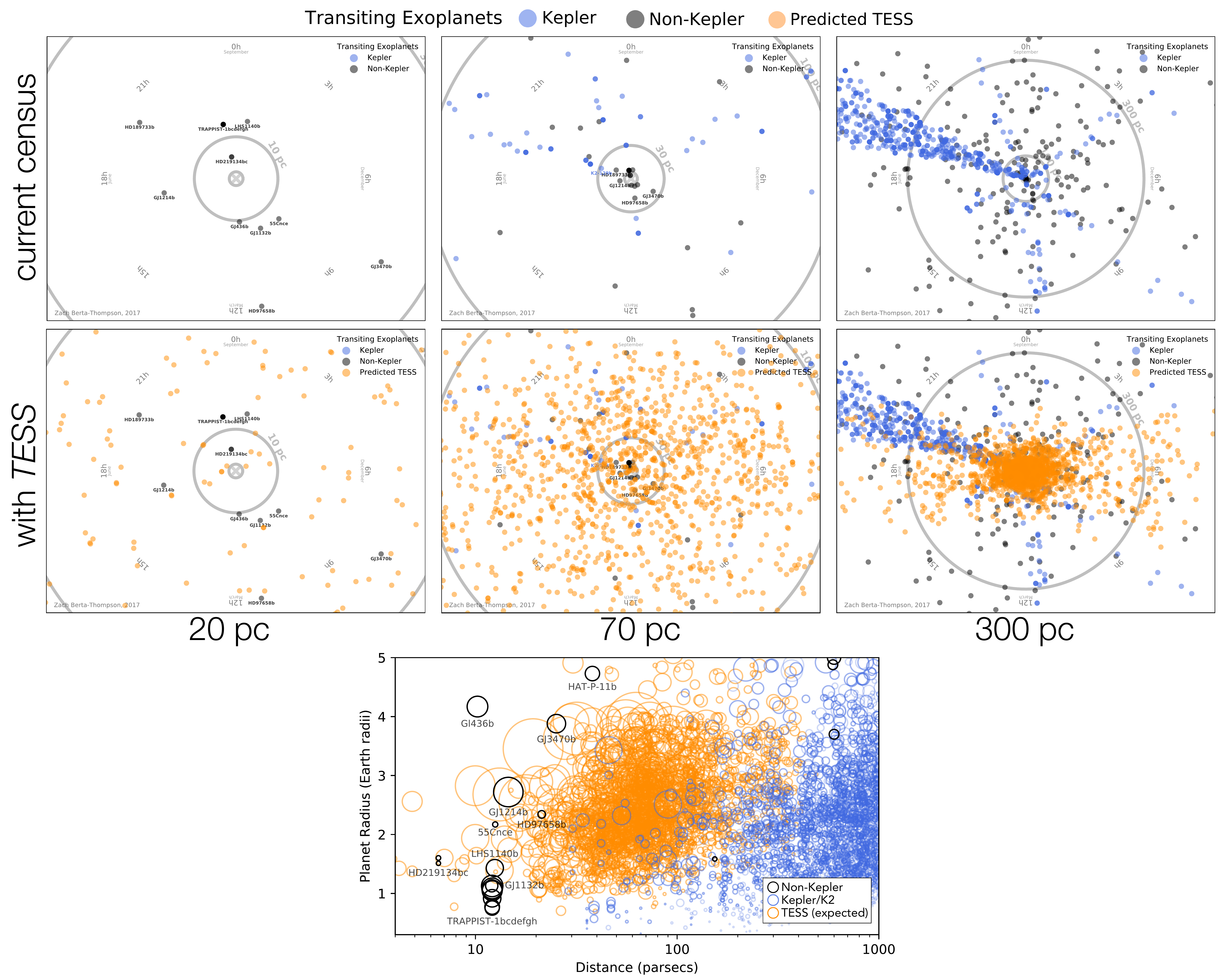} 
\caption{The three panels in the top row shows the current census of nearby transiting exoplanets and the middle row shows the expected census after the {\it TESS} mission, based on a simulation by \cite{sullivan2015}.  At nearby distances, the census of transiting exoplanets will be dominated by the 20,000 expected {\it TESS} discoveries (orange points). The blue points in the ``cone'' are {\it Kepler} planets.  The panel at the bottom shows the size-distance relation for these planets and the current census. The symbol {\it areas} are proportional to the transit depth.  Many of the nearby {\it TESS} discoveries will be bright (indicated by the size of the circle) and produce higher signals for {\it Webb} follow up. {\it Image Credit: Z.\ Berta-Thompson}} \label{Fig:TESSPlanetYield}
\end{figure*}

Over the past 20 years, exoplanet research has generally focused on discovery.  A multitude of techniques, such as radial velocity, transit photometry, direct imaging, microlensing, astrometry, and others have been used on ground and space-based telescopes to discover exoplanets orbiting other stars, and to begin inferring their bulk properties such as mass and radius and their orbital properties such as period, eccentricity, and inclination \citep[e.g.,][]{winn2015}.  These experiments have been highly successful, but have also been optimized in ways to maximize statistical metrics for studying planetary demographics.  As a result, our census of interesting planets for follow up detailed study, such as nearby rocky exoplanets orbiting bright stars, is highly incomplete.  The soon to launch {\it Transiting Exoplanet Survey Satellite} ({\it TESS}) will change this, by monitoring more than 200,000 nearby stars in an all-sky survey and discovering $>$300 planets with radius less than twice that of the Earth (i.e., super Earths) and 1000 nearby planets with radius less than four times that of the Earth \citep{ricker2015,sullivan2015}.  A schematic showing the expected census of nearby exoplanets in the {\it TESS} era, and their sizes, is shown in Figure~\ref{Fig:TESSPlanetYield}.

Both the {\it Hubble} and {\it Spitzer Space Telescopes}, as well as 6--10-m class ground-based telescopes such as {\it Magellan}, {\it VLT}, {\it Gemini}, {\it Keck}, {\it Gran Telescopio Canarias}, and others have begun to extend exoplanet research from the discovery phase to the characterization phase.  This breakthrough has been largely enabled by taking advantage of the geometry of systems where planets pass in front of and behind their host stars as seen from the observed line of sight \citep[e.g.,][]{seager2010}.  During the former (i.e., transit) phase, the light from the background star filters through the planet's atmosphere and is imprinted upon with spectral features of the planet's atmospheric composition.  The spectrum of the planet is simply the ratio of the combined spectrum during the transit to the spectrum of the star (either before or after transit).  Transit observations yield the planetary radius (measuring the drop in the amount of stellar flux during the transit) and the atmosphere's opacity (measuring the planet's apparent size at different wavelengths).  Together, these measurements yield atmospheric properties of the atmosphere (chemical composition, temperature structure, presence of clouds).  During the latter phase when the planet is about to pass behind the star (i.e., secondary eclipse), the total star and planet (thermal and reflected) light is measured.  The spectrum of the planet during this phase is simply the ratio of the stellar spectrum (during the eclipse) from that of the star-planet system (i.e., before or after the eclipse).  Secondary eclipse observations give information on the planet's atmospheric composition, temperature, and albedo. In between primary and secondary transit, the planet can be seen in different phases depending on its orbit and the phase curve can reveal the dynamics and thermal and chemical gradients in the planet's atmosphere.


The first atmospheric composition measurements of exoplanets through transit, eclipse, and phase curve techniques with {\it Hubble} spectroscopy and {\it Spitzer} photometry are now more than a decade old \citep[e.g.,][]{charbonneau2002,charbonneau2005,deming2005}.  The early studies successfully characterized sodium, water, and other features in the atmospheres of larger exoplanets, revealed winds, and also showed that spectral signatures at visible and near-infrared wavelengths may be diminished due to the presence of clouds and hazes \citep{knutson2007,sing2016}.  Recent {\it Hubble} and {\it Spitzer} observations have continued to take advantage of the transit, eclipse, and phase curve methods to yield initial constraints (e.g., presence of water vapor) on the atmospheres of larger exoplanets such as hot Jupiters \citep[e.g.,][]{deming2013}, and have also explored the spectra of some super Earths, which do not exhibit any features possibly due to clouds and/or hazes or not enough spectral precision \citep[e.g.,][]{kreidberg2014b}.

The next major frontier in exoplanet research will be brought to light by {\it Webb}, which is designed to measure high-precision atmospheric and orbital properties of exoplanets ranging from gas giants to Neptunes to the characterization of super Earths in the habitable zone of small, cool host stars \citep[e.g.,][and others]{beichman2014,cowan2015,stevenson2016,greene2016b,batalha2017}.  {\it Webb's} sensitivity to many molecules in the infrared spectrum also opens the door for challenging experiments to search for biosignature gases on nearby exoplanets in extremely favorable configurations \citep{seager2016}.  {\it Webb} can also make major advances to other themes of exoplanet research. For example, direct imaging with coronagraphy can enhance our understanding of the planet formation process by enabling exoplanet discoveries and imaging of newborn planets and disks.


\begin{figure}
\centering
\includegraphics[width=13cm, trim={0cm 0.2cm 0cm 0.5cm},clip]{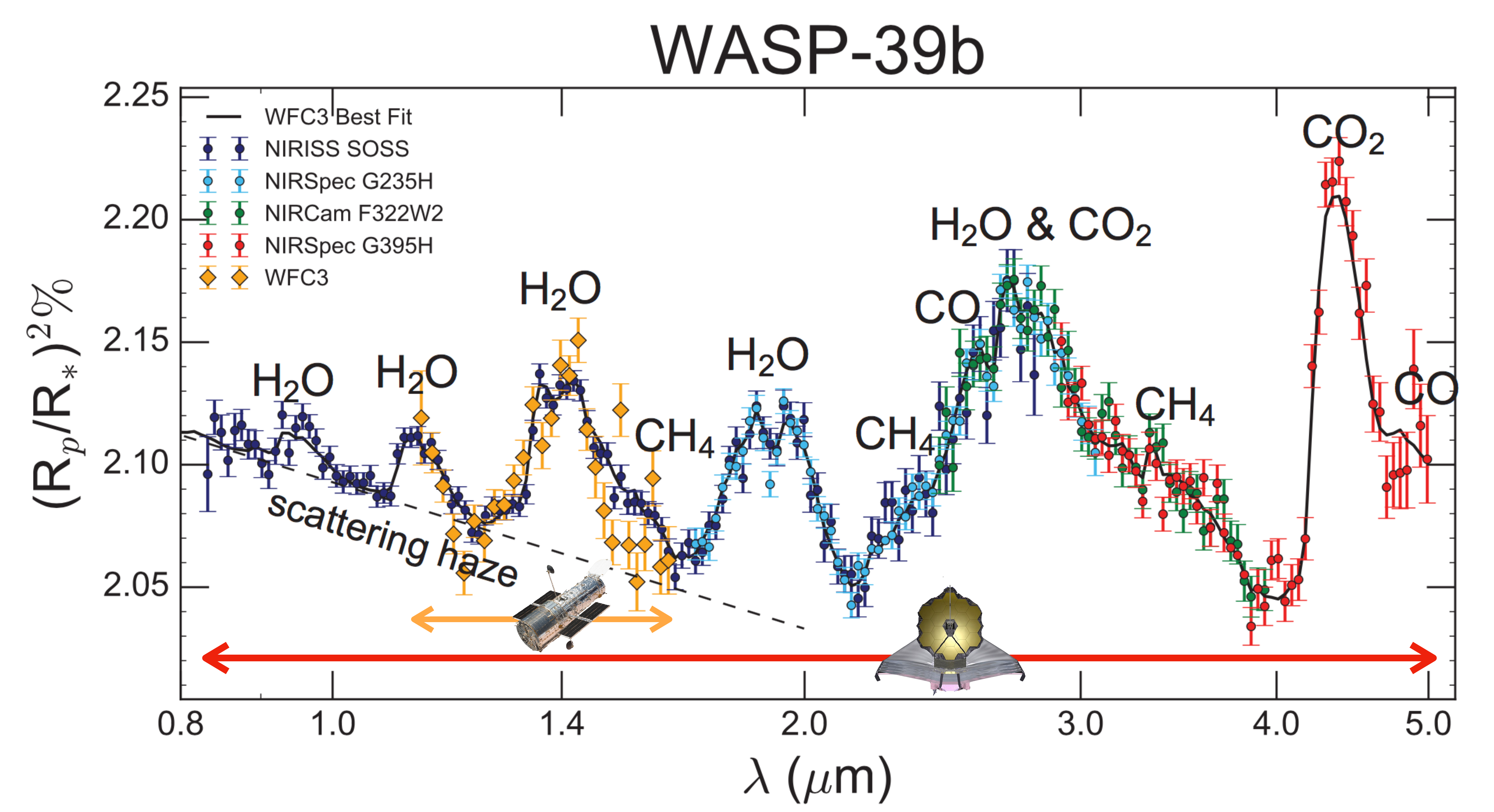} 
\caption{{\it Left} -- A simulated {\it Webb} spectrum using PandExo \citep{batalha2017} of the Saturn-mass planet WASP-39b from 0.8 to 5.0~microns using NIRISS, NIRSpec, and NIRCam.  The exquisite quality of the spectrum allows easy characterization of all of the molecular features across a much broader range than {\it Hubble} WFC3/G141 grism data (orange points; $\lambda$ = 1.1 -- 1.7~microns).  These future results can be enhanced further by combining the {\it Webb} infrared data with visible spectroscopy from {\it Hubble} and/or large ground-based telescopes \citep{sing2016,nikolov2016,wakeford2017}}. \label{Fig:WASP39b}
\end{figure}


\subsubsection{Transit and Eclipse Spectroscopy and Photometry}

The atmospheres of exoplanets contain imprints from many physical processes \citep[e.g.,][]{seager2010,fortney2013,madhusudhan2014,crossfield2015}; planetary formation (e.g., metallicity content), geology (e.g., exchanges and mixing with the surface), climate (e.g., weather patterns and clouds), environment (e.g., impact of stellar irradiation through pressure, density, and compositions), and potential habitability (e.g., signatures of biological processes).  However, disentangling physical effects from spectroscopic data has proven difficult due to the extremely limited wavelength range and spectral resolution of {\it Hubble} and {\it Spitzer}, which were not originally designed for such observations.

{\it Webb} will revolutionize exoplanet atmospheric research by enabling access to a multitude of absorption features in the infrared part of the spectrum, including water vapor, methane, carbon dioxide, carbon monoxide, ammonia, sodium, potassium, and others.  The spectral resolution will be an order of magnitude better than previous infrared studies of molecules, and the sensitivity beyond {\it Hubble's} $\lambda$ = 1.6~micron wavelength cutoff all the way out to 5~microns in the near-infrared (12~microns with slit spectroscopy in the mid-infrared) can help lift degeneracies and assumptions that plague current studies (e.g., haze properties, equilibrium chemistries, thermal structures). As one example, Figure~\ref{Fig:WASP39b} shows a simulated {\it Webb} spectrum of the Saturn-mass planet WASP-39b.  With such high-quality data, unknown atmospheric properties can be marginalized to derive constraints on the carbon-to-oxygen abundance ratio, metallicity, and other properties to over an order of magnitude better precision than with current infrared-only data \citep[e.g., see][]{greene2016b}.  In the {\it Webb} era, it is likely that such detailed information about exoplanet atmospheres for gas and ice giant planets will become routine.  This opens the door for extensive surveys to assess the demographics of planetary atmospheres and their relation with stellar environments, and also to relate measured metallicities of planets to formation scenarios \citep[e.g.,][]{fortney2013}.  The wide wavelength range of {\it Webb} can also characterize the general spectral shape and probe energy redistribution in the atmosphere.  More ambitious {\it Webb} observations can measure detailed atmospheric properties of smaller rocky exoplanets, thereby launching an entirely new area of research.  These projects will especially benefit from the discovery of nearby rocky exoplanets in favorable conditions from missions like {\it TESS}.


\begin{figure}[t]
\centering
\includegraphics[width=14cm]{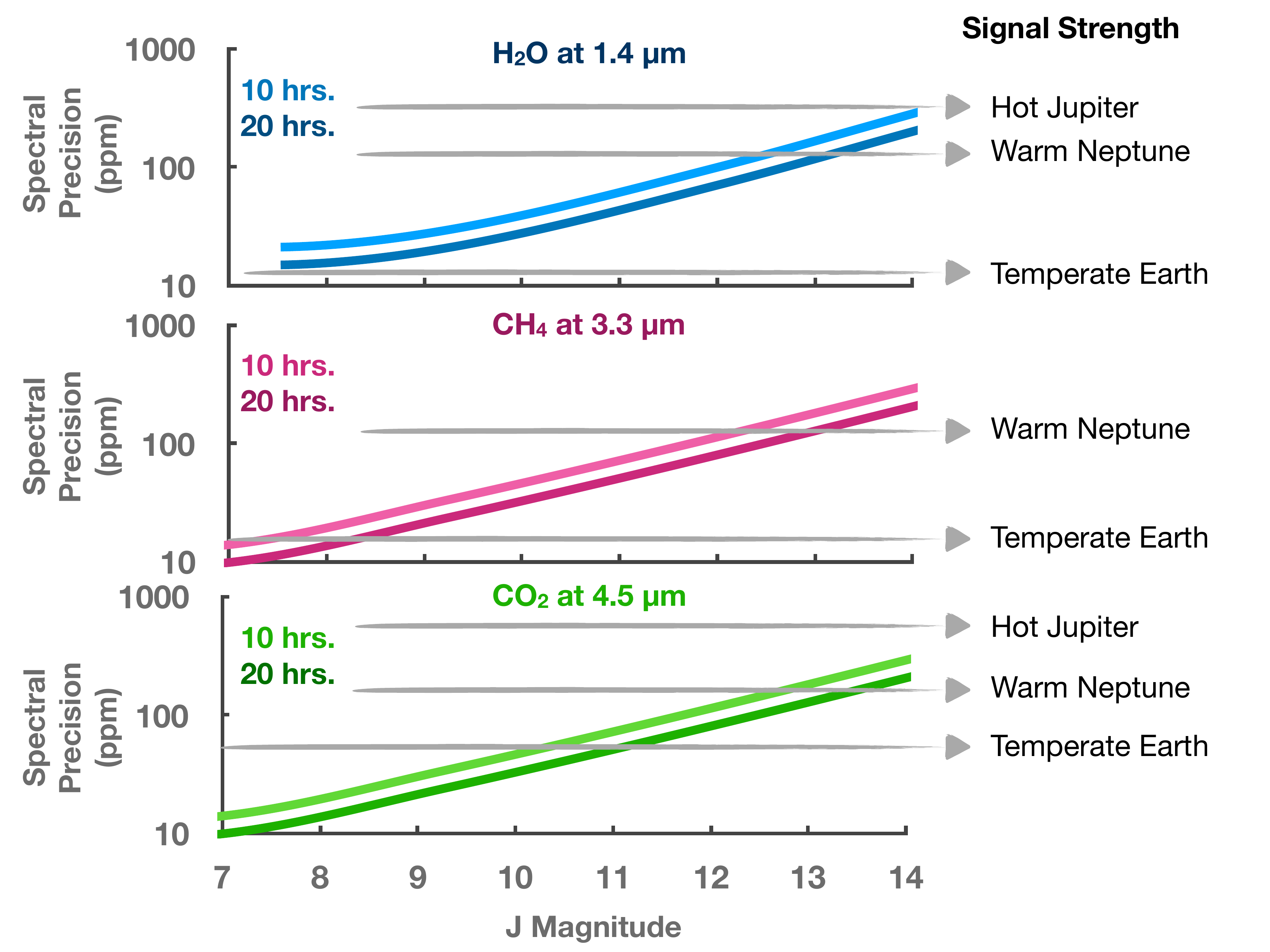} 
\caption{Each panel shows the expected precision that {\it Webb} can achieve on an exoplanetary transmission spectra as a function of stellar $J$ magnitude, down to faint limits of $J <$ 14 (i.e., above which {\it TESS} is expected to find {\it thousands} of exoplanets). Curves are computed with the {\it Webb} simulator, PandExo \citep{batalha2017} for either a 10 or 20 hour exposure, assuming no noise floor. The top panel shows the precision at 1.4~microns (location of water) with the NIRISS instrument's single object slitless spectroscopy mode \citep[see also][]{louie2017}. The middle panel shows the precision at 3.3~microns (location of methane) with NIRCam's long wavelength grism mode. The bottom panel shows the precision at 4.5~microns (location of carbon dioxide) with the NIRSpec grism. For comparison, all calculations were binned to a resolving power of $R$ = 100. The grey lines in each panel shows the typical molecular signal strengths for a hot Jupiter, warm Neptune, and a temperate (true) Earth analog orbiting an M dwarf (see Figure~\ref{Fig:JWSTPandExoSim2WithInsetonepage}). {\it Image Credit: N.\ Batalha}} \label{Fig:JWSTPandExoSim1}
\end{figure}


Beyond static measurements of atmospheres, {\it Webb's} increased efficiency can open a new dimension in temporal characterization of exoplanets.  Time-resolved spectroscopy tuned to different phases of an exoplanet as it orbits its host star can yield new insights on the global star-planet system by measuring both the day and night-side properties \citep[e.g.,][]{stevenson2014}.  For example, {\it Webb} data could establish correlations between stellar irradiance and changes in planetary winds, clouds, temperature structure, and chemistry.  Such studies can also obtain unprecedented spectroscopic data during the secondary eclipse phase, where current capabilities are largely photometric in nature (or very limited spectroscopy).  These observations are ideally suited to constrain the atmospheric structure (e.g., temperature-pressure profile), circulation (e.g., the large-scale movement of gas throughout the atmosphere), and assess the global energy budget \citep[e.g.,][]{deming2006,stevenson2010,stevenson2014}.  {\it Webb's} exquisite spectroscopic sensitivity at mid-infrared wavelengths from $\lambda$ = 5 -- 12~microns ($>$2 orders of magnitude better than {\it Spitzer}) will be especially important to map the 3D thermal structure and molecular composition of exoplanets.  For larger planets on eccentric orbits, phase mapping can provide high-precision tests to planetary models by constraining radiative, dynamical, and chemical processes in atmospheres \citep{showman2009,laughlin2009,lewis2017}.

\begin{figure}[t]
\centering
\includegraphics[width=13cm]{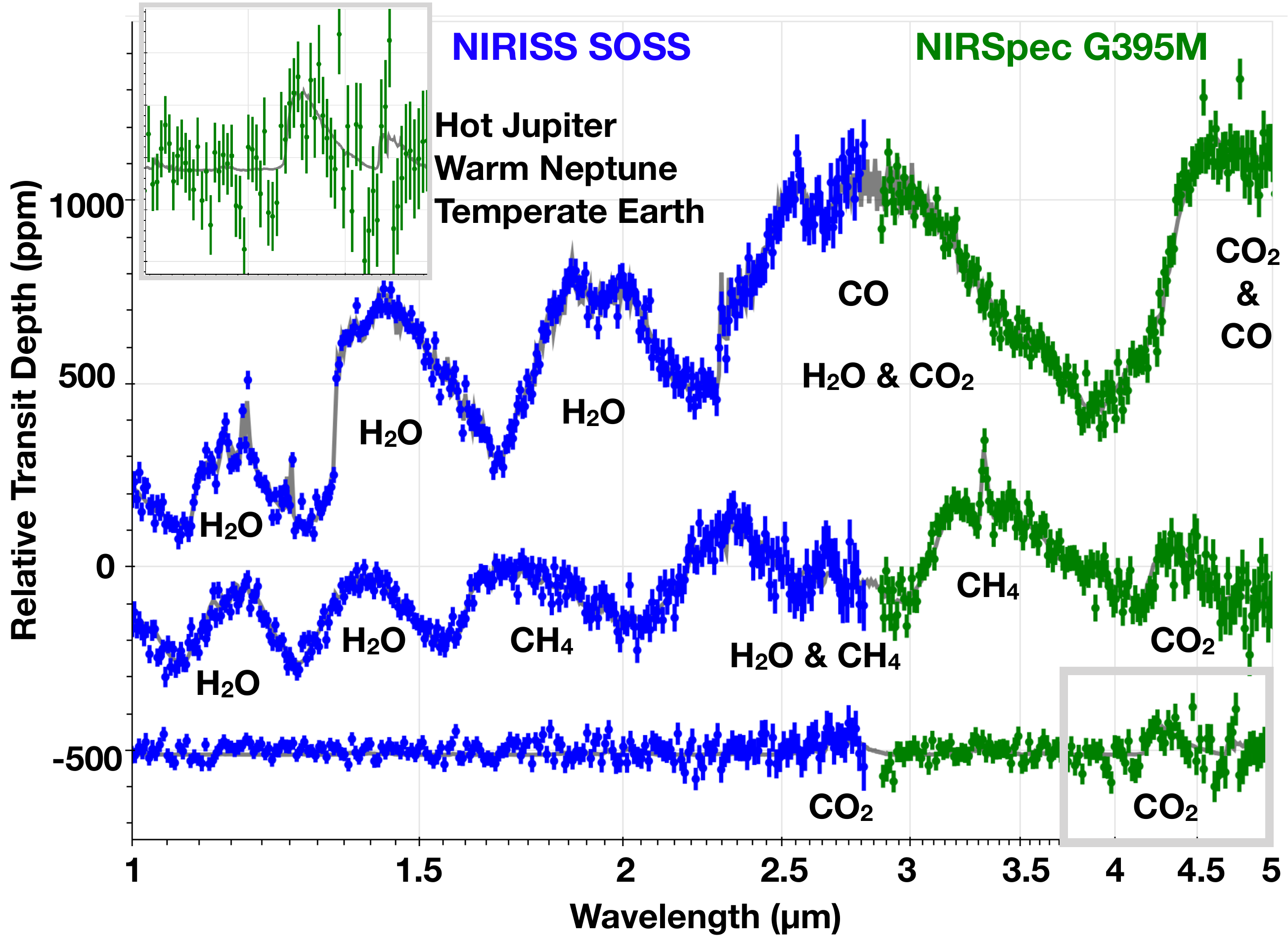} 
\caption{Simulated {\it Webb} observations of the transmission spectrum of a hot Jupiter (top), warm Neptune (middle) and temperature Earth (bottom) using PandExo \citep{batalha2017}.  Molecular features are labeled for reference.  All three cases assume a stellar $J$ magnitude = 8. The blue and green points show a 10 hour observation each with the NIRISS instrument's single object slitless spectroscopy mode and with NIRSpec's G395M grism, respectively. The hot Jupiter was computed using a solar metallicity, solar C/O model of WASP-62 from \cite{goyal2017}. The warm Neptune was computed using a 100$\times$ solar metallicity, solar C/O model of GJ-436b also from \cite{goyal2017}. The temperate Earth (orbiting an M dwarf) was computed by scaling an Earth spectrum \citep{robinson2011} to TRAPPIST-1e.  The carbon dioxide feature at 4.5~microns is blown up in the inset panel in the top-left corner of the Figure, and could be measured by{\it Webb} in this planet which has the same atmosphere and density as Earth. {\it Image Credit: N.\ Batalha}} \label{Fig:JWSTPandExoSim2WithInsetonepage}
\end{figure}

Simulations from \cite{batalha2017} in Figure~\ref{Fig:JWSTPandExoSim1} summarize the incredible ``speed'' of {\it Webb} for exoplanet atmosphere characterization.  At host star magnitudes as faint as $J \sim$ 13, a limit above which {\it TESS} will discover {\it thousands} of exoplanets, {\it Webb} can achieve 100 parts per million spectral precision for water vapor, methane, and carbon dioxide in just 10 -- 20~hours of total integration.  As transit events are discrete, this will involve multiple independent observations during the transiting phases (e.g., 3 -- 4 transits) and stacking of the resulting data.  This opens the door for quick and accurate characterization of hot Jupiters and warm Neptunes at these magnitudes.  At brighter magnitudes of $J <$ 11, {\it Webb} can achieve several 10s of parts per million precision for carbon dioxide measurement of temperate Earths (orbiting M dwarfs) in the same integration time.  Example simulated spectra from combining {\it Webb} NIRISS and NIRSpec data for a hot Jupiter, warm Neptune, and temperate Earth are shown in Figure~\ref{Fig:JWSTPandExoSim2WithInsetonepage}.  Other simulations in these and different instrument modes are presented in \cite{clampin2009,barstow2016,greene2016b,batalha2017,louie2017}.


\subsubsection{Direct Imaging}

Although extremely valuable, transit spectroscopy and photometry can only be applied to a small fraction of planets due to viewing geometry and the infrequent cadence of transits for systems with higher-mass stars.  Direct imaging observations of exoplanets are a complementary technique to study planetary atmospheres, and, like secondary eclipse and phase curve measurements, probe deeper into the planetary atmospheres than tranmission spectroscopy.  Such data can achieve higher S/N of interesting features \citep{morley2015,biller2017}.

\begin{figure}
\centering
\includegraphics[width=12cm]{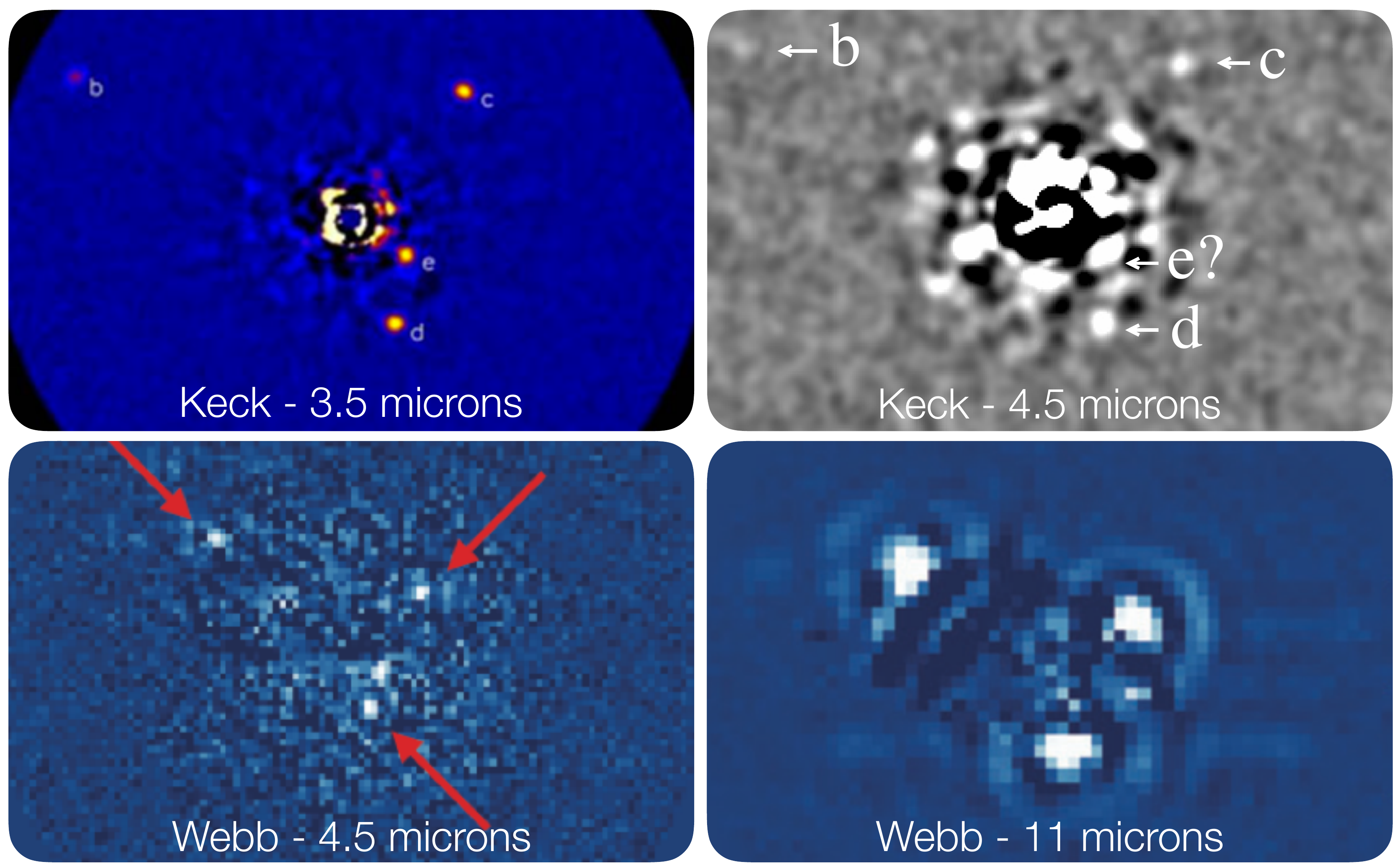} 
\caption{The top-left panel shows high-contrast imaging of the young main-sequence star HR8799 with the {\it Keck} 10-meter telescope. The 3.5~micron imaging reveals four massive exoplanets, labelled b, c, d, and e \citep{marois2008,marois2010}.  The top-right panel shows the same system observed at 4.5~microns \citep{galicher2011}.  At the longer wavelengths, the background due to the temperature of the telescope and Earth's atmosphere increases, and the planets become difficult to detect.  The bottom panels show simulated {\it Webb} coronagraphy observations of this system with NIRCam (left) and MIRI (right) \citep[see][]{lajoie2016}.  At the same wavelength and contrast measured by {\it Keck} (10$^{-4}$), the {\it Webb} detection of planet d has twice the signal-to-noise as the {\it Keck} observations (bottom-left).  The MIRI simulation at 11~microns (bottom-right) was made using the best fit models in \cite{marley2012}.  These observations are impossible from the ground due to the Earth's thermal background, and highlight the new discovery space that {\it Webb} will enable for exoplanet discovery.} \label{Fig:Coronagraphy}
\end{figure}

As described in Section~\ref{instruments}, {\it Webb's} instruments include coronagraphy modes at several wavelengths ranging from 1.7 to 22.75~microns.  These modes enable direct detection and characterization of exoplanets and allow for the study of their disk environments in the early formation process.  While coronagraphs on large ground-based telescopes can achieve smaller apparent separations at the same planet-to-star contrast ratio, {\it Webb} offers unique opportunities to image planets and debris disks at longer infrared wavelengths.  Here, sky backgrounds from the ground are high and so there are currently no directly observed exoplanets (i.e., at $\lambda$ $>$ 5~microns).  A simulation of the nearby HR8799 system, which contains four exoplanets discovered by high-contrast {\it Keck} imaging, is shown in Figure~\ref{Fig:Coronagraphy} \citep{lajoie2016}.  The planets are easily seen with both the NIRCam coronagraph at 4.5~microns and with MIRI at 11~microns. 


At near-infrared wavelengths, \cite{beichman2010} show that {\it Webb} can image planets with one-tenth Jupiter mass and separations of 50 AU in favorable cases.  More massive planets near Jupiter-mass can be seen orbiting M dwarfs out to a few tens of AU at mid-infrared wavelengths.  Direct imaging surveys with {\it Webb} are likely to characterize the distribution of planets as functions of mass and orbital distance in this regime and help constrain giant planet formation models \citep[e.g., ``core accretion'' vs ``hot start'';][]{marley2007,fortney2008}.  Specifically, {\it Webb} observations can measure both the carbon to hydrogen and carbon to oxygen ratio to characterize formation zones within planetary disks and test migration scenarios \citep[e.g.,][]{mordasini2009,oberg2011}

For debris disks (i.e., collisional material formed from early planetesimals, asteroids, and comets), mid-infrared imaging from {\it Webb} can spatially resolve thermal emission generated from dust in systems ranging in mass from brown dwarfs to solar-mass stars.  These maps are powerful clues to infer the presence of planets \citep[e.g., Beta Pic and Fomalhaut discoveries;][]{heap2000,kalas2005}, and to constrain theories of the middle to late formation process for terrestrial planets.  {\it Webb} can spatially resolve a large number of debris disks and measure their sizes and spectral energy distributions \citep{smith2010}, and help usher in a new generation of models that can distinguish grain properties (such as size, composition, and shape) in these systems.  {\it Webb's} study of debris disks in the mid-infrared is very synergistic with submillimeter maps of the thermal emission in the same systems from telescopes such as {\it ALMA}.  Taken together, these measurements can yield the albedo of the disk and the composition of the dust.  Related information on this research theme is described in \S\,\ref{starformation}.

\subsubsection{Summary of {\it Webb} Opportunities for Exoplanet Research}\label{summaryexoplanet}

{\it Webb's} capabilities for studying exoplanets include many methods such as transit light curves, phase light curves, transmission spectroscopy, emission spectroscopy, and direct imaging, and the targets include gas giants, intermediate mass objects, super Earths, and terrestrial planets \citep{clampin2009}.  The combination of these applications and targets make possible a multitude of science investigations ranging from physical structure studies, day-to-night emission mapping, spectral band diagnostics, atmospheric compositions, temperature measurements, and more.  An excellent quantitative summary of the major opportunities for advancing transiting exoplanet research with {\it Webb} is given in \cite{greene2016b}.


\subsection{Milky Way Stellar Populations and Nearby Galaxies}\label{stellarpops}

Astronomy is largely a study of the Universe through observations of light, and that light is dominated by populations of stars with different characteristics including evolutionary state, mass, and chemical composition.  It is only for nearby systems that we can resolve the individual stars in stellar populations and subject them to detailed analysis.  It is also for these populations that we can directly observe and characterize the formation environments (e.g., molecular clouds) and end products (e.g., planetary nebulae, supernova remnants) that complete the stellar lifecycle.  In this way, our understanding of light from different stellar populations in distant galaxies in the Universe is anchored on the detailed study of resolved stars in the Milky Way and Local Volume galaxies.

One of the most important diagnostic tools for stellar astrophysics research is the color-magnitude diagram (CMD).  At fixed age and metallicity, the relation between a star's color and magnitude represents the foundation upon which stellar models are tested and calibrated. In turn, these models are applied to understand resolved stellar populations in the Milky Way and nearby galaxies, and also serve as crucial ingredients to population synthesis models used to measure complex star formation and abundance histories in distant galaxies \citep[e.g.,][]{bruzual2003}.

Today, the gold standard for establishing benchmark color-magnitude relations is high-precision photometry of Milky Way star clusters and other co-spatial populations with {\it Hubble} \citep[e.g.,][]{sarajedini2007,dalcanton2009,dalcanton2012}.  {\it Webb} offers significant gains in general technical capabilities that will positively impact discovery and characterization of stellar populations through the infrared CMD.  For example, the better sensitivity and stability of {\it Webb} enables high-precision photometry across the full spectrum of stellar masses (from failed stars such as brown dwarfs to massive post-main sequence stars) and evolutionary stages (from newborn pre main sequence stars to dead white dwarfs).  By easily reaching 30th AB mag, {\it Webb} can image faint low-mass dwarfs out to the edge of the Milky Way stellar halo, and image brighter populations in more distant galaxies with different environments than the Local Group.  The higher resolution of {\it Webb} over current instruments can provide exquiste infrared image separation to overcome crowding in nearby stellar populations and enhance dynamical studies based on proper motions.  This also helps discriminate blended and confused sources to greater depths in more distant galaxies.  The significantly higher spectral resolving power of {\it Webb} (up to $R \sim$ 3000) over {\it Hubble} and {\it Spitzer} ($R <$ 200 at near-infrared and R $\sim$ 600 at mid-infrared wavelengths) can achieve better characterization of spectral energy distributions, abundances, and kinematics.  The broad infrared wavelength sensitivity of {\it Webb} allows penetration through extincted regions in the Milky Way plane (e.g., star forming environments) and the bulges of nearby galaxies for deeper studies of the luminosity function and improved characterization of the dust properties around newborn and evolved stars in the mid infrared.  The multiplexed spectroscopy of {\it Webb} enables the first population studies of chemical and dynamical stellar properties at high spatial resolution.

\subsubsection{The Formation of Stars}\label{starformation}

Space-based observatories with infrared sensitivity such as {\it Spitzer}, {\it Herschel}, and {\it WISE}, as well as ground based sub-millimeter telescopes, have established complete inventories and demographics of protostars out to several kpc in the Milky Way \citep[e.g.,][]{dunham2014}, and detected the first mid-infrared molecular features in individual protoplanetary disks \citep{carr2008,salyk2008,pontoppidan2010}.  However, many key questions on the detailed physics of the star formation process remain to be answered and require the full application of infrared diagnostics.  {\it Webb} can enable this at mid-infrared wavelengths that are not observable from the ground, by exceeding {\it Spitzer} (and all previous) capabilities by two orders of magnitude in sensitivity and several times higher spectral resolution.  Integral field spectroscopic data can resolve disks and envelopes and give tremendous insights into the physical structure of the warm dust, the geometry of the systems, the level of fragmentation in disks and cores, the properties (abundance, temperature, distribution) of the molecules, and much more.  {\it Webb} can resolve many infrared molecular features to map the abundance and spatial distribution of volatiles such as water and other organic materials \citep{pontoppidan2009}, and perform such analysis of disks spanning the entire stellar and substellar mass range within several kpc.  Key questions that {\it Webb} is poised to answer relate to the flow of matter from the inner core regions to the disk to the star, the origin and distribution of different chemicals, and the level of feedback of protostars on their surroundings.  {\it Webb} can also provide the data needed to understand how these overall characteristics change with stellar mass and evolutionary stage.  The study of star formation through {\it Webb} is highly complementary to observations of disks from the {\it ALMA}, which offers similar angular resolution and can survey the cold outer parts of young stellar systems.

On larger scales, {\it Webb} can assess how star formation impacts its broader environment.  In massive star forming environments of the Milky Way and Large and Small Magellanic Clouds \citep[such as the Tarantula Nebula; see][]{sabbi2016}, {\it Webb} can build large-scale and high-resolution mosaics of the gas and dust structure and catalog the huge populations of newborn stars embedded within the nebula.  With {\it Hubble} we are only able to see these stars after they have accreted most of their mass, and so we are missing a crucial piece of information.  {\it Webb} can open this window.  Specific diagnostics, including narrow and medium-band imaging, can isolate different chemical species while broadband imaging at different wavelengths can both penetrate and illuminate the dust content.  These observations can track the interplay of shocks, outflows, and cavities in a wide range of environments.  The study of star formation and its environments with {\it Webb} will be highly complementary to future high-resolution infrared surveys with missions such as {\it Euclid} and the {\it Wide Field Infrared Survey Telescope} ({\it WFIRST}).

\subsubsection{The Distribution of Stellar Masses}

\begin{figure*}[t]
\centering
\includegraphics[width=13cm, trim={1.2cm 1.5cm 0 0},clip]{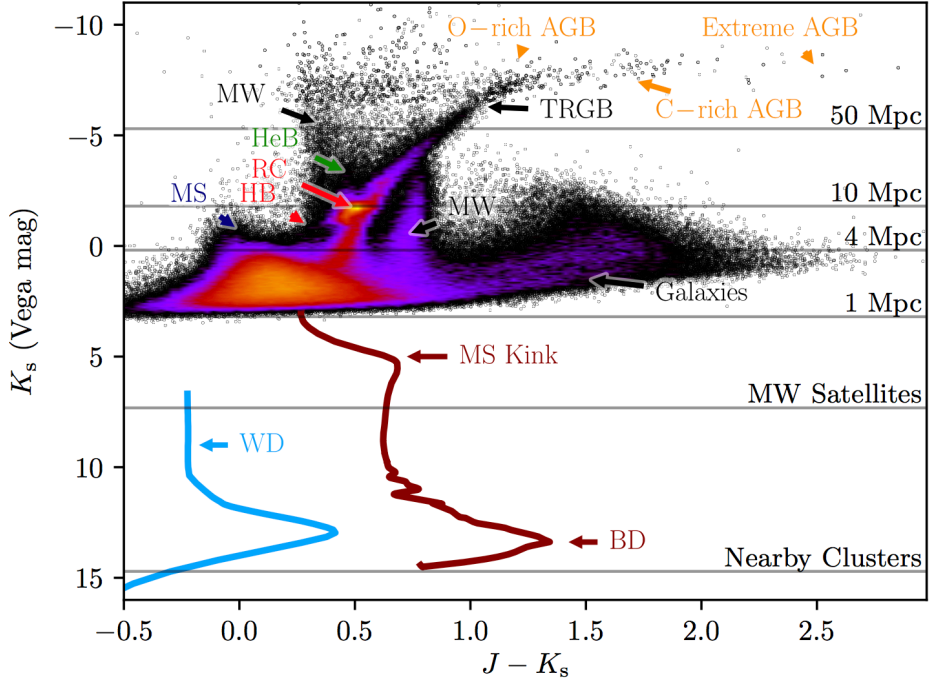} 
\caption{A mix of stellar populations in the near-infrared CMD, stretching from massive main-sequence (MS) stars and bright dust-enshrouded red giant branch (TRGB) and asymptotic giant branch (AGB) star phases down to lower mass objects.  This includes the lower main-sequence down to the expected location of brown dwarfs (BD) as a red sequence \citep{caiazzo2017,baraffe2015} and dead white dwarfs (WD) as a blue sequence \citep{fontaine2001}.  The observed stellar populations at $K_{s}$ $<$ 3 come from the VISTA Magellanic Cloud Survey and 2MASS \citep{cioni2011} and are color coded by typical stellar age, with redder labels indicating older features.  Horizontal grey bands mark the distance out to which {\it Webb} can measure the various populations in the CMD in ambitious programs (i.e., the tip of the red-giant branch in galaxies out to 50~Mpc, the old main-sequence turnoff in Sculptor Group galaxies at a few Mpc, M dwarfs below the MS kink out to the edge of the stellar halo of the Milky Way at 150~kpc (i.e., grey band labelled ``MW Satellites''), and the brightest brown dwarfs and faintest white dwarfs in globular clusters out to 5~kpc (i.e., grey band labelled ``Nearby Clusters'').  Other labels are Milky Way field stars (MW), helium burning stars (HeB), red clump (RC), horizontal branch (HB), background galaxies (Galaxies). {\it Image Credit: P.\ Rosenfield, D.\ Weisz, B.\ Williams, L.\ Girardi}}\label{Fig:StellarPopsWebbCMD}
\end{figure*}

The output of the star formation process is the distribution of masses forming in galaxies.  {\it Webb} can make the most precise measurements of this ``initial mass function'' \citep[e.g.,][]{bastian2010} to date, and test its uniformity across a wide spectrum of ages, chemical abundances, and interstellar medium densities and pressures.  In young nearby star-forming regions such as Orion, {\it Webb's} infrared sensitivity enables detection of free-floating Jupiter-mass objects.  Surveys with {\it Webb} can characterize the complete stellar-planetary census in star forming regions, including the study of thousands of disks in a variety of environments and evolutionary states.  {\it Webb} can also map the stellar main-sequence down to near the hydrogen burning limit in any Milky Way dwarf galaxy (see Figure~\ref{Fig:StellarPopsWebbCMD}).  These types of measurements can test for small variations in the initial mass function in systems that vary by over several orders of magnitude in age, metallicity, and environment \citep[e.g.,][]{kalirai2013,geha2013}.


\subsubsection{Resolving the Chronology of Galaxy Growth}

The halo of the Milky Way and other nearby galaxies are populated with some of the oldest generations of stars in the Universe.  Establishing high-precision ages for these stars is a major pursuit of stellar astrophysics, as it sets up a powerful and independent technique to measure the timescale of baryonic structure formation in the Universe and to test simulations of galaxy formation at high-redshift.  Most age constraints of the halo involve reconstructing features such as the main-sequence turnoff on the color-magnitude diagram of halo sightlines, either through shallow wide-field studies from ground-based surveys such as the Sloan Digital Sky Survey \citep[e.g.,][]{jofre2011} or through pencil-beam deep exposures of globular star clusters and ultra-faint dwarf galaxies \citep[e.g.,][]{dotter2010,brown2006,brown2014,weisz2014}.  These studies, largely based on {\it Hubble} observations of visible-light color-magnitude diagrams, have established old metal-poor populations as forming most of their stars 11 -- 13~Gyr ago.  However, the uncertainty on these measurements is still at the 10\% level, and largely driven by ``fitting'' uncertainties from taking models and applying them to the observational frame \citep{chaboyer2008}.

{\it Webb} can enable multiple breakthroughs in establishing cosmologically interesting ages for nearby stellar populations, based on fully exploiting the infrared color-magnitude diagram and its rich features.  For example, the dramatic inversion of the color-magnitude relation in infrared colors is orthogonal to the effects of distance, and therefore gives a new lever arm to shrink the uncertainty in age derivations to sub-Gyr ages \citep[e.g.,][]{correnti2016}.  {\it Webb} can greatly expand this technique due to its much higher sensitivity, greater extension into the infrared, higher resolution, larger field of view, and greater multiplexing.  {\it Webb} can exploit such diagnostics in systems that are not just the nearest ones to the Earth, but to target specific populations that are believed to be the most ancient fossils in the Milky Way galaxy. {\it Webb} can reach below the main-sequence ``kink'' in any population in the Milky Way halo and out to several hundred kpc in the Local Group (Figure~\ref{Fig:StellarPopsWebbCMD}).  {\it Webb's} infrared imaging can also ``stretch'' the color-magnitude relation of brighter stars that are just exhausting their hydrogen supply (i.e., the ``turnoff feature'') from $\lesssim$1~micron to $\gtrsim$2~micron, and therefore establish a much better mapping between small changes in luminosity and temperature and observed differences in magnitude and color.  For old stellar populations, near-infrared color-magnitude diagrams with {\it Webb} should result in a factor of three improvement in stellar ages.

Beyond these standard techniques to establish the chronology of stellar populations in galaxies, {\it Webb} can vastly improve and also open up new diagnostics that are only possible for very cool and low luminous stars.  For example, theory predicts a ``gap'' between the luminosity of the lowest mass hydrogen burning stars and brown dwarfs (i.e., failed stars with $M$ $\lesssim$ 0.08~$M_\odot$ are not massive enough to ignite hydrogen in their cores).  This ``gap'' is a direct age indicator as the brown dwarfs cool and fade over time \citep{burrows1997}.  {\it Webb's} exquisite sensitivity at near-infrared wavelengths is the ideal tool to photometrically characterize the gap in nearby stellar populations (e.g., see Figure~\ref{Fig:StellarPopsWebbCMD} -- noting that the brown dwarfs will be much brighter in redder {\it Webb} filters).  {\it Webb} can also exploit the white dwarf cooling method in more distant clusters \citep[e.g.,][]{hansen2013}.  These stellar remnants have exhausted all of their hydrogen supply, and cool by simply radiating their stored thermal energy into space.  Over an age range between 10 and 13~Gyr, the older and cooler white dwarfs will be 2.5$\times$ fainter.  {\it Webb} can measure the cooling function of these stars out to greater distances than currently possible, and target truly ancient and metal-poor clusters.

In the {\it Webb} era, individual stellar populations in the nearby Milky Way galaxy can be studied using {\it all} of these age diagnostics.  When combined with Gaia constraints on distances from the brightest giants, and a wealth of spectroscopic data from large ground-based telescopes for abundances, ages of the oldest stars in the Milky Way can be established to a precision of well under 0.5~Gyr, providing firm limits on when such populations formed in the Universe.  For ultra-faint dwarf galaxies and other interesting sightlines in the Milky Way halo, the improved resolution of these age diagnostics can lead to stringent tests of the effect of cosmic reionization on star formation histories \citep[e.g.,][]{brown2014,boylan-kolchin2015}.  With future {\it WFIRST} observations, these types of diagnostics can be extended to large contiguous areas of the Milky Way halo.

\subsubsection{Other {\it Webb} Opportunities for Stellar Astrophysics Research}

Separate from the few examples discussed above, {\it Webb} can make a transformative impact on many other stellar population themes.  For example, astrometry with {\it Webb's} high-resolution imagers over single fields can measure the positions of thousands of stars to precisions of 0.3 milliarcsec. {\it Webb} will push existing studies into the crowded centers of stellar populations and also extend baselines established with {\it Hubble} \citep[e.g.,][]{bellini2014,watkins2015a}.  These surveys will advance studies of the dynamical states and shapes of star clusters, reveal the reveal the mass-to-light ratios, test formation and dynamical models, probe for the presence of intermediate mass black holes (core), and search for tidal effects (outskirts).  When combined with next generation radial velocities from future large ground-based telescopes, the joint data sets can yield full 3D intrinsic velocities.  This not only resolves the anisotropy by providing both the tangential and radial components, but also introduces independent distance measurements.  For more distant systems, {\it Webb} can establish the orbits, dynamical masses, and shapes of stellar halos \citep[e.g.,][]{vandermarel2012,deason2013,boylan-kolchin2013}, which, when combined with star formation histories, can be used to perform high-precision studies of the influence of environment on galaxy evolution \citep{kallivayalil2015}.

As shown in Figure~\ref{Fig:StellarPopsWebbCMD}, {\it Webb} can also image galaxies through their bright and red stellar populations out to much further distances than {\it Hubble}.  This can enable spatially-dependent star formation histories across disks, bulges, halos, and streams in galaxies out to many 10s of Mpc.  Combined with synergistic multiobject spectroscopic programs from {\it Webb} and ground-based 8--10~meter and future 20--40~meter telescopes, these observations will establish the most accurate relations between stellar age and metallicity to date.  {\it Webb} can also provide exquisite characterization of detected substructure in the halos of nearby galaxies \citep[e.g.,][]{ferguson2002}  Such relations, over a broad parameter space, inform simulations of hierarchical galaxy formation processes that inform the formation and assembly timescales of different components of Milky Way type galaxies \citep[e.g.,][]{bullock2005}.  Such investigations will be greatly aided in the future by {\it WFIRST} wide-field infrared surveys of the full extent of all nearby galaxies.

In nearby galaxies, {\it Webb} can resolve and image evolved stars on the red giant and asymptotic giant branch in the infrared to characterize their dusty envelopes, yielding compositional studies, measurements of the dust-to-gas ratios, and mass loss rates.  These stars are among the most important contributors to dust in galaxies \citep[e.g.,][]{matsuura2009,boyer2012}, and play a vital role in shaping the infrared luminosities that are used to infer total masses and star formation rates \citep{maraston2005}.  To date, most such studies have been confined to the brightest stars in the nearby Magellanic Clouds \citep[e.g., the {\it Spitzer} Surveying the Agents of Galaxy's Evolution -- SAGE survey;][]{meixner2006,gordon2011}.  With {\it Webb}, for the first time, these studies can be extended to fainter stars in the Magellanic Clouds and to different galaxies in the Local Volume out to 10s of Mpc (e.g., see limits in Figure~\ref{Fig:StellarPopsWebbCMD}) to enable fundamental tests on stellar evolution theory in post main-sequence phases at different ages and metallicities.  For example, multi-color and spectroscopic near-infrared surveys with {\it Webb} can easily separate carbon-rich and oxygen-rich populations, and mid-infrared bands can efficiently sample peaks in the spectrum caused by dust grains with specific compositions.  As these evolved stars represent a large fraction of the infrared light seen in distant galaxies \citep{maraston2005}, {\it Webb's} near-field observations will have numerous applications to refine measurements of the properties of distant galaxies.

{\it Webb} can uniquely map to high-precision the cold interstellar medium in Local Volume galaxies.  By using the spectral energy distribution of stars as a tracer, the amount and properties of foreground dust can be measured \citep[e.g.,][]{gordon2016}.  This not only serves as an input ingredient into star formation models (i.e., it impacts heating and cooling rates), but also calibrates our methods of interpreting colors of distant galaxies.  At the ideal $\sim$10~pc scale for probing the interstellar medium \citep[e.g.,][]{roman-duval2014}, {\it Spitzer} and {\it Herschel} were only able to survey the Magellanic Clouds at high resolution \citep{draine2007,gordon2014}.  {\it Webb} can produce higher quality quality dust maps for all galaxies in the Local Group, through sensitive near and mid-infrared diagnostics that ideally sample features in the spectrum of different dust grains.


\subsection{Galaxy Formation and Evolution Across Cosmic Time}

Understanding the history of the Universe is a fundamental pursuit of science, and galaxies are the visible building blocks that allow us to reconstruct its past events and evolution.  By observing the sizes, shapes, spectral energy, and spatial distribution of galaxies at different distances, we can measure the fossil history of the Universe across different eras and test theoretical models and simulations that predict how galaxies form and evolve.  Today, our fundamental picture is one of hierarchical merging, where smaller systems form first and then are accreted due to the influence of dark matter and gravity into larger galaxies \citep[e.g.,][]{searle1978,blumenthal1984}.  As this bulk galaxy assembly proceeds, generations of star formation and their resulting feedback shapes the observed properties of galaxies.  As a result, the mean properties of galaxies vary by large amounts at different look back times and measuring and correlating these changes with galaxy type, evolutionary state, and environment represents key observables in understanding galaxy formation and evolution \citep{conselice2014}.

\begin{figure}
\centering
\includegraphics[width=10.5cm, trim={1.8cm 2.5cm 1.9cm 2.6cm},clip]{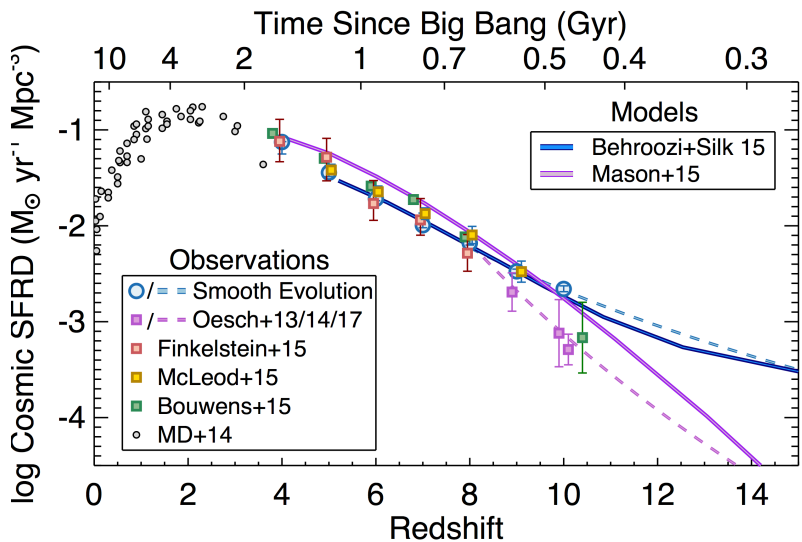} 
\caption{The shape of cosmic star formation rate density across cosmic time shows an increase from early times up to a redshift of 2, and then a rapid decline to present times \citep{madau2014}.  The slope of the relation at higher redshifts of $>$8, largely derived from {\it Hubble} deep field observations as referenced in the Figure, has large uncertainties and is based on a small numbers of galaxies.  The lack of observational constraints provides little leverage to test models with different assumptions on star formation efficiency and feedback.  As an example, the models of \cite{behroozi2015} and \cite{mason2015} are shown and diverge by large amounts at higher redshifts.  The first {\it Webb} extragalactic surveys will resolve the differences in these models by extending measurements of the cosmic star formation rate to higher redshifts and with robust samples of galaxies. Figure adapted from the compilation by \cite{finkelstein2016}. {\it Image Credit: S.\ Finkelstein}} \label{Fig:GalaxiesSFRD}
\end{figure}

Many ground and space-based observatories have made important contributions to understanding galaxy assembly out to redshifts of $\sim$6 (i.e., over the last 12.5~Gyr of the Universe's evolution).  For example, we know that 75\% of the Universe's stars formed 1 -- 5~Gyr after the Big Bang and we've tracked the peak of the star formation rate to $\sim$3.5~Gyr after the Big Bang \citep[e.g.,][]{madau2014}.  The slope of the early ramp up and more recent ramp down of the star formation rate on both sides of this peak has also been measured \citep[see][and Figure~\ref{Fig:GalaxiesSFRD}]{shapley2011}, and the correlation and (small) scatter between the rate and the stellar masses of galaxies has been established at different look back times \citep[e.g.,][]{noeske2007,salmon2015}.  The cause of the recent decline of the star formation rate in galaxies has been linked to decreases in the amount of fuel available \citep[i.e., the molecular gas fraction;][]{genzel2010} and the resulting fading and quenching of galaxies over this epoch has been directly observed \citep[e.g.,][]{faber2007}.  These types of diagnostics have been correlated with quantitative measurements of galaxy structures and morphologies \citep[e.g.,][]{lotz2006}. Nearby, we've found that systems with higher star formation generally exhibit distinct spiral arms, knots with young populations, and central starbursts.  The diagnostics have also been used to measure the growth rate of different classes of galaxies.  For example, we've observed that massive, but, compact, red galaxies show rapid growth at late times \citep{vanderwel2014} and that it's the smallest star-forming galaxies at a given mass that quench first.  


Despite the tremendous achievements in our understanding of galaxy assembly through countless extragalactic observations and surveys \citep[e.g., see][]{blanton2009,shapley2011}, many fundamental questions in this vast field remain open \citep{silk2012} and have served as a primary motivation for a telescope with the capabilities of {\it Webb}.  The superb sensitivity and resolution of {\it Webb} can detect and characterize the smallest building blocks of nearby galaxies and also greatly enhance morphological studies at larger distances (i.e., earlier times).  Deep field observations can reach two magnitudes fainter than current {\it Hubble} studies and yield excellent constraints on the luminosity function of high-redshift galaxies.   The spectroscopic capabilities can provide much more sensitive characterization of spectral energy distributions and emission lines for redshift and chemical enrichment measurements, and the broad infrared coverage introduces far greater diagnostics in systems out to larger redshifts and also probe dust and gas content that is not possible from previous generation visible-light observatories and/or lower resolution infrared instruments.  Taken together, {\it Webb's} observations of galaxy properties can disentangle the complex interplay of stars, gas, dust, metals, dark matter, and active nuclei that shape the basic foundation of galaxy assembly and evolution from the earliest systems in the Universe to those in the present era.

\subsubsection{Searching for the Universe's First Light}

The first stars and galaxies in the Universe seeded everything we see today, and represent the initial conditions for the study of galaxy evolution.  Unfortunately, detecting this ``first light'' has proven to be very challenging, and has led to a gap in our view of the Universe between the early epochs probed by cosmological studies of the cosmic microwave background by {\it COBE}, {\it WMAP}, and {\it Planck} \citep{smoot1992,spergel2003,planckcollaboration2014} and UV-optical-infrared observations of distant galaxies (see below).  Our insights into this epoch of the Universe are now guided entirely by simulations that attempt to track the evolution of baryons from the end of the recombination phase at redshift of 30 (100~Myr after the Big Bang) to the reionization phase \citep{bromm2011}.  Theory predicts that the first stars formed during this time in mini dark matter halos with masses of $\sim$10$^6$~$M_\odot$ \citep{couchman1986,tegmark1997}.  These stars themselves were likely quite massive \citep[10 -- $>$100~$M_\odot$;][]{bromm2004} and would have contained purely hydrogen and helium as the Universe itself had no heavy metals at this time.  As these stars completed their life cycles, they fused the first heavy elements in the Universe and then enriched their surroundings through the first supernova explosions.  This process not only began to ionize the Universe through UV radiation, but also may have sparked fragmentation of collapsing gas clouds and eventual fall back of enriched gas into the formation of clusters of stars and the first population II objects.  As the dark matter halos grew to the 10${^8}$~$M_\odot$ range, the first galaxies are believed to have been born with stellar masses of $\gtrsim$10${^6}$~$M_\odot$.  Detecting these objects and understanding their nature represents a key link into our knowledge of structure formation in the Universe (e.g., how and when it happened).

The Great Observatories {\it Hubble} and {\it Spitzer} have led the way in finding the earliest galaxies by successively pushing the study of ``deep fields'' to new limits.  This includes reaching fainter depths with better detectors, probing earlier times with increased infrared sensitivity, adding statistical significance with survey observations, and developing new strategies that leverage natural cosmic lenses to magnify distant background sources.  Some of the key studies have been the Hubble Deep Field \citep{williams1996,ferguson2000}, Hubble Deep Field South \citep{williams2000}, Great Observatories Origins Deep Survey \citep{giavalisco2004}, the Hubble Ultra Deep Field \citep{beckwith2006}, the COSMOS field \citep{scoville2007}, the Extended Groth Strip survey \citep{davis2007}, the CANDELS survey \citep{grogin2011,koekemoer2011}, the Brightness of Reionization Galaxies Survey \citep{trenti2011}, the CLASH survey \citep{postman2012}, and the Frontier Fields program \citep{lotz2017}.  These UV-optical-IR space-based projects have been aided by synergistic observations at other wavelengths (e.g., with the {\it Chandra X-Ray Telescope}), and also with large investments of deep imaging and spectroscopy from 8 -- 10~meter ground based observatories including {\it Subaru}, {\it VLT}, {\it Keck}, and others \citep[e.g.,][]{ouchi2009,tilvi2013}.

Altogether, the current census of redshift $\sim$4 -- 10 galaxies from these legacy surveys (based on 1000 arcmin$^2$) is now $>$10,000 \citep{bouwens2015}.  Using techniques such as searching for Lyman breaks with combined {\it Hubble} and longer-wavelength {\it Spitzer} data, this census includes a half dozen redshift 10 galaxies \citep[e.g.,][]{ellis2013,coe2013,oesch2013,oesch2014,bouwens2015,bouwens2016,mcleod2015}.  Most of these highest redshift galaxies with redshift $>$ 9 lack spectroscopic confirmation.  

Analysis of the spectral energy distributions of the earliest detected galaxies in the current census shows that they are not representative of the primordial or very low metallicity first light objects, and rather have likely undergone a generation or more of star formation \citep[i.e., they are the $\sim$10$^{7-8}$~$M_\odot$ descendants of the first galaxies;][]{gonzalez2010,labbe2010,dunlop2013}.  The sparse sample of these galaxies is also inadequate to distinguish model-based predictions of the high-redshift galaxy luminosity function \citep[e.g.,][]{behroozi2015}.  This has led to large uncertainties in our knowledge of key physical inputs (e.g., the efficiency of the conversion of gas into stars) as well as the broader characteristics of reionization as understood through the number of ionizing photons from the faint-end galaxy luminosity function \citep{robertson2013}.  Unfortunately, exploring the more distant Universe with {\it Hubble} is difficult.  The faintness of the galaxies aside, the Lyman $\alpha$ feature used to discover high-redshift galaxies moves out to $>$1.3~microns at a redshift of 10, limiting the number of filters in which such observations can be performed to one or two that are not well separated in wavelength.

{\it Webb's} tremendous infrared sensitivity can lift the veil on the redshift $>$9 Universe by discovering large samples of galaxies in an era when the Universe was only 200 -- 500~Myr old \citep{windhorst2006}.  {\it Webb} can also spectroscopically confirm them and characterize their properties.  Among the most exciting projects will be those in search of truly primordial first light objects, systems for which we currently have no observations.  Theory predicts that such first light galaxies may exist over a range of redshifts depending on their luminosities, masses, and environment \citep[e.g.,][]{zackrisson2012}.  To find these, {\it Webb} can use multiple diagnostics \citep[see][]{stiavelli2009}; 1.) detection of strong evolution in the luminosity function predicted by theory \citep{wyithe2006}, 2.) measurement of very low metallicity in individual objects, and 3.) characterizing spectral energy distributions that show no evidence for a generation of stars other than the first burst of star formation.

\begin{figure}[t]
\centering
\includegraphics[width=12cm]{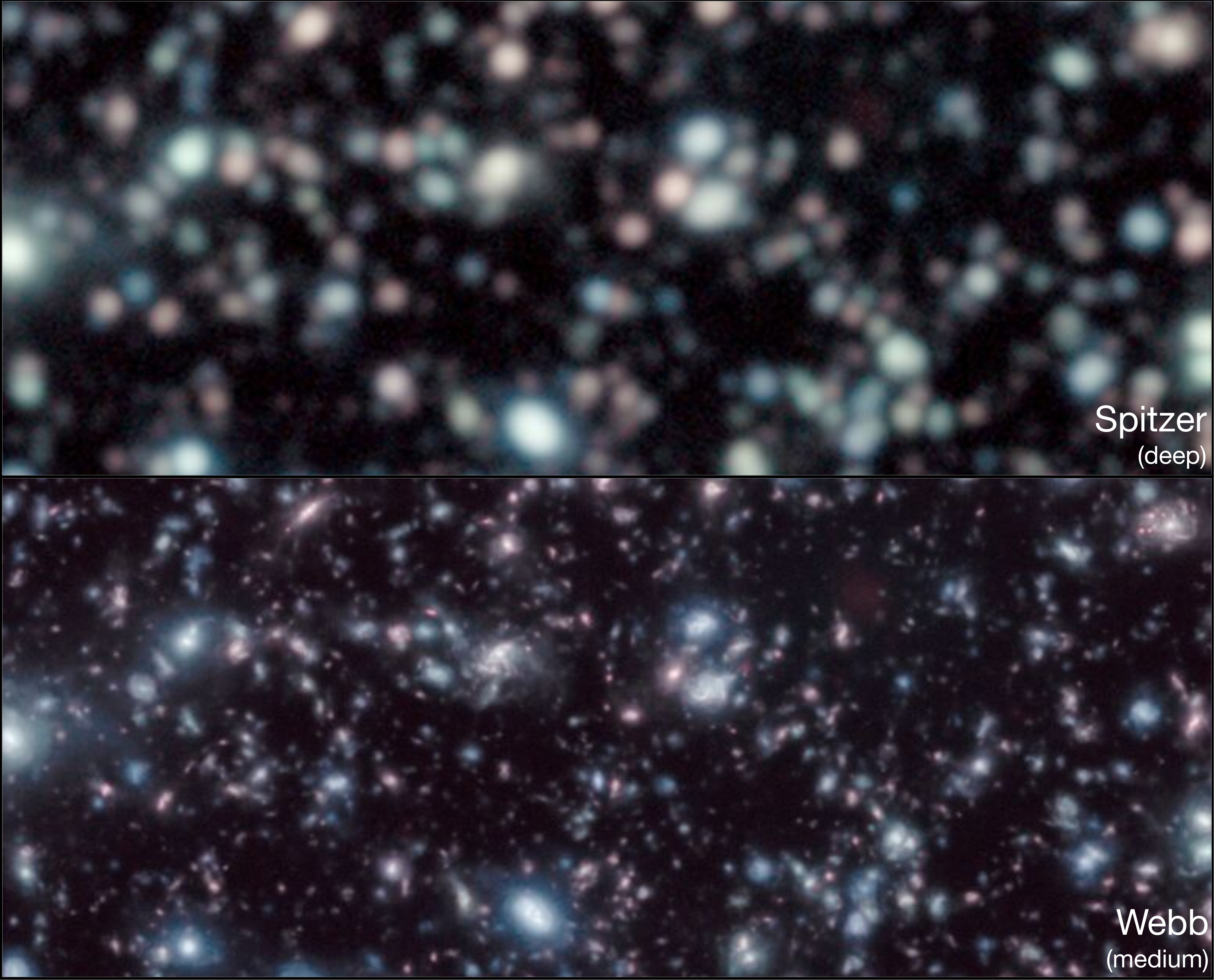} 
\caption{A simulated deep field image from the cosmological simulation Illustris \citep{vogelsberger2014,snyder2017}, over a 2.3 $\times$ 0.9~arcmin area.  Illustris applied a comprehensive model for galaxy formation to simulate the evolution through time of a large slice of the Universe. This shows a projected line of sight through this volume, which yields a simulated survey of distant galaxies.  The top panels show what {\it Spitzer} would measure at 3.5, 4.5, and 5.7~microns and the bottom panel shows what {\it Webb} would measure at comparable wavelengths of 3.5, 4.4, and 5.6~microns.  The {\it Webb} image reaches several magnitudes deeper than {\it Spitzer} in much less integration time, and exhibits more than an order of magnitude more faint galaxies. {\it Image Credit: G.\ Snyder}} \label{Fig:DeepFields}
\end{figure}

At near-infrared wavelengths, {\it Webb's} imaging sensitivity could easily lead to 100s of galaxies with $M >$ 10$^{8}$~$M_\odot$ at redshift = 9 -- 10 through shallow mosaic observations.  This mass corresponds to the upper limit of first generation galaxies in some theoretical studies; \cite{bromm2011}.  With large samples, {\it Webb} could spectroscopically study the brightest of these systems in detail (see below).  Deep field surveys can extend down to much lower luminosities than {\it Hubble} and much higher resolution than {\it Spitzer} (see Figure~\ref{Fig:DeepFields}), enabling detection of dozens of faint galaxies with redshift $>$ 10 and, for the first time, a multi-band color characterization through NIRCam's high-resolution short wavelength channel. {\it Webb} can see Lyman-$\alpha$ out to redshift 20 in this imaging mode.  These data can test for the detection of first light by distinguishing between divergent theoretical luminosity functions that predict the early redshift distribution of galaxies (e.g., see Figure~\ref{Fig:GalaxiesSFRD}) and also through galaxy color analysis of individual objects to explore their metal content.  As successfully demonstrated on {\it Hubble}, {\it Webb} will also take advantage of natural gravitational lensing by massive galaxy clusters to explore the very high-redshift galaxy census \citep{livermore2017,mason2015}.  At redshifts $>$15, deep integrations with the aid of lensing can reveal an order of magnitude more galaxies than standard blank fields.  In the mid 2020s, the precision of the overall galaxy luminosity function will be greatly increased through projects such as the High Latitude Survey from the {\it WFIRST} mission.  The most exciting discoveries from this census will be prime targets for detailed {\it Webb} follow up.


{\it Webb's} high-precision spectroscopy over a broad wavelength range from $\lambda$ = 1 -- 5~microns with NIRSpec can simultaneously detect redshifted UV and optical metal emission lines -- such as CIV, CIII, NeV, and OIII -- which probe the ionizing spectrum and gas phase metallicity to redshift of 9.  The OIII line specifically can establish a connection between metallicity, luminosity, and line intensity, and can be probed in dozens of high-redshift galaxies simultaneously using {\it Webb's} multiplexing capabilities.  {\it Webb} can also measure H$\beta$ and exploit classical diagnostics in high-redshift galaxies, to aid in the interpretation of chemical compositions and early nucleosynthesis of potential first light galaxies at redshift $>$ 10.  Observations of Lyman-$\alpha$ with {\it Webb} can reach much lower flux limits than current studies and vastly improve the robustness of the high-redshift galaxy luminosity function through spectroscopic confirmation.  Spectroscopic observations of these early galaxies at $\lambda$ = 2.5 -- 5.0~microns are beyond the reach of current and future ground-based telescopes.

Present deep field observations of galaxies in the optical and near-infrared can only sample the rest-frame UV, and therefore may be biased towards very young, massive stars populations.  Such star formation may be masking an older (more dominant) component.  {\it Webb's} MIRI instrument can enable the first high-precision studies of the rest-frame near-infrared spectral energy distribution for high-redshift galaxies, and effectively lift degeneracies between age and dust to yield robust stellar masses.  These mid-infrared observations can also characterize H-$\alpha$ emission in redshift $>$ 10 galaxies to derive instantaneous star formation rates \citep{paardekooper2013}.  For example, MIRI's integral field unit can measure a star formation rate of 8~$M_\odot$/yr on a redshift = 10 galaxy in 30 hours \citep{rieke2015a}.  The combined near and mid-infrared colors can establish the amount of metals and dust in these highest redshift galaxies to test whether they are formed from pristine material \citep[e.g., see][]{finkelstein2012,dunlop2013}.

Separate from the detection of first light from galaxies, {\it Webb} may also explore detection of the first stars in the Universe as they exploded as pair-instability supernovae at redshifts of 10 -- 25.  The lack of metals and, therefore, efficient cooling, results in these first stars being very massive (and short lived).  Due to time dilation, the luminous explosions and resulting afterglow of these stars persists for several hundred years at the present time.  The rate of such explosions and the maximum luminosities that they reach depend on many factors including the unknown masses of the stars, so it is still unclear whether they are bright enough (and common enough) for {\it Webb} to see \citep[e.g.,][]{whalen2013,hartwig2017}.  Interestingly, since the shape of the high-redshift galaxy luminosity function is affected by the strength of the energetic feedback that the first stars imparted to the gas in their host halos, measurements of high-redshift galaxies with {\it Webb} may shed light on the properties of the first stars such as their masses \citep{haiman2009}.  The properties of these stars can also be inferred by {\it Webb} from studying accretion disks around the black holes that they form \citep{windhorst2018}.

\subsubsection{Galaxy Growth and Evolution}

Anchored on first-time measurements of the first galaxies in the Universe, {\it Webb's} full range of capabilities can be used to tackle forefront research themes in the study of galaxy assembly across different times and environments.  There are many methods and diagnostics to assess and quantify how the dynamical buildup of small irregular galaxies at early times to larger systems occurred, as well as what the scatter in galaxy properties at a given epoch is.  These include analysis of the shapes of galaxy luminosity functions through different redshift slices as well as the interpretation of a large number of fundamental relations that define trends between galaxy properties such as mass, size, luminosity, surface brightness, metallicity, and velocity dispersion (e.g., scaling relations.  {\it Webb} can provide state-of-the-art imaging and spectroscopic observations of all types of galaxies across all redshifts, and enable major breakthroughs in many of these diagnostics. 

The properties of galaxies are set both by the merging of structure in the hierarchical paradigm and by internal processes within galaxies.  One of key drivers for the latter is the interplay between the level of ongoing star formation (i.e., which consumes a galaxy's fuel) and the amount of energetic feedback from the galaxy's populations to the circumgalactic medium \citep[e.g., through stellar evolution, supernovae, and active galactic nuclei; see][]{tumlinson2017}.  Theoretical models predict specific changes in the bulk galaxy luminosity function at different epochs depending on the strength of these processes \citep[e.g.,][]{hopkins2012,somerville2015}, an observable that {\it Webb} can test with high-precision photometric and spectroscopic data.  For example, over the redshift = 1 -- 3 range, the majority of the growth of stellar populations results from infrared luminous galaxies that are dusty and contain super massive black holes.  Multi-band mid-infrared photometry with {\it Webb's} MIRI instrument can increase sample sizes of these galaxies by many factors over current {\it Spitzer} and {\it Herschel} catalogs, due to an order of magnitude gain in both depth {\it and} resolution.  These data can isolate the features that are caused by dust from those of star formation \citep{kirkpatrick2017}, and assess whether the powerful black holes at the centers of these galaxies are responsible for heating and expelling the gas at recent times and subsequently shutting off their star formation.  {\it Webb} can also extend the luminosity function of galaxies with supermassive black holes to higher redshifts through measurements of high ionization species in near-infrared spectra.  Such studies can chart the growth of black holes from the earliest times and extend scaling relations between galaxy mass and black hole mass \citep[e.g.,][]{ferrarese2000} to smaller systems.

The mid-infrared capabilities of {\it Webb} can also provide breakthroughs in the study of the structure and star formation of local galaxies, such as those explored by the Spitzer Infrared Nearby Galaxy Survey \citep[SINGS;][]{kennicutt2003,calzetti2007}.  Here, the integral field units on MIRI can spatially resolve dust obscured star clusters and spiral structure to constrain the timescale of dust-enshrouded star formation.  These observations will be highly synergistic with {\it ALMA} which can measure the properties of molecular clouds in the same systems.  {\it Webb} can also resolve HII and photodissociation regions in galaxies out 10~Mpc and thereby explore the physics of dust in a range of different physical and chemical conditions. These detailed studies in the nearby Universe directly impact how well we can apply diagnostics to the distant Universe \citep{calzetti2007,calzetti2010}.

For galaxies where dust attenuation is not a major concern, {\it Webb's} multiplexed near-infrared capabilities will be transformative to our understanding of galaxy assembly.  The NIRSpec multiobject spectrograph can achieve grism-like multiplexing at the higher resolution needed to characterize a large set of emission lines over $\lambda$ = 1 -- 5~microns (i.e., much broader than the wavelength range of {\it Hubble's} WFC3/IR instrument and its $\lambda$ = 1.67~micron cutoff).  {\it Webb's} survey of large samples of galaxies can establish strong links between the bursty or smooth nature of star formation and how it is regulated by feedback processes, can map similarities and differences in the chemical composition (e.g., O/H and C/O ratios) of galaxies at different epochs to constrain nucleosynthesis, can establish how much dust galaxies have at increasing redshifts (out to redshift $>$ 6), and can characterize UV metals to indirectly set limits on the ionizing spectrum.  As an example, H$\alpha$ emission spectroscopy with {\it Webb} can yield star formation rates down to 1~$M_\odot$/yr in galaxies out at redshift of 5 or 6, opening up a sensitive new understanding of the scale of early galaxy growth just 1~Gyr after the Big Bang.  It was during this time that the interstellar medium of galaxies was being heavily polluted by metals and dust through new generations of star formation and stellar evolution.

Deriving masses for these galaxies can benefit from sampling the peak of the spectral energy distribution, which, at redshift = 6, occurs at 11~microns.  {\it Webb's} sensitivity at mid-infrared wavelengths can provide masses of galaxies to an order of magnitude lower limits than explored in the deepest {\it Spitzer} observations \citep{caputi2011,rieke2015a}.  Across a broad wavelength range, these high quality spectroscopic data can be subjected to well established techniques to derive accurate masses, metallicities, and dust content.  This research can refine and extend galaxy scaling relations to new limits.


\begin{figure*}[t]
\centering
\includegraphics[width=14cm]{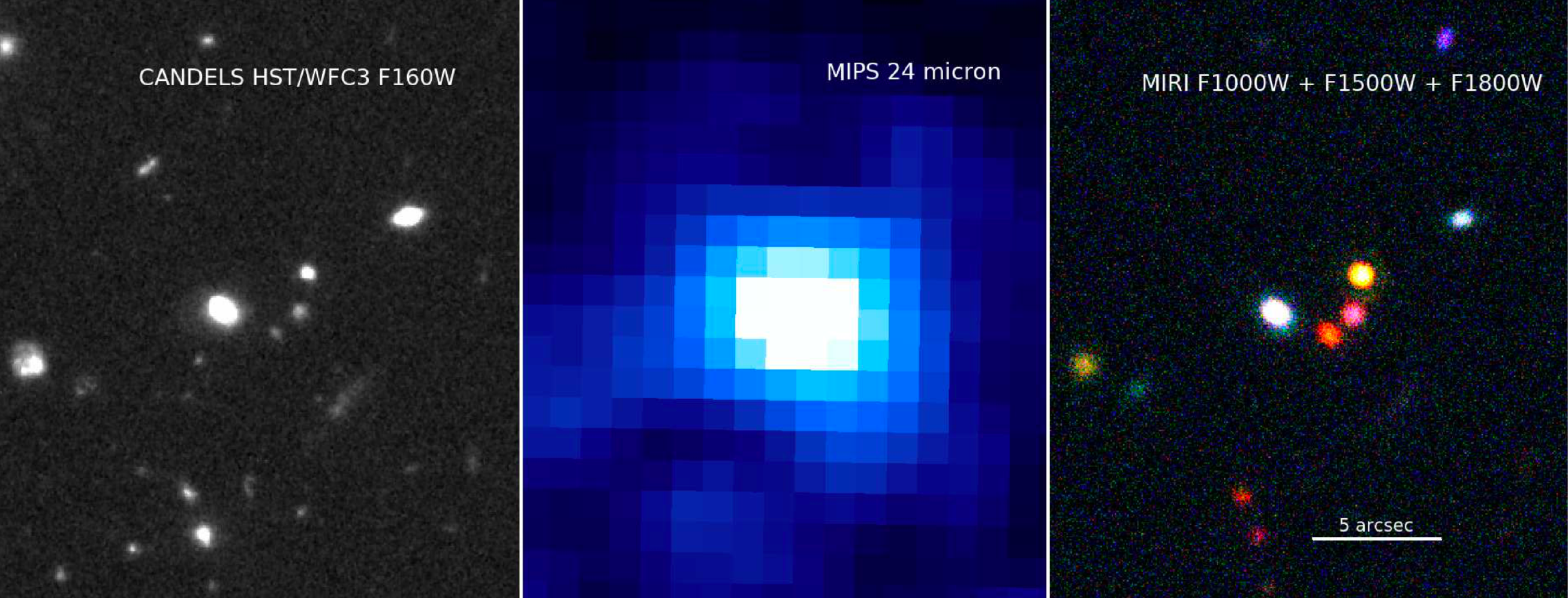} 
\caption{A comparison between an image from the {\it Hubble}/WFC3 CANDELS survey in the near-infrared (left) and {\it Spitzer}/MIPS in the mid-infrared (middle) to the expected simulated mid-infrared image from {\it Webb}/MIRI (right).  The simulation is based on predicting and scaling the infrared spectral energy distribution of galaxies based on {\it Hubble}, {\it Spitzer}, and {\it Herschel} colors (where available), and convolving the relevant image shape through MIRI with derived morphologies from the {\it Hubble} data (and adding appropriate noise).  The MIRI image is more than an order of magnitude deeper and an order of magnitude higher resolution than previous infrared surveys, and easily distinguishes sources and establishes their infrared emission from stellar populations, dust, and active galactic nuclei. {\it Image Credit: C.\ Papovich and the CEERS and CANDELS Team}} \label{Fig:SimulatedMIRIImage}
\end{figure*}


At lower spectral resolutions, {\it Webb} can also be used for extensive wide-field grism surveys to discover and characterize emission line galaxies and study metal enrichment and dust content through spectral energy distributions.  Relative to similar {\it Hubble} programs \citep[e.g.,][]{atek2010}, {\it Webb} spectroscopy has much higher resolution and covers a redder and broader wavelength range.

An important aspect of understanding galaxy assembly is the rate at which galaxy mergers occur over cosmic time.  Although simulations predict mergers to be common at high redshift \citep[e.g.,][]{springel2005,wuyts2010}, the lower resolution of {\it Hubble} and {\it Spitzer} infrared imaging compared to {\it Hubble} visible data leads to less discriminating power to measure galaxy morphologies and search for tidal structure (Figure~\ref{Fig:SimulatedMIRIImage}).  {\it Webb's} high-resolution cameras can resolve galaxy morphologies down to kpc scales out to redshift of 7, and likely enable the first direct evidence of how and when disks and bulges form.  Combining these higher redshift measurements with those of more evolved galaxies at redshifts = 1 -- 7 can establish how the structure of a wide variety of galaxies, including star-forming, irregular, merging, and quiescent systems, evolves over cosmic time and with environment \citep[e.g.,][]{lotz2011}.  Correlating these detailed morphologies with new measures of the star formation rate with {\it Webb} may also provide new insights on the interplay of the ``external'' and ``internal'' processes that drive galaxy assembly \citep{vandokkum2010}.  Taken over different redshift slices, these high-precision imaging results from {\it Webb} can help us understand how and when the Hubble sequence formed, and, more generally, what controls galaxy shapes and why some of them quench \citep[as observed through massive red galaxies that are bulge-dominated at lower redshifts;][]{kriek2010}.  This research theme will be greatly aided by adaptive optics imaging on the next generation large ground-based telescopes, which can achieve higher angular resolution than {\it Webb} (over much smaller fields of view).

\subsubsection{Dark Energy, Dark Matter, and Gravitational Waves}

Separate from the study of first light and galaxy evolution, {\it Webb} can enable new breakthroughs in many areas of extragalactic astronomy due to its increased sensitivity and imaging and spectral resolution.  {\it Webb} can extend the rate of type Ia supernovae discovery through wider area, deeper, and redder surveys compared to current {\it Hubble} studies.  Such samples can reduce the uncertainty in measurements of the Hubble constant and therefore constrain dark energy \citep{riess1998,perlmutter1999}.  {\it Webb's} supernovae surveys out to redshift $>$ 2 can also better constrain a number of potential systematic effects in the measurements \citep{riess2006}.  For dark matter, {\it Webb} can image gravitational clusters to exquisite depths and at higher spatial resolution than {\it Hubble} and {\it Spitzer}.  The resulting lensing maps from such studies provide constraints on the dark matter distribution within the clusters \citep[e.g.,][]{treu2015}.  More details on these and other examples of dark energy and dark matter research with {\it Webb} can be found in \cite{gardner2012}.

{\it Webb} can also be a powerful facility for emerging scientific fields throughout the 2020s.  For example, {\it Webb} can follow up the large number of expected gravitational wave signal detections.  With superb imaging and spectroscopic sensitivity, {\it Webb} can characterize properties of the sources responsible for these events. \citep[e.g.,][]{levan2017}.


\section{Observing Opportunities}\label{observing}

As the next Great Observatory, {\it Webb} will make possible higher precision photometric, spectroscopic, and coronagraphic characterization of astrophysical objects throughout the Universe.  The specific scientific motivation flows from the ``Astronomy and Astrophysics in the New Millennium'' US Decadal Survey \citep{mckee2001} \citep[see also][]{gardner2006}, which was written more than a decade ago.  To ensure that the observatory is executing the most compelling science in today's astrophysical landscape (e.g., see examples in \S\,\ref{science}), STScI has developed a science operations framework for {\it Webb} that closely follows the philosophy that has successfully guided the operation of NASA's previous Great Observatories.  Within this framework, {\it Webb} is fully expected to answer the challenges posed by its foundational documents, such as characterizing the atmospheres of rocky exoplanets, revealing newborn stars and planets in the Milky Way, mapping the assembly history of galaxies across cosmic time, and finding the Universe's first light.

Anybody in the world can write a proposal to use {\it Webb} and all of the data collected by the mission will be publicly available to everyone through STScI's Mikulski Archive for Space Telescopes (MAST).  The process of selecting the best {\it Webb} proposals will be based on a peer-review competition each year, where approximately 150 scientists evaluate and rank the scientific merits of observations proposed by the worldwide community.  This process inspires creativity and collaboration through community engagement, enables a diversity of science ideas to be developed and tackled using innovative observing and analysis techniques, and allows for adaptability in the review and selection process based on the changing astrophysical landscape as {\it Webb} and other concurrent missions enable new scientific discovery (e.g., {\it Hubble}, {\it Gaia}, {\it TESS}, {\it Euclid}, {\it LSST}, {\it WFIRST}, multiple 30-meter ground-based telescopes).  The ``open forum'' proposal process will engage a large community and enable science at all scales, ranging from small projects at individual universities to large 100$+$ hour programs to international teams.  The goal of this framework is to ensure that the observatory returns the best short and long-term science.  The first {\it Webb} Call for Proposals will be released soon.

Information geared towards future {\it Webb} users on observing opportunities, proposal processes and timelines, and policies is provided in Appendix~\ref{observing2} and \ref{policies}.


\section{Conclusion}

The Great Observatories are our deepest space exploration vehicles, allowing us to explore and understand the Universe on all scales at a high level of precision.  Their scientific discovery has sparked a world-wide following of astrophysics and inspired people from all cultures to seek answers to the most fundamental question about our origins, ``What is our place in the cosmos?''.  The tremendous scientific return from these programs, as well as that from other space and ground-based telescopes, has led to breakthrough research in many scientific disciplines from discovering planets orbiting other stars to understanding the global evolution of the Universe.  Equally exciting, the scientific results from these projects reveal new puzzles, the resolution of which requires more powerful future telescopes.

{\it Webb} is the next step in the evolution of space telescopes; an engineering marvel built from new technologies to enable us to see deeper and sharper into the Universe than ever before.  It is one of the biggest and most visible research programs in the world and represents the top scientific priority of the astronomical community.  {\it Webb} has been motivated, designed, and built over 30 years by three space agencies (NASA, ESA, and CSA), and several thousand engineers and scientists.  The observatory is now (early 2018) in the final stages of its comprehensive integration and test phase on the ground and will soon be lifted into space to open its golden eye on the cosmos.

The design of {\it Webb} includes state-of-the-art capabilities such as a 6.5~meter segmented and adjustable primary mirror, cryogenic operations in a thermally stable orbit, and high-precision near- and mid-infrared imagers, spectrographs, and coronagraphs with a high degree of multiplexing.  The observatory will operate for a mission goal of 10~years, and almost all of the observing time ($>$90\%) has intentionally not been allocated to allow astronomers to openly compete for access throughout the mission lifetime.  Anybody in the world can submit a proposal to use {\it Webb} and to access the data that it collects.

The scientific breakthroughs that {\it Webb} can enable span almost all of modern-day astrophysics.  This includes exquisite characterization of nearby objects in our own solar system, measurements of the composition of the atmospheres of other exoplanet worlds, analysis of the properties of nearby stellar populations that are at the heart of defining fundamental astrophysical relations to interpret light from the Universe, the tracking of how the components of galaxies grow over cosmic time, the search for the Universe's earliest visible light, and much more.  These discoveries are expected to reshape our understanding of the Universe, and introduce entirely new research themes that motivate future generations of scientists and engineers to build even bolder observatories to advance space science.


\section*{Acknowledgements}
{\it Webb} is the collective outcome of decades of dedication by more than two thousand scientists and engineers in 14 different countries.  This incredible facility would not be possible without the sacrifice that so many have made for the purpose of advancing science and engineering.  As the project transitions into its science operations phase, we all remain deeply in debt to those that overcame countless challenges to build this incredible scientific facility, and to those who continue to manage and operate it over its lifetime.

The author is extremely grateful to the many hundreds of scientists with whom he has had the pleasure of discussing {\it Webb} capabilities, and to those in the community who have simulated its performance.  The author wishes to thank J.\ Mather, J.\ Gardner, E.\ Smith, and R.\ Ellis for reviewing an early draft of this paper and providing very important and useful feedback. The author is also thankful to several groups who have advised the {\it Webb} project and documented its design and capabilities through many reports and papers.  This includes the members of the {\it Webb} Science Working Group (SWG),\footnote{\url https://www.jwst.nasa.gov/workinggroup.html} {\it Webb} Advisory Group (JSTAC),\footnote{\url https://jwst.stsci.edu/science-planning/user-committees/jwst-advisory-committee-jstac} and {\it Webb} Users Committee (JSTUC).\footnote{\url https://jwst.stsci.edu/science-planning/user-committees/jwst-users-committee-jstuc}

The author also thanks several individuals who helped in the writing of this paper; N.\ Batalha, Z.\ Berta-Thompson, J.\ Brown, D.\ Coe, A.\ Conti, A.\ Feild, S.\ Finkelstein, A.\ Fullerton, D.\ Hines, J.\ Heyl, N.\ Lewis, J.\ Lotz, S.\ Milam, J.\ Muzerolle, P.\ Oesch, C.\ Papovich, M.\ Perrin, K.\ Pontoppidan, L.\ Pueyo, I.\ N.\ Reid, M.\ Robberto, P.\ Rosenfield, K.\ Sembach, G.\ Snyder, J.\ Stansberry, M.\ Stiavelli, G.\ Villanueva, D.\ Weisz, B.\ Williams, and the numerous individuals at STScI and GSFC who developed the {\it Webb} documentation system.\footnote{\url https://jwst-docs.stsci.edu}


\clearpage

\appendix

\section{Observing Opportunities}\label{observing2}

As the next Great Observatory, {\it Webb} will make possible higher precision photometric, spectroscopic, and coronagraphic characterization of astrophysical objects throughout the Universe.  The specific scientific motivation flows from the ``Astronomy and Astrophysics in the New Millennium'' US Decadal Survey (McKee \& Taylor 2001 -- see also Gardner et~al.\ 2006), which was written more than a decade ago.  To ensure that the observatory is executing the most compelling science in today's astrophysical landscape (e.g., see examples in \S\,\ref{science}), STScI has developed a science operations framework for {\it Webb} that closely follows the philosophy that has successfully guided the operation of NASA's previous Great Observatories.  Within this framework, {\it Webb} is fully expected to answer the challenges posed by its foundational documents, such as characterizing the atmospheres of rocky exoplanets, revealing newborn stars and planets in the Milky Way, mapping the assembly history of galaxies across cosmic time, and finding the Universe's first light.

\subsection{Access and Proposal Process}

Anybody in the world can write a proposal to use {\it Webb} and all of the data collected by the mission will be publicly available to everyone through STScI's Mikulski Archive for Space Telescopes (MAST).  Such an open forum enables access to a diverse community and invites a wide range of science ideas of varying scales.  The process of selecting the best {\it Webb} proposals will be based on a peer-review competition, where approximately 150 scientists evaluate and rank the scientific merits of observations proposed by the worldwide community.  As on {\it Hubble}, the time allocation process will involve multiple mirrored panels covering broad subject areas to minimize bias and ensure expertise across diverse astrophysics disciplines.  The time allocation process will be repeated once every year (``cycles'') to ensure that knowledge from the changing astronomical landscape (due to {\it Webb's} discoveries, and those of other concurrent missions) is continuously incorporated into the review and selection process.

For {\it Webb} specifically, the proposal process ensures that the observatory returns the best short and long-term science, and is built on several principles.

\begin{itemize}
\item \underline{\it Creativity through Community Engagement} -- {\it Webb} is a much more complex telescope than previous Great Observatories, and offers a wide range of new observing modes at both near and mid-infrared wavelengths (see \S\,\ref{instruments}).  An oversubscribed and open competition model inspires creativity and collaboration in the broadest scientific community.  Successful {\it Webb} users will develop novel ideas and strategies built on robust theoretical frameworks, and will use innovative observing techniques and analysis algorithms to maximize scientific return.  {\it Webb} observing teams are expected to be diverse and include experts from many research and technical areas.

\item \underline{\it Adaptability} -- The original science ideas that seeded {\it Webb} were formulated almost two decades ago.  Major themes of astrophysics have evolved since the initial science case was established (e.g., the discovery of exoplanets).  Annual competitions for {\it Webb} observations ensure flexibility to tackle a diverse set of leading questions that are at the forefront of astrophysics in that era.

\item \underline{\it Enabling Science at all Scales} -- The {\it Webb} observing time allocation process will enable a range of science that includes small, medium, and large programs through dedicated categories.  Hundreds of programs in each year of the mission will involve focused scientific investigations requiring observing times of a few to dozens of hours each.  These ``small'' programs represent the backbone of individual principal investigator research projects at universities.  Complementing these smaller projects, the observing program will solicit dedicated large projects that often involve international collaborations of many dozens of investigators.  These types of programs can be allocated 100 hours or more for each scientific investigation, and have previously resulted in some of the most visible scientific discoveries that helped define the legacy of the Great Observatories.

\item \underline{\it Maximizing Synergies} -- Within the {\it Webb} 10-year science mission goal, major new ground and space-based observatories will see first light.  This includes the TESS exoplanet mission, the Euclid dark energy mission, the LSST synoptic survey telescope, multiple 30-meter class telescopes in the U.S. and Europe, the Wide Field Infrared Survey Telescope (WFIRST), and others.  The flexible time allocation process with annual cycles allows the scientific community to plan synergistic {\it Webb} observations with these other facilities, and to quickly respond to new discoveries that require higher resolution or higher throughput {\it Webb} follow up.
\end{itemize}

\noindent Proposals submitted for {\it Webb} observations will use STScI's ``Astronomer's Proposal Tool'',\footnote{See online article by B.\ Blair -- {\url http://newsletter.stsci.edu/webb-proposing-support-tools}} the same software used for {\it Hubble} and similar to other Great Observatories.  As on all Great Observatories, only a small fraction of the proposals submitted to {\it Webb} are likely to be selected given the demand for telescope time and resulting oversubscription.  At least initially, for many proposals,\footnote{Some classes of proposals have zero exclusive access period, such as large, treasure, and Director's discretionary proposals.} the successful team will be given one year exclusive access to the data.  STScI is adopting a {\it Spitzer}-like proposal process for {\it Webb}, including both single-stream submission a framework based on pre-defined templates for individual instrument modes.\footnote{See online article by A.\ Moro-Mart{\'{\i}}n -- {\url http://newsletter.stsci.edu/proposal-submission-process}}  This is designed to minimize the time between each year's proposal deadline and the start of the next observing cycle, thereby increasing the amount of {\it Webb} data that is available in the archive for users to plan and learn from.  The change will require most proposers to specify full lists of targets, specifications (filters, exposure times, dithers, observational sequence), and other user-specified scheduling constraints such as roll angle and timing at the initial proposal submission.  Of course, some proposals, such as those that require pre-imaging for follow up spectroscopy or involve target of opportunity observations, will be exempt from this requirement.   Proposers can request help in preparing their observations through STScI's help desk and related knowledgebase.  More information on planning {\it Webb} proposals is available in the online STScI documentation system.\footnote{\url https://jwst-docs.stsci.edu/display/JPP/JWST+Observation+Planning+Documentation}


\subsection{Types of Programs and Science Timeline}\label{sciencetimeline}

{\it Webb} observing time will be scheduled against wall-clock time in hours.  In each cycle (i.e., year), the total amount of time available is therefore 8760 hours for science observations, calibration programs, and overheads involved in executing observations and maintaining the observatory (see \S\,\ref{overheads}).

\subsubsection{Director's Discretionary Proposals}\label{dd}

Similar to {\it Hubble}, up to 10\% of the available total time on {\it Webb} will be at the discretion of the STScI Director for strategic projects.  These can include projects for rapid follow up of in-cycle opportunities through the mission lifetime, opportunities to seed high-impact legacy programs, support for other NASA missions that doesn't come through the regular process, and other related opportunities.  For {\it Webb's} Cycle 1, a large fraction of the Director's Discretionary time is being used for the Early Release Science program (see below).

\subsubsection{General Observer and Guaranteed Time Observer Proposals}\label{goandgto}

Other than time for calibration programs to characterize {\it Webb's} instrumental signatures and overall performance ($<$10\% each cycle), the remaining observing time is divided between General Observer (GO) programs and time already allocated to Guaranteed Time Observers (GTO).  The latter is an international community that includes the Principle Investigators of the {\it Webb} science instruments, a telescope scientist, and several interdisciplinary scientists. A total multi-year allocation of $\sim$4020 hours of {\it Webb} observing time was made to the GTOs in recognition of their decade-long contributions to developing {\it Webb}.  The GTO time represents 10\% of the total observing time available in the 5-year prime mission.  The remaining (large fraction) of time on {\it Webb} is available to General Observers through the annual Call for Proposals.


In the spring of 2017, the GTOs requested $\sim$95\% of their available time in Cycle~1 of {\it Webb}.  The overall program includes 20 proposals totaling 2100 observations by almost 200 worldwide astronomers.  The GTO programs include observations of the solar system, exoplanet spectroscopy and direct imaging with coronagraphy, low mass stars and brown dwarfs, protostellar disks and young stellar objects, debris disks, star clusters and stellar populations, nearby galaxies, galaxy clusters, distant galaxies, quasars, deep fields, and more.  As a result, the GO and GTO allocation will be approximately the same in Cycle 1 with 3800~hours going to each pool.  With very limited GTO observations remaining, Cycle~2$+$ of {\it Webb} will offer nearly $>$7500~hours for the GO community.  In comparison, {\it Hubble} offers 3500~orbits (3100~hours) of observing time in a typical cycle.  The specifications of all of the GTO observations have been made public on the STScI website and Astronomer's Proposal Tool templates based on these programs were released with the {\it Webb} Cycle 1 Call for Proposals to General Observers.\footnote{\url https://jwst-docs.stsci.edu/display/JSP/James+Webb+Space+Telescope+Call+for+Proposals+for+Cycle+1} 

\subsubsection{Early Release Observations}\label{ero}

Following launch, the commissioning phase of {\it Webb} will last $\sim$6 months (see \S\,\ref{commissioning}) after which the Cycle~1 calibration and science program begins.  As with every Great Observatory, one of the first science programs on {\it Webb} will be designed to engage the broadest public audience by showcasing beautiful images and spectroscopy captured through all of the observatory's instruments. Considering the goal, these Early Release Observations (EROs) are not selected through rigorous scientific evaluation as is the case for science programs.  The ERO program is led by NASA through a committee that includes NASA, ESA, CSA, STScI, and instrument team representatives with a range of backgrounds in science, engineering, and public outreach.  The total size of the ERO program is expected to be $\sim$100 hours.

\subsubsection{Early Release Science Proposals}\label{ers}

The first opportunity to propose for Cycle~1 {\it Webb} observations for the GO community was announced by STScI at the Jan 2017 American Astronomical Society Meeting.  Following the advice of the JSTAC, the STScI Director allocated most of his Cycle 1 Director's Discretionary time (up to 500~hours) to make possible an Early Release Science program on {\it Webb} (DD~ERS).\footnote{See online article by J.\ Lee -- {\url http://newsletter.stsci.edu/webb-directors-discretionary-early-release-science-dders}}  The DD~ERS program is distinct from the first GO call for proposals in several ways.  The DD~ERS program will be executed very early in Cycle 1, and all of the observations will be immediately available with no exclusive access period in STScI's MAST archive.  The program is designed to populate the {\it Webb} archive with scientifically compelling data from {\it Webb's} primary observing modes, and will serve as a training set for the community to prepare for observations in subsequent cycles.  The successful teams for {\it Webb} DD~ERS programs will be required to design and deliver science-enabling products to help the community understand the observatory's capabilities.  The program allows the entire community to engage with {\it Webb} observations early in the mission, as the Cycle~2 call for proposals will occur just 5 -- 6 months after Cycle~1 begins (and most {\it Webb} programs will have 1~year exclusive access period).  The DD~ERS program will also yield among the first scientific discoveries from the mission.  In Aug 2017, a total of 106 proposals with more than 3000 scientists from 38 countries were submitted to STScI for the DD~ERS program.  Thirteen of these proposals were selected by a peer-review committee and allocated 460 hours, and cover a broad set of science topics including solar system, exoplanet, stellar, galactic, and extragalactic astrophysics.\footnote{\url https://jwst.stsci.edu/news-events/news/News\%20items/selections-made-for-the-jwst-directors-discretionary-early-release-science-program}  As intended by the program's goals, the DD~ERS projects cover observations in all of {\it Webb's} major observing modes.


\subsubsection{Cycle 1 Timeline for General Observers}\label{timeline}

The {\it Webb} Cycle 1 GO call for proposals will be released several months before launch, and is expected to engage a large fraction of the worldwide astronomical community. To help the community prepare proposals, template files based on the GTO and successful DD~ERS programs have been made available through the STScI website.  Theory proposals related to {\it Webb} observations and/or science themes and funded archival studies based on available data in MAST (e.g., the DD~ERS program) can also be requested.  The time allocation committee will meet after the proposal deadline to evaluate the GO proposals and select the Cycle 1 observing programs.  The time allocation committee will include appropriate representation from the USA, ESA countries, and Canada, and the observing time allocation for Europe and Canada is expected to meet or exceed their 15\% and 5\% respective contributions to {\it Webb}.

For all successful {\it Webb} proposals, STScI will assign both a program coordinator and contact scientist.  These individuals are experts on {\it Webb's} planning and scheduling system and on the capabilities of the instrument(s) being used for the observations.  The staff will work with the principal investigators to get the most out of their approved program by answering technical questions and by suggesting ways to improve efficiency and on-source integration (i.e. vs spending time on overheads).  Requests from the principal investigators for other types of modifications, for example different strategies to achieve the science goals or adjustments in the allocated time because of new insights on instrument performance will have to be approved by a Telescope Time Review Board and the STScI Director for all programs.  All such modifications will need to be scientifically justified and have minimal impact on the {\it Webb} long range schedule.

The different {\it Webb} observing opportunities are summarized in Table~\ref{Tab:ScienceTimeline}.\footnote{See online article by I.\ N.\ Reid \& J.\ Lee -- {\url http://newsletter.stsci.edu/the-webb-science-timeline}}  More information on proposal opportunities is available in the online STScI documentation system.\footnote{\url https://jwst-docs.stsci.edu/display/JSP/JWST+Cycle+1+Proposal+Opportunities}



\begin{table}
  \tabcolsep 3pt

\tbl{{\it Webb} Cycle 1 Schedule of Observing Opportunities and Cycle 1 and 2 Observing Time Availability}
{\begin{tabular}{lccccc} \toprule
    Program Type & Call for  & Proposal &   Announcement of      &    Cycle 1        &    Cycle 2      \\
                 & Proposals\textsuperscript{a} & Deadline\textsuperscript{a} & Successful Proposals\textsuperscript{a}   & Hours Available   & Hours Available \\ \midrule
 Calibration                     & Various\textsuperscript{b} & Before Launch & Before Launch & 750 (8.5\%) & 750 (8.5\%) \\
 Early Release Obs.      & 2017\textsuperscript{b} & TBD & Early in Cycle 1 & $\sim$100 (1.1\%) & 0 (0\%) \\
 Early Release Sci.\textsuperscript{c} & May 19th 2017 & Aug 18th 2017 & Nov 2017 & $\sim$500 (5.7\%) & 0 (0\%) \\
 Director's Discret.            & Various & Various & Various & 376 (4.3\%) & 876 (10\%) \\
 Guaranteed Time Obs.        & Jan 6th 2017 & Apr 1st 2017 & Jun 15th 2017\textsuperscript{d} & 3800 (43.4\%) & 220 (2.5\%) \\
 General Obs.                & TBD & TBD & TBD & 3234\textsuperscript{e} (36.9\%)\textsuperscript{e} & 6914 (78.9\%) \\ 
 {\bf Total}                     & .....         & ......       & ......   & {\bf 8760 (100\%)} & {\bf 8760 (100\%)} \\ \bottomrule
\end{tabular}}
\begin{flushleft}
\tabnote{\textsuperscript{a}All dates are for Cycle 1.}
\tabnote{\textsuperscript{b}These programs do not have a public call for proposals.  This date marks the beginning of the planning process.}
\tabnote{\textsuperscript{c}The Early Release Science program is a Cycle 1 program supported entirely by Director's Discretionary Time.}
\tabnote{\textsuperscript{d}GTO time allocation was pre-approved by NASA, ESA, and CSA.  This date marks the public release of the specifications of their planned observation.}
\tabnote{\textsuperscript{e}Efficient scheduling of {\it Webb} observations requires an oversubscription of available targets.  Therefore, the first call for proposals will offer up to 6000~hours for General Observers, and some programs will be executed in the second year of {\it Webb} operations.}
\end{flushleft}
\label{Tab:ScienceTimeline}
\end{table}


\section{Science Policies for the Astronomical Community}\label{policies}

A number of science policies for {\it Webb} have been carefully crafted by NASA, ESA, CSA, and STScI.  Their application impacts the scientific community proposing to use {\it Webb}.  Complete information on the science policies is available in the online STScI {\it Webb} documentation system.\footnote{\url https://jwst-docs.stsci.edu/display/JSP/JWST+General+Science+Policies}  A brief summary of a few of these policies that most users should be aware of is given below.

\subsection{Overheads}\label{overheads}

Every science program has an associated set of telescope slews, mechanism motions and detector preparations associated with the observations. In addition, like all other telescopes, {\it Webb} must devote a fraction of the time to housekeeping activities, including momentum management and calibration observations. Those activities are essential to supporting all observational programs. Consequently, every program is assigned an indirect overhead associated with those activities.

The total wall-clock time allocated to {\it Webb} programs will be inclusive of 1.) the time spent collecting photons from astronomical objects, 2.) the direct overheads associated with making these observations (e.g, guide star acquisition, filter changes, dithers, detector readouts), 3.) other overheads related to preparing the observatory for the observations (e.g., slews to the target from the previous target), and 4.) indirect overheads related to maintaining the observatory (e.g., wavefront sensing, angular momentum unloads, and calibration).

The direct overheads from executing a science program flow from the specific observational set up and can be easily calculated in the Astronomer's Proposal Tool.  Overheads related to observatory slews to get to the target are more complicated, and depend on the overall {\it Webb} schedule.  These overheads will be calculated based on the expected number of slews and their duration, for each observing program.  The ``smart accounting'' will factor in the likelihood that observations will be scheduled together. For Cycle 1, the average slew time for different slews will be based on simulations and factored into a statistical overhead.  In subsequent cycles, the overhead will be based on data from previously executed programs.  These direct and indirect overheads will be calculated in the Astronomer's Proposal Tool and a detailed summary will be provided to each user.  The overheads related to observatory maintenance are unrelated to specific science programs.  To calculate this, a statistical estimate will be made by adding the two overheads above and dividing them by (1-$f$), where $f$ is the fraction of wall-clock time that is previously spent on these activities.  For Cycle 1 proposals, $f$ will be based on simulations of a year-long {\it Webb} long-range plan that covers the range of science that is expected in Cycle~1.  This indirect overhead rate will be communicated to all proposers through the call for proposals.

The efficiency of different {\it Webb} science programs will vary greatly depending on the integration times and observational set up.  Based on an analysis of {\it Webb's} design reference mission, the overall {\it Webb} efficiency is expected to be $\sim$65\%.

More information on {\it Webb} overhead policy and its implementation is available in Lee et~al.\ (2014) and in the {\it Webb} science policies documentation.


\subsection{Parallel Observing}\label{parallels}

Parallel observations with multiple science instruments simultaneously have proven to increase the scientific productivity of {\it Hubble} and to lead to unique research.  There have been $>$6000 parallel orbits allocated since Cycle 11.  This is over 2 years of observing over a 10~year baseline, for a 20\% efficiency gain.  Many of {\it Hubble's} legacy programs, such as recent deep fields (e.g., Bouwens et~al.\ 2010; Illingworth et~al.\ 2013), the Frontier Fields program (Lotz et~al.\ 2017), the Panchromatic Hubble Andromeda Treasury (PHAT) program (Dalcanton et~al.\ 2012), the Cosmic Assembly Near-infrared Deep Extragalactic Legacy Survey (CANDELS) program (Grogin et~al.\ 2011), the Cluster Lensing And Supernova survey with Hubble (CLASH) program (Postman et~al.\ 2012), and others would not be possible (or would be significantly more time consuming) in the absence of parallel capabilities.



{\it Webb} was initially designed to use parallel observations only for calibration purposes.  This is not surprising, as parallel imaging on {\it Hubble} became very popular only after the recent availability of the two wide-format imaging instruments WFC3 and ACS (e.g., all of the legacy programs cited above).  Recognizing the importance of science return on a limited life mission with multiple wide-field and high-resolution instruments, STScI, NASA, ESA, and CSA developed a concept to enable science parallels beginning with Cycle~1 on {\it Webb}.  Most parallel programs will be ``coordinated parallels'', where the same proposal team aims to use multiple instruments in a single, coherent science program (e.g., to increase field of view, observe at multiple wavelengths, or multiplex imaging and spectroscopy).  The location of {\it Webb's} instruments in the focal plane span $\sim$13~arcmin, so each instrument will observe a different field on the sky.  For extended or intrinsically large objects such as nearby star forming regions, star clusters, galaxies, and galaxy clusters, parallel observations can therefore effectively target multiple fields to explore different environments.  Parallel observations of extragalactic deep fields also increase the statistical significance of the measurement being made.  Such proposals can be constructed in ways that a second epoch of the initial observations would execute at 180 degree orientation, so that the two instruments are ``flipped''.  In this way, the primary and parallel instruments would each observe both fields.  Such programs can be designed in the Astronomer's Proposal Tool using pre-defined templates.  Both the primary and parallel observing requests must be scientifically justified in relation to the core objectives of the proposal, and the parallel observing must be approved by the time allocation committee and the STScI Director.  For {\it Webb} Cycle 1, five different combinations of parallel observing will be allowed; 1.) NIRCam imaging $+$ MIRI imaging, 2.) NIRCam imaging $+$ NIRISS wide-field slitless spectroscopy, 3.) MIRI imaging $+$ NIRISS wide-field slitless spectroscopy, 4.) NIRCam imaging $+$ NIRISS imaging, 5.) NIRSpec multi-object spectroscopy $+$ NIRCam imaging.  Additional parallel combinations will be developed for Cycle~2 and beyond. 

A second mode of parallel observations on {\it Webb} will be called ``pure parallels''.  These are observations where the prime and parallel observing requests come from separate programs with separate teams.  In this case, the scientific goals of the parallel observations are unrelated to that of the primary observations, and the parallel observation cannot dictate how the primary will be structured.  The time allocation committee will review all pure parallel observing requests and select a subset that STScI will then match up to primary opportunities. 

Within the framework described above, parallel observations will have several limitations.  For example, certain types of observations (even with one instrument) over long periods of time can fill {\it Webb's} data volume limits (58.8~GB) and therefore not allow for parallel instrument use.  The Astronomer's Proposal Tool will calculate the data rate for observations and alert users of such scenarios.  Pure parallel science observations will also have a lower priority than {\it Webb} calibration programs.  More information on parallel observing is available in the {\it Webb} science policies documentation.

\subsection{Duplications}\label{duplications}

The duplication policy for {\it Webb} is designed to maximize the scientific productivity of the observatory by flagging and evaluating any requested observations that closely match ones that are already allocated or available in the archive.  Formally, a duplication on {\it Webb} is an observation of the same target using the same instrument mode with an on-target exposure time within a factor of four of the previously scheduled observation.  For cases where a duplication is identified, the time allocation committee will evaluate the justification for the repeat observations and make a recommendation to the STScI Director.  The Director must explicitly approve all duplications.  Note, the duplication policy is not intended to protect specific science programs; different observations are frequently used to address similar science goals.  

The sequence of {\it Webb} observing programs defines how duplications will be identified.  For example, Director's Discretionary Early Release Science and Cycle 1 GO programs can not duplicate Cycle 1 GTO programs (all of which are published online).  Similarly, Cycle 2 GTO programs can not duplicate Cycle 1 GO programs, and so on.  More information on the duplication policy, including example cases of scientifically motivated duplications and the application of the policy to modes such as NIRSpec's multi-object spectrograph, are described in the {\it Webb} science policies documentation.


\subsection{Time Critical and Target of Opportunity Observations}\label{ToOs}

Two classes of {\it Webb} proposals that require non-standard setups are time critical and target of opportunity observations.

Time critical refers to an observation of a known object that must occur within a specific temporal window.  For example, this can include an observation of an eclipsing binary star at a specific phase, a transiting exoplanet during transit, a solar system object at a favorable location, etc.  These observations need to be specified as regular targets in a {\it Webb} observing proposal, but with specific timing requirements.  Additional overheads will be applied to situations that force ``dead time'' of the observatory to get ready for such an observation.  For example, observations with timing within a 1 hour window will have an additional overhead.

Target of opportunity refers to an observation of an object that must be executed rapidly due to an astrophysical change in its characteristics.  This can include the serendipitous discovery of new objects that will evolve quickly (e.g., a comet, supernova, gamma ray burst, or gravitational wave source) or an object that is already known but is seen to go through an unpredictable change (e.g., a dwarf nova).  Most target of opportunity observations are likely to involve follow up of interesting sources that are predicted to be discovered from ongoing wide-field and transient surveys during a {\it Webb} cycle. The proposals for such observations will be evaluated by the regular time allocation committee, and successful proposers will be able to ``activate'' an approved target of opportunity at any time throughout the {\it Webb} cycle using on online STScI-provided interface.  Requests with a turnaround time more than 14 days are considered ``non-disruptive'' and easier to schedule without impacting {\it Webb's} schedule.  More rapid turnaround, to within 2~days, can also be scheduled but such opportunities will be limited to the most deserving science cases to avoid significant impacts to {\it Webb's} schedule and efficiency.  Such requests will be assessed additional overheads as they will likely disrupt a visit that is being executed.  These programs will need to be explicitly approved by the STScI Director.  Requests for target of opportunity observations where an approved proposal does not already exist can also be submitted for Director's Discretionary time, and STScI will arrange for a rapid evaluation and decision to make possible frontier science.

More information on time critical and target of opportunity observations is available in the {\it Webb} science policies documentation.  This includes the rules and responsibilities that the proposal team must follow to successfully trigger a target of opportunity and information on tools that can be used to quickly predict visibility windows of such targets.


\subsection{Funding}\label{funding}

One of the foundational principles behind NASA's Great Observatories is to ``fund the science''.  US investigators that are successful in the highly competitive time allocation process are given robust funding to dedicate resources for analysis, publication, and communication of their scientific data. This funding is a part of the NASA operations budget for the observatories. For {\it Hubble}, one-third of the \$98 million annual budget goes back to the astronomical community in the form of research grants to get the science out.  Over its 28~year life, over \$750~million has been invested into the community.  This funding has provided training for over 1000 graduate students and has led to more than 600 PhD thesis.

{\it Webb's} operations budget will include research grants to fund US-based teams on scientific analysis and publication of new observations, analysis of archival observations for new science, and theoretical support of {\it Webb} programs.  Similar to the single-stream proposal process discussed above, the allocation of funding for successful {\it Webb} science programs will be streamlined relative to the current paradigm on {\it Hubble}.  In Cycle~1, all successful proposers will submit a separate budget to a financial review committee that outlines the resources needed to turn their scientific data into published results.  The budget request can include funding for students, postdocs, and more senior researchers, equipment such as computers and other computational resources, travel to scientific conferences to present results from the program, journal publication charges, and other eligible expenses.  The financial review committee will evaluate all budget proposals and work plans and assign a level of funding to each one based on the complexity of the science program.  After Cycle~1, STScI will calibrate the decisions made by the financial review committee into a formula to assign a base level of funding for successful proposals in future cycles.  Requests for additional funding over the base level will be allowed, but will be subject to a full budget proposal review by a financial review committee.  The goal of this change is to allocate budgets to successful programs shortly after proposal acceptance so teams can rapidly begin building up resources to analyze {\it Webb} data.

The specific level of grant funding for {\it Webb} has yet to be decided and will depend on the cost of operations.

\section{Instrument Modes on {\it Webb}} \label{instrumentmodes}




\begin{table}
\tabcolsep -1pt
\tbl{Specifications of {\it Webb's} Instrument Modes}
{\begin{tabular}{lccccc} \toprule
              & $\lambda$ & Pixel Scale & Field of View & {$\lambda$}$_{\rm Nyquist}$ $_{\rm Sampling}$  \\
 {\bf Imaging}       & (microns) &   (arcsec)  &    (arcmin) & (microns)  \\ \midrule
 NIRCam\textsuperscript{a} & 0.6 -- 2.3 & 0.032 & 2.2 $\times$ 4.4   & $>$2 \\
 NIRCam\textsuperscript{a} & 2.4 -- 5.0 & 0.065 & 2.2 $\times$ 4.4   & $>$4 \\
 NIRISS                    & 0.8 -- 5.0 & 0.065 & 2.2 $\times$ 2.2   & $>$4 \\
 MIRI                      & 5.0 -- 27.5 & 0.11 & 1.23 $\times$ 1.88 & $>$6.25 \\ \bottomrule

 \multicolumn{0}{l}{} \\
              & $\lambda$ & Pixel Scale & Inner Working Angle\textsuperscript{b} & Field of View  \\
 {\bf Coronagraphy} & (microns) &   (arcsec) &    (arcsec)  & (arcsec)  \\ \midrule
 NIRCam 210R (round)     & 2.1         & 0.032 & 0.40         & 20 $\times$ 20 \\
 NIRCam SW (bar)         & 1.7 -- 2.2  & 0.032 & 0.13 -- 0.40 & 20 $\times$ 20 \\
 NIRCam 335R (round)     & 3.35        & 0.065 & 0.63, 0.81   & 20 $\times$ 20 \\
 NIRCam 430R (round)     & 4.3         & 0.065 & 0.63, 0.81   & 20 $\times$ 20 \\
 NIRCam LW (bar)         & 2.4 -- 5.0  & 0.065 & 0.29 -- 0.88 & 20 $\times$ 20 \\
 MIRI 4QPM1              & 10.65       & 0.11  & 0.33         & 24 $\times$ 24 \\
 MIRI 4QPM2              & 11.4        & 0.11  & 0.36         & 24 $\times$ 24 \\
 MIRI 4QPM3              & 15.5        & 0.11  & 0.49         & 24 $\times$ 24 \\
 MIRI Lyot Spot          & 22.75       & 0.11  & 2.16         & 30 $\times$ 30 \\ \bottomrule

 \multicolumn{0}{l}{} \\
             & $\lambda$ & Pixel Scale & Inner Working Angle  \\
 {\bf Aper. Mask Interfer.} & (microns) &   (arcsec)  &    (milliarcsec)    \\ \midrule
 NIRISS & 2.41 -- 3.14, 3.73 -- 3.93, & 0.065 & 70 -- 400 \\
        & 4.18 -- 4.40, 4.67 -- 4.97 &  &  \\ \bottomrule

 \multicolumn{0}{l}{} \\

                       & $\lambda$ & Resolving Power & Slit Size  \\
 {\bf Single Slit Spect.} & (microns) & ($\lambda$/${\Delta}{\lambda}$)& (arcsec) \\ \midrule
 NIRSpec & 0.6 -- 5.3 & 100, 1000, 2700 & 0.4 $\times$ 3.65, 0.2 $\times$ 3.2, \\
         &            &                 & 1.6 $\times$ 1.6 \\
 MIRI & 5 -- $\sim$12 & $\sim$100 at 7.5 microns & 0.51 $\times$ 4.7 \\ \bottomrule

 \multicolumn{0}{l}{} \\

                       & $\lambda$ & Resolving Power & Field of View  \\
 {\bf Slitless Spect.} & (microns) & ($\lambda$/${\Delta}{\lambda}$)& (arcmin) \\ \midrule
 NIRISS & 0.8 -- 2.2 &  150 & 2.2 $\times$ 2.2 \\
 NIRISS & 0.6 -- 2.8 &  700 & single object \\
 NIRCam & 2.4 -- 5.0 & 1130 -- 1680 & 2.2 $\times$ 2.2 \\
 MIRI & 5 -- $\sim$12 & $\sim$100 at 7.5 microns & single object \\ \bottomrule

 \multicolumn{0}{l}{} \\
                       & $\lambda$ & Resolving Power & Field of View & Multiplexing & Aperture Size  \\
 {\bf Multi-object Spect.} & (microns) & ($\lambda$/${\Delta}{\lambda}$) & (arcmin) & (\# shutters) &  (arcsec) \\ \midrule
 NIRSpec & 0.6 -- 5.3 & 100, 1000, 2700 & 3.6 $\times$ 3.4 & 250,000 & 0.2 $\times$ 0.46 \\ \bottomrule

 \multicolumn{0}{l}{} \\
                       & $\lambda$ & Resolving Power & Field of View & Image Slices & Slice Width  \\
 {\bf Integral Field Spect.} & (microns) & ($\lambda$/${\Delta}{\lambda}$)& (arcsec) & & (arcsec) \\ \midrule
 NIRSpec                     & 0.6 -- 5.3     & 100, 1000, 2700 & 3.0 $\times$ 3.0 & 30 & 0.1 \\
 MIRI Ch1\textsuperscript{c} & 4.89 -- 7.66   & 3100 -- 3750    & 3.70 $\times$ 3.70 & 21 & 0.176 \\
 MIRI Ch2\textsuperscript{c} & 7.49 -- 11.71  & 2750 -- 3300    & 4.71 $\times$ 4.52 & 17 & 0.277 \\
 MIRI Ch3\textsuperscript{c} & 11.53 -- 18.05 & 1790 -- 2880    & 6.19 $\times$ 6.14 & 16 & 0.387 \\
 MIRI Ch4\textsuperscript{c} & 17.66 -- 28.45 & 1330 -- 1930    & 7.74 $\times$ 7.95 & 12 & 0.645 \\ \bottomrule
\end{tabular}}\label{Tab:InstrumentModes}
\begin{flushleft}
\tabnote{\textsuperscript{a}The short and long-wavelength NIRCam channels can observe the same field simultaneously.}
 \tabnote{\textsuperscript{b}Quoted values for round masks are the radii and for bar masks are the half-width half-maximum (HWHM).}
 \tabnote{\textsuperscript{c}All four MIRI integral field units can observe simultaneously over one-third of their wavelength range \citep{wells2015}.}
\end{flushleft}

 \label{sample-table}
\end{table}





\clearpage


\bibliographystyle{tfnlm}
\bibliography{references}

\begin{thebibliography}{100}
\providecommand{\url}[1]{\normalfont{#1}}
\providecommand{\urlprefix}{Available from: }

\bibitem{delacaille1763}
{de La Caille}~NL. {Coelum australe stelliferum}. Paris, France; 1763.

\bibitem{messier1764}
{Messier}~C. {A Table of the Places of the Comet of 1764 Discovered at the
  Observatory of the Marine at Paris, the 3d of January, about 8 O'Clock in the
  Evening, in the Constellation of the Dragon, Concluded from Its Situation
  Observed with Regard to the Stars: By Monsieur Charles Messier, Astronomer at
  the Depot of the Plans of the Marine of France, at Paris}. Philosophical
  Transactions of the Royal Society of London Series I.
  1764;\hspace{0pt}54:68--68.

\bibitem{messier1781}
{Messier}~C. {Catalogue des N{\'e}buleuses et des Amas d'{\'E}toiles (Catalog
  of Nebulae and Star Clusters)}; 1781.

\bibitem{hubble1926}
{Hubble}~EP. {Extragalactic nebulae.} \apj. 1926 Dec;\hspace{0pt}64:321--369.

\bibitem{sandage1958}
{Sandage}~A. {Current Problems in the Extragalactic Distance Scale.} \apj. 1958
  May;\hspace{0pt}127:513--526A.

\bibitem{schmidt1968}
{Schmidt}~M. {Space Distribution and Luminosity Functions of Quasi-Stellar
  Radio Sources}. \apj. 1968 Feb;\hspace{0pt}151:393--410.

\bibitem{spitzer1990}
{Spitzer}~L~Jr. {Report to project rand: Astronomical advantages of an
  extra-terrestrial observatory}. Astronomy Quarterly.
  1990;\hspace{0pt}7:131--142.

\bibitem{williams1996}
{Williams}~RE, {Blacker}~B, {Dickinson}~M, et~al. {The Hubble Deep Field:
  Observations, Data Reduction, and Galaxy Photometry}. \aj. 1996
  Oct;\hspace{0pt}112:1335.

\bibitem{ferguson2000}
{Ferguson}~HC, {Dickinson}~M, {Williams}~R. {The Hubble Deep Fields}. \araa.
  2000;\hspace{0pt}38:667--715.

\bibitem{beckwith2006}
{Beckwith}~SVW, {Stiavelli}~M, {Koekemoer}~AM, et~al. {The Hubble Ultra Deep
  Field}. \aj. 2006 Nov;\hspace{0pt}132:1729--1755.

\bibitem{seery2003}
{Seery}~BD. {Next Generation Space Telescope(NGST): Hubble's scientific and
  technological successor}. In: {Mather}~JC, editor. IR Space Telescopes and
  Instruments; (\procspie; Vol. 4850); Mar.; 2003. p. 170--178.

\bibitem{sabelhaus2004}
{Sabelhaus}~PA, {Decker}~JE. {An overview of the James Webb Space Telescope
  (JWST) project}. In: {Mather}~JC, editor. Optical, Infrared, and Millimeter
  Space Telescopes; (\procspie; Vol. 5487); Oct.; 2004. p. 550--563.

\bibitem{gardner2006}
{Gardner}~JP, {Mather}~JC, {Clampin}~M, et~al. {The James Webb Space
  Telescope}. \ssr. 2006 Apr;\hspace{0pt}123:485--606.

\bibitem{feinberg2006}
{Feinberg}~LD, {Hagopian}~JG, {Diaz}~C. {New approach to cryogenic optical
  testing the James Webb Space Telescope}. In: Society of Photo-Optical
  Instrumentation Engineers (SPIE) Conference Series; (\procspie; Vol. 6265);
  Jun.; 2006. p. 62650P.

\bibitem{kimble2012}
{Kimble}~RA, {Davila}~PS, {Diaz}~CE, et~al. {The integration and test program
  of the James Webb Space Telescope}. In: Space Telescopes and Instrumentation
  2012: Optical, Infrared, and Millimeter Wave; (\procspie; Vol. 8442); Sep.;
  2012. p. 84422K.

\bibitem{nella2004}
{Nella}~J, {Atcheson}~PD, {Atkinson}~CB, et~al. {James Webb Space Telescope
  (JWST) Observatory architecture and performance}. In: {Mather}~JC, editor.
  Optical, Infrared, and Millimeter Space Telescopes; (\procspie; Vol. 5487);
  Oct.; 2004. p. 576--587.

\bibitem{clampin2008}
{Clampin}~M. {The James Webb Space Telescope (JWST)}. Advances in Space
  Research. 2008;\hspace{0pt}41:1983--1991.

\bibitem{menzel2010}
{Menzel}~MT, {Bussman}~M, {Davis}~M, et~al. {Systems engineering on the James
  Webb Space Telescope}. In: Modeling, Systems Engineering, and Project
  Management for Astronomy IV; (\procspie; Vol. 7738); Jul.; 2010. p. 77380X.

\bibitem{lightsey2012}
{Lightsey}~PA, {Atkinson}~C, {Clampin}~M, et~al. {James Webb Space Telescope:
  large deployable cryogenic telescope in space}. Optical Engineering. 2012
  Jan;\hspace{0pt}51(1):011003--011003--20.

\bibitem{glassman2016}
{Glassman}~T, {Levi}~J, {Liepmann}~T, et~al. {Alignment of the James Webb Space
  Telescope optical telescope element}. In: Space Telescopes and
  Instrumentation 2016: Optical, Infrared, and Millimeter Wave; (\procspie;
  Vol. 9904); Jul.; 2016. p. 99043Z.

\bibitem{feinberg2012}
{Feinberg}~LD, {Clampin}~M, {Keski-Kuha}~R, et~al. {James Webb Space Telescope
  optical telescope element mirror development history and results}. In: Space
  Telescopes and Instrumentation 2012: Optical, Infrared, and Millimeter Wave;
  (\procspie; Vol. 8442); Sep.; 2012. p. 84422B.

\bibitem{martinez2007}
{Martinez}~K, {Sullivan}~J, {Barto}~A, et~al. {Cryogenic design and predicted
  performance of the James Webb space telescope beryllium aft optics subsystem
  optical bench}. In: Society of Photo-Optical Instrumentation Engineers (SPIE)
  Conference Series; (\procspie; Vol. 6666); Sep.; 2007. p. 66660S.

\bibitem{atkinson2012}
{Atkinson}~C, {Texter}~S, {Keski-Kuha}~R, et~al. {Status of the JWST optical
  telescope element}. In: Space Telescopes and Instrumentation 2012: Optical,
  Infrared, and Millimeter Wave; (\procspie; Vol. 8442); Sep.; 2012. p. 84422E.

\bibitem{arneson2010}
{Arneson}~A, {Alongi}~C, {Bernier}~R, et~al. {Successful production of the
  engineering development unit (EDU) primary mirror segment and flight unit
  tertiary mirror for JWST}. In: Space Telescopes and Instrumentation 2010:
  Optical, Infrared, and Millimeter Wave; (\procspie; Vol. 7731); Jul.; 2010.
  p. 77310I.

\bibitem{ostaszewski2007}
{Ostaszewski}~M, {Vermeer}~W. {Fine steering mirror for the James Webb Space
  Telescope}. In: Society of Photo-Optical Instrumentation Engineers (SPIE)
  Conference Series; (\procspie; Vol. 6665); Sep.; 2007. p. 66650D.

\bibitem{keski-kuha2012}
{Keski-Kuha}~RA, {Bowers}~CW, {Quijada}~MA, et~al. {James Webb Space Telescope
  optical telescope element mirror coatings}. In: Space Telescopes and
  Instrumentation 2012: Optical, Infrared, and Millimeter Wave; (\procspie;
  Vol. 8442); Sep.; 2012. p. 84422J.

\bibitem{arenberg2016}
{Arenberg}~J, {Flynn}~J, {Cohen}~A, et~al. {Status of the JWST sunshield and
  spacecraft}. In: Space Telescopes and Instrumentation 2016: Optical,
  Infrared, and Millimeter Wave; (\procspie; Vol. 9904); Jul.; 2016. p. 990405.

\bibitem{johns2008}
{Johns}~A, {Seaton}~B, {Gal-Edd}~J, et~al. {James Webb Space Telescope: L2
  communications for science data processing}. In: Observatory Operations:
  Strategies, Processes, and Systems II; (\procspie; Vol. 7016); Jul.; 2008. p.
  70161D.

\bibitem{acton2004}
{Acton}~DS, {Atcheson}~PD, {Cermak}~M, et~al. {James Webb Space Telescope
  wavefront sensing and control algorithms}. In: {Mather}~JC, editor. Optical,
  Infrared, and Millimeter Space Telescopes; (\procspie; Vol. 5487); Oct.;
  2004. p. 887--896.

\bibitem{barto2008}
{Barto}~AA, {Atkinson}~C, {Contreras}~J, et~al. {Optical performance
  verification of the James Webb Space Telescope}. In: Space Telescopes and
  Instrumentation 2008: Optical, Infrared, and Millimeter; (\procspie; Vol.
  7010); Jul.; 2008. p. 70100P.

\bibitem{acton2012}
{Acton}~DS, {Knight}~JS, {Contos}~A, et~al. {Wavefront sensing and controls for
  the James Webb Space Telescope}. In: Space Telescopes and Instrumentation
  2012: Optical, Infrared, and Millimeter Wave; (\procspie; Vol. 8442); Sep.;
  2012. p. 84422H.

\bibitem{perrin2016}
{Perrin}~MD, {Acton}~DS, {Lajoie}~CP, et~al. {Preparing for JWST wavefront
  sensing and control operations}. In: Space Telescopes and Instrumentation
  2016: Optical, Infrared, and Millimeter Wave; (\procspie; Vol. 9904); Jul.;
  2016. p. 99040F.

\bibitem{contos2006}
{Contos}~AR, {Acton}~DS, {Atcheson}~PD, et~al. {Aligning and maintaining the
  optics for the James Webb Space Telescope (JWST) on-orbit: the wavefront
  sensing and control concept of operations}. In: Society of Photo-Optical
  Instrumentation Engineers (SPIE) Conference Series; (\procspie; Vol. 6265);
  Jun.; 2006. p. 62650X.

\bibitem{kinzel2012}
{Kinzel}~WM, {Douglas}~RE. {JWST observation specification and expansion to
  support planning and scheduling}. In: Observatory Operations: Strategies,
  Processes, and Systems IV; (\procspie; Vol. 8448); Sep.; 2012. p. 84480Z.

\bibitem{rowlands2004}
{Rowlands}~N, {Aldridge}~D, {Allen}~R, et~al. {The JWST fine guidance sensor}.
  In: {Mather}~JC, editor. Optical, Infrared, and Millimeter Space Telescopes;
  (\procspie; Vol. 5487); Oct.; 2004. p. 664--675.

\bibitem{rowlands2016}
{Rowlands}~N, {Beaton}~A, {Chayer}~P, et~al. {Updated cryogenic performance
  test results for the flight model JWST fine guidance sensor}. In: Space
  Telescopes and Instrumentation 2016: Optical, Infrared, and Millimeter Wave;
  (\procspie; Vol. 9904); Jul.; 2016. p. 99044D.

\bibitem{knight2012}
{Knight}~JS, {Acton}~DS, {Lightsey}~P, et~al. {Observatory alignment of the
  James Webb Space Telescope}. In: Space Telescopes and Instrumentation 2012:
  Optical, Infrared, and Millimeter Wave; (\procspie; Vol. 8442); Sep.; 2012.
  p. 84422C.

\bibitem{rohrbach2016}
{Rohrbach}~SO, {Kubalak}~DA, {Gracey}~RM, et~al. {Critical science instrument
  alignment of the James Webb Space Telescope (JWST) Integrated Science
  Instrument Module (ISIM)}. In: Optical System Alignment, Tolerancing, and
  Verification X; (\procspie; Vol. 9951); Sep.; 2016. p. 995106.

\bibitem{lightsey2014b}
{Lightsey}~PA, {Wei}~Z, {Skelton}~DL, et~al. {Stray light performance for the
  James Webb Space Telescope}. In: Space Telescopes and Instrumentation 2014:
  Optical, Infrared, and Millimeter Wave; (\procspie; Vol. 9143); Aug.; 2014.
  p. 91433P.

\bibitem{wei2006}
{Wei}~Z, {Lightsey}~PA. {Stray light from galactic sky and zodiacal light for
  JWST}. In: Society of Photo-Optical Instrumentation Engineers (SPIE)
  Conference Series; (\procspie; Vol. 6265); Jun.; 2006. p. 62653C.

\bibitem{elliott2013}
{Elliott}~E, {Mountain}~M, {Postman}~M, et~al. {Space Telescopes in the
  Ultraviolet, Optical, and Infrared (UV/O/IR)}; 2013. p. 361--429.

\bibitem{lightsey2014a}
{Lightsey}~PA, {Knight}~JS, {Golnik}~G. {Status of the optical performance for
  the James Webb Space Telescope}. In: Space Telescopes and Instrumentation
  2014: Optical, Infrared, and Millimeter Wave; (\procspie; Vol. 9143); Aug.;
  2014. p. 914304.

\bibitem{perrin2014}
{Perrin}~MD, {Soummer}~R, {Choquet}~{\'E}, et~al. {James Webb Space Telescope
  Optical Simulation Testbed I: overview and first results}. In: Space
  Telescopes and Instrumentation 2014: Optical, Infrared, and Millimeter Wave;
  (\procspie; Vol. 9143); Aug.; 2014. p. 914309.

\bibitem{greenhouse2011}
{Greenhouse}~MA, {Balzano}~V, {Davila}~P, et~al. {Status of the James Webb
  Space Telescope integrated science instrument module system}. In: Society of
  Photo-Optical Instrumentation Engineers (SPIE) Conference Series; (\procspie;
  Vol. 8146); Sep.; 2011. p. 814606.

\bibitem{davila2004}
{Davila}~PS, {Bos}~BJ, {Contreras}~J, et~al. {The James Webb Space Telescope
  science instrument suite: an overview of optical designs}. In: {Mather}~JC,
  editor. Optical, Infrared, and Millimeter Space Telescopes; (\procspie; Vol.
  5487); Oct.; 2004. p. 611--627.

\bibitem{greenhouse2004}
{Greenhouse}~MA, {Sullivan}~PC, {Boyce}~LA, et~al. {The James Webb Space
  Telescope integrated science instrument module}. In: {Mather}~JC, editor.
  Optical, Infrared, and Millimeter Space Telescopes; (\procspie; Vol. 5487);
  Oct.; 2004. p. 754--764.

\bibitem{greenhouse2016}
{Greenhouse}~MA. {The JWST science instrument payload: mission context and
  status}. In: Space Telescopes and Instrumentation 2016: Optical, Infrared,
  and Millimeter Wave; (\procspie; Vol. 9904); Jul.; 2016. p. 990406.

\bibitem{rieke2005}
{Rieke}~MJ, {Kelly}~D, {Horner}~S. {Overview of James Webb Space Telescope and
  NIRCam's Role}. In: {Heaney}~JB, {Burriesci}~LG, editors. Cryogenic Optical
  Systems and Instruments XI; (\procspie; Vol. 5904); Aug.; 2005. p. 1--8.

\bibitem{beichman2010}
{Beichman}~CA, {Krist}~J, {Trauger}~JT, et~al. {Imaging Young Giant Planets
  From Ground and Space}. \pasp. 2010 Feb;\hspace{0pt}122:162--200.

\bibitem{beichman2012}
{Beichman}~CA, {Rieke}~M, {Eisenstein}~D, et~al. {Science opportunities with
  the near-IR camera (NIRCam) on the James Webb Space Telescope (JWST)}. In:
  Space Telescopes and Instrumentation 2012: Optical, Infrared, and Millimeter
  Wave; (\procspie; Vol. 8442); Sep.; 2012. p. 84422N.

\bibitem{rieke2011}
{Rieke}~M. {The JWST-NIRCam View of Galaxy Evolution}. In: {Wang}~W, {Lu}~J,
  {Luo}~Z, et~al., editors. Galaxy Evolution: Infrared to Millimeter Wavelength
  Perspective; (Astronomical Society of the Pacific Conference Series; Vol.
  446); Oct.; 2011. p. 331.

\bibitem{greene2017}
{Greene}~TP, {Kelly}~DM, {Stansberry}~J, et~al. {{$\lambda$} = 2.4 to 5
  {$\mu$}m spectroscopy with the James Webb Space Telescope NIRCam instrument}.
  Journal of Astronomical Telescopes, Instruments, and Systems. 2017
  Jul;\hspace{0pt}3(3):035001.

\bibitem{greene2007}
{Greene}~T, {Beichman}~C, {Eisenstein}~D, et~al. {Observing exoplanets with the
  JWST NIRCam grisms}. In: Techniques and Instrumentation for Detection of
  Exoplanets III; (\procspie; Vol. 6693); Sep.; 2007. p. 66930G.

\bibitem{greene2016a}
{Greene}~TP, {Chu}~L, {Egami}~E, et~al. {Slitless spectroscopy with the James
  Webb Space Telescope Near-Infrared Camera (JWST NIRCam)}. In: Space
  Telescopes and Instrumentation 2016: Optical, Infrared, and Millimeter Wave;
  (\procspie; Vol. 9904); Jul.; 2016. p. 99040E.

\bibitem{krist2007}
{Krist}~JE, {Beichman}~CA, {Trauger}~JT, et~al. {Hunting planets and observing
  disks with the JWST NIRCam coronagraph}. In: Techniques and Instrumentation
  for Detection of Exoplanets III; (\procspie; Vol. 6693); Sep.; 2007. p.
  66930H.

\bibitem{krist2010}
{Krist}~JE, {Balasubramanian}~K, {Muller}~RE, et~al. {The JWST/NIRCam
  coronagraph flight occulters}. In: Space Telescopes and Instrumentation 2010:
  Optical, Infrared, and Millimeter Wave; (\procspie; Vol. 7731); Jul.; 2010.
  p. 77313J.

\bibitem{ferruit2012}
{Ferruit}~P, {Bagnasco}~G, {Barho}~R, et~al. {The JWST near-infrared
  spectrograph NIRSpec: status}. In: Space Telescopes and Instrumentation 2012:
  Optical, Infrared, and Millimeter Wave; (\procspie; Vol. 8442); Sep.; 2012.
  p. 84422O.

\bibitem{birkmann2016}
{Birkmann}~SM, {Ferruit}~P, {Rawle}~T, et~al. {The JWST/NIRSpec instrument:
  update on status and performances}. In: Space Telescopes and Instrumentation
  2016: Optical, Infrared, and Millimeter Wave; (\procspie; Vol. 9904); Jul.;
  2016. p. 99040B.

\bibitem{dorner2016}
{Dorner}~B, {Giardino}~G, {Ferruit}~P, et~al. {A model-based approach to the
  spatial and spectral calibration of NIRSpec onboard JWST}. \aap. 2016
  Aug;\hspace{0pt}592:A113.

\bibitem{teplate2016}
{Te Plate}~M, {Birkmann}~S, {Sirianni}~M, et~al. {JWST's near infrared
  spectrograph status and performance overview}. In: Infrared Remote Sensing
  and Instrumentation XXIV; (\procspie; Vol. 9973); Sep.; 2016. p. 99730E.

\bibitem{rauscher2014}
{Rauscher}~BJ, {Boehm}~N, {Cagiano}~S, et~al. {New and Better Detectors for the
  JWST Near-Infrared Spectrograph}. \pasp. 2014 Aug;\hspace{0pt}126:739.

\bibitem{ferruit2014}
{Ferruit}~P, {Birkmann}~S, {B{\"o}ker}~T, et~al. {Observing transiting
  exoplanets with JWST/NIRSpec}. In: Space Telescopes and Instrumentation 2014:
  Optical, Infrared, and Millimeter Wave; (\procspie; Vol. 9143); Aug.; 2014.
  p. 91430A.

\bibitem{closs2008}
{Closs}~MF, {Ferruit}~P, {Lobb}~DR, et~al. {The Integral Field Unit on the
  James Webb Space Telescope's Near-Infrared Spectrograph}. In: Space
  Telescopes and Instrumentation 2008: Optical, Infrared, and Millimeter;
  (\procspie; Vol. 7010); Jul.; 2008. p. 701011.

\bibitem{purll2010}
{Purll}~DJ, {Lobb}~DR, {Barnes}~AR, et~al. {Flight model performance of the
  integral field unit for the James Webb Space Telescope's near-infrared
  spectrograph}. In: Modern Technologies in Space- and Ground-based Telescopes
  and Instrumentation; (\procspie; Vol. 7739); Jul.; 2010. p. 773917.

\bibitem{moseley2004}
{Moseley}~SH, {Arendt}~RG, {Boucarut}~RA, et~al. {Microshutters arrays for the
  JWST near-infrared spectrometer}. In: {Mather}~JC, editor. Optical, Infrared,
  and Millimeter Space Telescopes; (\procspie; Vol. 5487); Oct.; 2004. p.
  645--652.

\bibitem{kutyrev2008}
{Kutyrev}~AS, {Collins}~N, {Chambers}~J, et~al. {Microshutter arrays: high
  contrast programmable field masks for JWST NIRSpec}. In: Space Telescopes and
  Instrumentation 2008: Optical, Infrared, and Millimeter; (\procspie; Vol.
  7010); Jul.; 2008. p. 70103D.

\bibitem{li2010}
{Li}~MJ, {Brown}~AD, {Kutyrev}~AS, et~al. {JWST microshutter array system and
  beyond}. In: MOEMS and Miniaturized Systems IX; (\procspie; Vol. 7594); Feb.;
  2010. p. 75940N.

\bibitem{doyon2012}
{Doyon}~R, {Hutchings}~JB, {Beaulieu}~M, et~al. {The JWST Fine Guidance Sensor
  (FGS) and Near-Infrared Imager and Slitless Spectrograph (NIRISS)}. In: Space
  Telescopes and Instrumentation 2012: Optical, Infrared, and Millimeter Wave;
  (\procspie; Vol. 8442); Sep.; 2012. p. 84422R.

\bibitem{louie2017}
{Louie}~DR, {Deming}~D, {Albert}~L, et~al. {Simulated JWST/NIRISS Transit
  Spectroscopy of Anticipated TESS Planets Compared to Select Discoveries from
  Space-Based and Ground-Based Surveys}. ArXiv e-prints. 2017 Nov;\hspace{0pt}.

\bibitem{monnier2003}
{Monnier}~JD. {Optical interferometry in astronomy}. Reports on Progress in
  Physics. 2003 May;\hspace{0pt}66:789--857.

\bibitem{sivaramakrishnan2012}
{Sivaramakrishnan}~A, {Lafreni{\`e}re}~D, {Ford}~KES, et~al. {Non-redundant
  Aperture Masking Interferometry (AMI) and segment phasing with JWST-NIRISS}.
  In: Space Telescopes and Instrumentation 2012: Optical, Infrared, and
  Millimeter Wave; (\procspie; Vol. 8442); Sep.; 2012. p. 84422S.

\bibitem{artigau2014}
{Artigau}~{\'E}, {Sivaramakrishnan}~A, {Greenbaum}~AZ, et~al. {NIRISS aperture
  masking interferometry: an overview of science opportunities}. In: Space
  Telescopes and Instrumentation 2014: Optical, Infrared, and Millimeter Wave;
  (\procspie; Vol. 9143); Aug.; 2014. p. 914340.

\bibitem{greenbaum2015}
{Greenbaum}~AZ, {Pueyo}~L, {Sivaramakrishnan}~A, et~al. {An Image-plane
  Algorithm for JWST's Non-redundant Aperture Mask Data}. \apj. 2015
  Jan;\hspace{0pt}798:68--81.

\bibitem{sivaramakrishnan2010}
{Sivaramakrishnan}~A, {Lafreni{\`e}re}~D, {Tuthill}~PG, et~al. {Planetary
  system and star formation science with non-redundant masking on JWST}. In:
  Space Telescopes and Instrumentation 2010: Optical, Infrared, and Millimeter
  Wave; (\procspie; Vol. 7731); Jul.; 2010. p. 77313W.

\bibitem{ford2014}
{Ford}~KES, {McKernan}~B, {Sivaramakrishnan}~A, et~al. {Active Galactic Nucleus
  and Quasar Science with Aperture Masking Interferometry on the James Webb
  Space Telescope}. \apj. 2014 Mar;\hspace{0pt}783:73--89.

\bibitem{wright2010}
{Wright}~GS, {Rieke}~G, {Boeker}~T, et~al. {Progress with the design and
  development of MIRI, the mid-IR instrument for JWST}. In: Space Telescopes
  and Instrumentation 2010: Optical, Infrared, and Millimeter Wave; (\procspie;
  Vol. 7731); Jul.; 2010. p. 77310E.

\bibitem{rieke2015a}
{Rieke}~GH, {Wright}~GS, {B{\"o}ker}~T, et~al. {The Mid-Infrared Instrument for
  the James Webb Space Telescope, I: Introduction}. \pasp. 2015
  Jul;\hspace{0pt}127:584--594.

\bibitem{wright2015}
{Wright}~GS, {Wright}~D, {Goodson}~GB, et~al. {The Mid-Infrared Instrument for
  the James Webb Space Telescope, II: Design and Build}. \pasp. 2015
  Jul;\hspace{0pt}127:595--611.

\bibitem{bouchet2015}
{Bouchet}~P, {Garc{\'{\i}}a-Mar{\'{\i}}n}~M, {Lagage}~PO, et~al. {The
  Mid-Infrared Instrument for the James Webb Space Telescope, III: MIRIM, The
  MIRI Imager}. \pasp. 2015 Jul;\hspace{0pt}127:612--622.

\bibitem{rieke2015b}
{Rieke}~GH, {Ressler}~ME, {Morrison}~JE, et~al. {The Mid-Infrared Instrument
  for the James Webb Space Telescope, VII: The MIRI Detectors}. \pasp. 2015
  Jul;\hspace{0pt}127:665--674.

\bibitem{kendrew2015}
{Kendrew}~S, {Scheithauer}~S, {Bouchet}~P, et~al. {The Mid-Infrared Instrument
  for the James Webb Space Telescope, IV: The Low-Resolution Spectrometer}.
  \pasp. 2015 Jul;\hspace{0pt}127:623--632.

\bibitem{wells2015}
{Wells}~M, {Pel}~JW, {Glasse}~A, et~al. {The Mid-Infrared Instrument for the
  James Webb Space Telescope, VI: The Medium Resolution Spectrometer}. \pasp.
  2015 Jul;\hspace{0pt}127:646--664.

\bibitem{boccaletti2015}
{Boccaletti}~A, {Lagage}~PO, {Baudoz}~P, et~al. {The Mid-Infrared Instrument
  for the James Webb Space Telescope, V: Predicted Performance of the MIRI
  Coronagraphs}. \pasp. 2015 Jul;\hspace{0pt}127:633--645.

\bibitem{pontoppidan2016}
{Pontoppidan}~KM, {Pickering}~TE, {Laidler}~VG, et~al. {Pandeia: a
  multi-mission exposure time calculator for JWST and WFIRST}. In: Observatory
  Operations: Strategies, Processes, and Systems VI; (\procspie; Vol. 9910);
  Jul.; 2016. p. 991016.

\bibitem{mckee2001}
{McKee}~C, {Taylor}~J. National research council; astronomy and astrophysics in
  the new millennium. Washington, DC: The National Academies Press; 2001.
  \urlprefix\url{https://www.nap.edu/catalog/9839/astronomy-and-astrophysics-in-the-new-millennium}.

\bibitem{grotzinger2012}
{Grotzinger}~JP, {Crisp}~J, {Vasavada}~AR, et~al. {Mars Science Laboratory
  Mission and Science Investigation}. \ssr. 2012 Sep;\hspace{0pt}170:5--56.

\bibitem{glassmeier2007}
{Glassmeier}~KH, {Boehnhardt}~H, {Koschny}~D, et~al. {The Rosetta Mission:
  Flying Towards the Origin of the Solar System}. \ssr. 2007
  Feb;\hspace{0pt}128:1--21.

\bibitem{stern2015}
{Stern}~SA, {Bagenal}~F, {Ennico}~K, et~al. {The Pluto system: Initial results
  from its exploration by New Horizons}. Science. 2015
  Oct;\hspace{0pt}350:aad1815.

\bibitem{bolton2017}
{Bolton}~SJ, {Adriani}~A, {Adumitroaie}~V, et~al. {Jupiter's interior and deep
  atmosphere: The initial pole-to-pole passes with the Juno spacecraft}.
  Science. 2017 May;\hspace{0pt}356:821--825.

\bibitem{weaver2006}
{Weaver}~HA, {Stern}~SA, {Mutchler}~MJ, et~al. {Discovery of two new satellites
  of Pluto}. \nat. 2006 Feb;\hspace{0pt}439:943--945.

\bibitem{showalter2013}
{Showalter}~MR, {de Pater}~I, {French}~RS, et~al. {The Neptune System
  Revisited: New Results on Moons and Rings from the Hubble Space Telescope}.
  In: AAS/Division for Planetary Sciences Meeting Abstracts; (AAS/Division for
  Planetary Sciences Meeting Abstracts; Vol.~45); Oct.; 2013. p. 206.01.

\bibitem{verbiscer2009}
{Verbiscer}~AJ, {Skrutskie}~MF, {Hamilton}~DP. {Saturn's largest ring}. \nat.
  2009 Oct;\hspace{0pt}461:1098--1100.

\bibitem{roth2014}
{Roth}~L, {Saur}~J, {Retherford}~KD, et~al. {Transient Water Vapor at Europa's
  South Pole}. Science. 2014 Jan;\hspace{0pt}343:171--174.

\bibitem{sparks2016}
{Sparks}~WB, {Hand}~KP, {McGrath}~MA, et~al. {Probing for Evidence of Plumes on
  Europa with HST/STIS}. \apj. 2016 Oct;\hspace{0pt}829:121.

\bibitem{schlichting2009}
{Schlichting}~HE, {Ofek}~EO, {Wenz}~M, et~al. {A single sub-kilometre Kuiper
  belt object from a stellar occultation in archival data}. \nat. 2009
  Dec;\hspace{0pt}462:895--897.

\bibitem{villanueva2012}
{Villanueva}~GL, {Mumma}~MJ, {Bonev}~BP, et~al. {Water in planetary and
  cometary atmospheres: H$_{2}$O/HDO transmittance and fluorescence models}.
  \jqsrt. 2012 Feb;\hspace{0pt}113:202--220.

\bibitem{villanueva2017}
{Villanueva}~GL, {Mandell}~A, {Protopapa}~S, et~al. {Planetary Spectrum
  Generator (PSG): An Online Tool to Synthesize Spectra of Comets, Small
  Bodies, and (Exo)Planets}. In: Planetary Science Vision 2050 Workshop; (LPI
  Contributions; Vol. 1989); Feb.; 2017. p. 8006.

\bibitem{protopapa2017}
{Protopapa}~S, {Grundy}~WM, {Reuter}~DC, et~al. {Pluto's global surface
  composition through pixel-by-pixel Hapke modeling of New Horizons Ralph/LEISA
  data}. \icarus. 2017 May;\hspace{0pt}287:218--228.

\bibitem{thomas2016}
{Thomas}~CA, {Abell}~P, {Castillo-Rogez}~J, et~al. {Observing Near-Earth
  Objects with the James Webb Space Telescope}. \pasp. 2016
  Jan;\hspace{0pt}128(1):018002.

\bibitem{rivkin2016}
{Rivkin}~AS, {Marchis}~F, {Stansberry}~JA, et~al. {Asteroids and the James Webb
  Space Telescope}. \pasp. 2016 Jan;\hspace{0pt}128(1):018003.

\bibitem{jones2016}
{Jones}~RL, {Juri{\'c}}~M, {Ivezi{\'c}}~{\v Z}. {Asteroid Discovery and
  Characterization with the Large Synoptic Survey Telescope}. In: {Chesley}~SR,
  {Morbidelli}~A, {Jedicke}~R, et~al., editors. Asteroids: New Observations,
  New Models; (IAU Symposium; Vol. 318); Jan.; 2016. p. 282--292.

\bibitem{villanueva2016}
{Villanueva}~GL, {Altieri}~F, {Clancy}~RT, et~al. {Unique Spectroscopy and
  Imaging of Mars with the James Webb Space Telescope}. \pasp. 2016
  Jan;\hspace{0pt}128(1):018004.

\bibitem{villanueva2013}
{Villanueva}~GL, {Mumma}~MJ, {Novak}~RE, et~al. {A sensitive search for
  organics (CH$_{4}$, CH$_{3}$OH, H$_{2}$CO, C$_{2}$H$_{6}$, C$_{2}$H$_{2}$,
  C$_{2}$H$_{4}$), hydroperoxyl (HO$_{2}$), nitrogen compounds (N$_{2}$O,
  NH$_{3}$, HCN) and chlorine species (HCl, CH$_{3}$Cl) on Mars using
  ground-based high-resolution infrared spectroscopy}. \icarus. 2013
  Mar;\hspace{0pt}223:11--27.

\bibitem{norwood2016b}
{Norwood}~J, {Moses}~J, {Fletcher}~LN, et~al. {Giant Planet Observations with
  the James Webb Space Telescope}. \pasp. 2016 Jan;\hspace{0pt}128(1):018005.

\bibitem{sanchez-lavega2011}
{S{\'a}nchez-Lavega}~A, {del R{\'{\i}}o-Gaztelurrutia}~T, {Hueso}~R, et~al.
  {Deep winds beneath Saturn's upper clouds from a seasonal long-lived
  planetary-scale storm}. \nat. 2011 Jul;\hspace{0pt}475:71--74.

\bibitem{moses2005}
{Moses}~JI, {Fouchet}~T, {B{\'e}zard}~B, et~al. {Photochemistry and diffusion
  in Jupiter's stratosphere: Constraints from ISO observations and comparisons
  with other giant planets}. Journal of Geophysical Research (Planets). 2005
  Aug;\hspace{0pt}110:E08001.

\bibitem{dobrijevic2010}
{Dobrijevic}~M, {Cavali{\'e}}~T, {H{\'e}brard}~E, et~al. {Key reactions in the
  photochemistry of hydrocarbons in Neptune's stratosphere}. \planss. 2010
  Oct;\hspace{0pt}58:1555--1566.

\bibitem{sayanagi2013}
{Sayanagi}~KM, {Dyudina}~UA, {Ewald}~SP, et~al. {Dynamics of Saturn's great
  storm of 2010-2011 from Cassini ISS and RPWS}. \icarus. 2013
  Mar;\hspace{0pt}223:460--478.

\bibitem{clarke1996}
{Clarke}~JT, {Ballester}~GE, {Trauger}~J, et~al. {Far-Ultraviolet Imaging of
  Jupiter's Aurora and the lo ``Footprint''}. Science. 1996
  Oct;\hspace{0pt}274:404--409.

\bibitem{hammel1995}
{Hammel}~HB, {Beebe}~RF, {Ingersoll}~AP, et~al. {HST Imaging of Atmospheric
  Phenomena Created by the Impact of Comet Shoemaker-Levy 9}. Science. 1995
  Mar;\hspace{0pt}267:1288--1296.

\bibitem{sparks2017}
{Sparks}~WB, {Schmidt}~BE, {McGrath}~MA, et~al. {Active Cryovolcanism on
  Europa?} \apjl. 2017 Apr;\hspace{0pt}839:L18.

\bibitem{kivelson2002}
{Kivelson}~MG, {Khurana}~KK, {Volwerk}~M. {The Permanent and Inductive Magnetic
  Moments of Ganymede}. \icarus. 2002 Jun;\hspace{0pt}157:507--522.

\bibitem{stofan2007}
{Stofan}~ER, {Elachi}~C, {Lunine}~JI, et~al. {The lakes of Titan}. \nat. 2007
  Jan;\hspace{0pt}445:61--64.

\bibitem{soderblom1990}
{Soderblom}~LA, {Becker}~TL, {Kieffer}~SW, et~al. {Triton's geyser-like plumes
  - Discovery and basic characterization}. Science. 1990
  Oct;\hspace{0pt}250:410--415.

\bibitem{morabito1979}
{Morabito}~LA, {Synnott}~SP, {Kupferman}~PN, et~al. {Discovery of currently
  active extraterrestrial volcanism}. Science. 1979 Jun;\hspace{0pt}204:972.

\bibitem{keszthelyi2016}
{Keszthelyi}~L, {Grundy}~W, {Stansberry}~J, et~al. {Observing Outer Planet
  Satellites (Except Titan) with the James Webb Space Telescope: Science
  Justification and Observational Requirements}. \pasp. 2016
  Jan;\hspace{0pt}128(1):018006.

\bibitem{nixon2016}
{Nixon}~CA, {Achterberg}~RK, {{\'A}d{\'a}mkovics}~M, et~al. {Titan Science with
  the James Webb Space Telescope}. \pasp. 2016 Jan;\hspace{0pt}128(1):018007.

\bibitem{mcewen1998}
{McEwen}~AS, {Keszthelyi}~L, {Geissler}~P, et~al. {Active Volcanism on Io as
  Seen by Galileo SSI}. \icarus. 1998 Sep;\hspace{0pt}135:181--219.

\bibitem{thatte2016}
{Thatte}~D, {Greenbaum}~A, {McGruder}~C, et~al. {JWST NIRISS Simulations and
  Analysis of Vulcanism on Io}. In: Lunar and Planetary Science Conference;
  (Lunar and Planetary Inst.~Technical Report; Vol.~47); Mar.; 2016. p. 3005.

\bibitem{keszthelyi2007}
{Keszthelyi}~L, {Jaeger}~W, {Milazzo}~M, et~al. {New estimates for Io eruption
  temperatures: Implications for the interior}. \icarus. 2007
  Dec;\hspace{0pt}192:491--502.

\bibitem{kelley2016}
{Kelley}~MSP, {Woodward}~CE, {Bodewits}~D, et~al. {Cometary Science with the
  James Webb Space Telescope}. \pasp. 2016 Jan;\hspace{0pt}128(1):018009.

\bibitem{weidenschilling1997}
{Weidenschilling}~SJ. {The Origin of Comets in the Solar Nebula: A Unified
  Model}. \icarus. 1997 Jun;\hspace{0pt}127:290--306.

\bibitem{emery2007}
{Emery}~JP, {Dalle Ore}~CM, {Cruikshank}~DP, et~al. {Ices on (90377) Sedna:
  confirmation and compositional constraints}. \aap. 2007
  Apr;\hspace{0pt}466:395--398.

\bibitem{tegler2010}
{Tegler}~SC, {Cornelison}~DM, {Grundy}~WM, et~al. {Methane and Nitrogen
  Abundances on Pluto and Eris}. \apj. 2010 Dec;\hspace{0pt}725:1296--1305.

\bibitem{parker2016}
{Parker}~A, {Pinilla-Alonso}~N, {Santos-Sanz}~P, et~al. {Physical
  Characterization of TNOs with the James Webb Space Telescope}. \pasp. 2016
  Jan;\hspace{0pt}128(1):018010.

\bibitem{stansberry2008}
{Stansberry}~J, {Grundy}~W, {Brown}~M, et~al. {Physical Properties of Kuiper
  Belt and Centaur Objects: Constraints from the Spitzer Space Telescope}.
  Tucson, Arizona; 2008. p. 161--179.

\bibitem{lellouch2013}
{Lellouch}~E, {Santos-Sanz}~P, {Lacerda}~P, et~al. {``TNOs are Cool'': A survey
  of the trans-Neptunian region. IX. Thermal properties of Kuiper belt objects
  and Centaurs from combined Herschel and Spitzer observations}. \aap. 2013
  Sep;\hspace{0pt}557:A60.

\bibitem{lellouch2017}
{Lellouch}~E, {Moreno}~R, {M{\"u}ller}~T, et~al. {The thermal emission of
  Centaurs and trans-Neptunian objects at millimeter wavelengths from ALMA
  observations}. \aap. 2017 Dec;\hspace{0pt}608:A45.

\bibitem{noll2008}
{Noll}~KS, {Grundy}~WM, {Chiang}~EI, et~al. {Binaries in the Kuiper Belt}.
  Tucson, Arizona; 2008. p. 345--363.

\bibitem{parker2017}
{Parker}~AH, {Buie}~MW, {Zangari}~AM, et~al. {Multiplicity of the New Horizons
  Extended Mission Target (486958) 2014 MU69}. In: AAS/Division for Planetary
  Sciences Meeting Abstracts; (AAS/Division for Planetary Sciences Meeting
  Abstracts; Vol.~49); Oct.; 2017. p. 504.04.

\bibitem{tiscareno2016}
{Tiscareno}~MS, {Showalter}~MR, {French}~RG, et~al. {Observing Planetary Rings
  and Small Satellites with the James Webb Space Telescope: Science
  Justification and Observation Requirements}. \pasp. 2016
  Jan;\hspace{0pt}128(1):018008.

\bibitem{sicardy2011}
{Sicardy}~B, {Ortiz}~JL, {Assafin}~M, et~al. {A Pluto-like radius and a high
  albedo for the dwarf planet Eris from an occultation}. \nat. 2011
  Oct;\hspace{0pt}478:493--496.

\bibitem{santos-sanz2016}
{Santos-Sanz}~P, {French}~RG, {Pinilla-Alonso}~N, et~al. {James Webb Space
  Telescope Observations of Stellar Occultations by Solar System Bodies and
  Rings}. \pasp. 2016 Jan;\hspace{0pt}128(1):018011.

\bibitem{milam2016}
{Milam}~SN, {Stansberry}~JA, {Sonneborn}~G, et~al. {The James Webb Space
  Telescope's Plan for Operations and Instrument Capabilities for Observations
  in the Solar System}. \pasp. 2016 Jan;\hspace{0pt}128(1):018001.

\bibitem{mayor1995}
{Mayor}~M, {Queloz}~D. {A Jupiter-mass companion to a solar-type star}. \nat.
  1995 Nov;\hspace{0pt}378:355--359.

\bibitem{batalha2013}
{Batalha}~NM, {Rowe}~JF, {Bryson}~ST, et~al. {Planetary Candidates Observed by
  Kepler. III. Analysis of the First 16 Months of Data}. \apjs. 2013
  Feb;\hspace{0pt}204:24--44.

\bibitem{winn2015}
{Winn}~JN, {Fabrycky}~DC. {The Occurrence and Architecture of Exoplanetary
  Systems}. \araa. 2015 Aug;\hspace{0pt}53:409--447.

\bibitem{ricker2015}
{Ricker}~GR, {Winn}~JN, {Vanderspek}~R, et~al. {Transiting Exoplanet Survey
  Satellite (TESS)}. Journal of Astronomical Telescopes, Instruments, and
  Systems. 2015 Jan;\hspace{0pt}1(1):014003.

\bibitem{sullivan2015}
{Sullivan}~PW, {Winn}~JN, {Berta-Thompson}~ZK, et~al. {The Transiting Exoplanet
  Survey Satellite: Simulations of Planet Detections and Astrophysical False
  Positives}. \apj. 2015 Aug;\hspace{0pt}809:77--105.

\bibitem{seager2010}
{Seager}~S, {Deming}~D. {Exoplanet Atmospheres}. \araa. 2010
  Sep;\hspace{0pt}48:631--672.

\bibitem{charbonneau2002}
{Charbonneau}~D, {Brown}~TM, {Noyes}~RW, et~al. {Detection of an Extrasolar
  Planet Atmosphere}. \apj. 2002 Mar;\hspace{0pt}568:377--384.

\bibitem{charbonneau2005}
{Charbonneau}~D, {Allen}~LE, {Megeath}~ST, et~al. {Detection of Thermal
  Emission from an Extrasolar Planet}. \apj. 2005 Jun;\hspace{0pt}626:523--529.

\bibitem{deming2005}
{Deming}~D, {Seager}~S, {Richardson}~LJ, et~al. {Infrared radiation from an
  extrasolar planet}. \nat. 2005 Mar;\hspace{0pt}434:740--743.

\bibitem{knutson2007}
{Knutson}~HA, {Charbonneau}~D, {Allen}~LE, et~al. {A map of the day-night
  contrast of the extrasolar planet HD 189733b}. \nat. 2007
  May;\hspace{0pt}447:183--186.

\bibitem{sing2016}
{Sing}~DK, {Fortney}~JJ, {Nikolov}~N, et~al. {A continuum from clear to cloudy
  hot-Jupiter exoplanets without primordial water depletion}. \nat. 2016
  Jan;\hspace{0pt}529:59--62.

\bibitem{deming2013}
{Deming}~D, {Wilkins}~A, {McCullough}~P, et~al. {Infrared Transmission
  Spectroscopy of the Exoplanets HD 209458b and XO-1b Using the Wide Field
  Camera-3 on the Hubble Space Telescope}. \apj. 2013
  Sep;\hspace{0pt}774:95--111.

\bibitem{kreidberg2014b}
{Kreidberg}~L, {Bean}~JL, {D{\'e}sert}~JM, et~al. {Clouds in the atmosphere of
  the super-Earth exoplanet GJ1214b}. \nat. 2014 Jan;\hspace{0pt}505:69--72.

\bibitem{beichman2014}
{Beichman}~C, {Benneke}~B, {Knutson}~H, et~al. {Observations of Transiting
  Exoplanets with the James Webb Space Telescope (JWST)}. \pasp. 2014
  Dec;\hspace{0pt}126:1134.

\bibitem{cowan2015}
{Cowan}~NB, {Greene}~T, {Angerhausen}~D, et~al. {Characterizing Transiting
  Planet Atmospheres through 2025}. \pasp. 2015 Mar;\hspace{0pt}127:311.

\bibitem{stevenson2016}
{Stevenson}~KB, {Lewis}~NK, {Bean}~JL, et~al. {Transiting Exoplanet Studies and
  Community Targets for JWST's Early Release Science Program}. \pasp. 2016
  Sep;\hspace{0pt}128(9):094401.

\bibitem{greene2016b}
{Greene}~TP, {Line}~MR, {Montero}~C, et~al. {Characterizing Transiting
  Exoplanet Atmospheres with JWST}. \apj. 2016 Jan;\hspace{0pt}817:17--38.

\bibitem{batalha2017}
Batalha~NE, Mandell~A, Pontoppidan~K, et~al. Pandexo: A community tool for
  transiting exoplanet science with jwst \& hst. Publications of the
  Astronomical Society of the Pacific. 2017;\hspace{0pt}129(976):064501.

\bibitem{seager2016}
{Seager}~S, {Bains}~W, {Petkowski}~JJ. {Toward a List of Molecules as Potential
  Biosignature Gases for the Search for Life on Exoplanets and Applications to
  Terrestrial Biochemistry}. Astrobiology. 2016 Jun;\hspace{0pt}16:465--485.

\bibitem{nikolov2016}
{Nikolov}~N, {Sing}~DK, {Gibson}~NP, et~al. {VLT FORS2 Comparative Transmission
  Spectroscopy: Detection of Na in the Atmosphere of WASP-39b from the Ground}.
  \apj. 2016 Dec;\hspace{0pt}832:191--199.

\bibitem{wakeford2017}
{Wakeford}~HR, {Sing}~DK, {Deming}~D, et~al. {The Complete transmission
  spectrum of WASP-39b with a precise water constraint}. ArXiv e-prints. 2017
  Nov;\hspace{0pt}.

\bibitem{fortney2013}
{Fortney}~JJ, {Mordasini}~C, {Nettelmann}~N, et~al. {A Framework for
  Characterizing the Atmospheres of Low-mass Low-density Transiting Planets}.
  \apj. 2013 Sep;\hspace{0pt}775:80--92.

\bibitem{madhusudhan2014}
{Madhusudhan}~N, {Knutson}~H, {Fortney}~JJ, et~al. {Exoplanetary Atmospheres}.
  Protostars and Planets VI. 2014;\hspace{0pt}:739--762.

\bibitem{crossfield2015}
{Crossfield}~IJM. {Observations of Exoplanet Atmospheres}. \pasp. 2015
  Oct;\hspace{0pt}127:941--960.

\bibitem{stevenson2014}
{Stevenson}~KB, {D{\'e}sert}~JM, {Line}~MR, et~al. {Thermal structure of an
  exoplanet atmosphere from phase-resolved emission spectroscopy}. Science.
  2014 Nov;\hspace{0pt}346:838--841.

\bibitem{deming2006}
{Deming}~D, {Harrington}~J, {Seager}~S, et~al. {Strong Infrared Emission from
  the Extrasolar Planet HD 189733b}. \apj. 2006 Jun;\hspace{0pt}644:560--564.

\bibitem{stevenson2010}
{Stevenson}~KB, {Harrington}~J, {Nymeyer}~S, et~al. {Possible thermochemical
  disequilibrium in the atmosphere of the exoplanet GJ 436b}. \nat. 2010
  Apr;\hspace{0pt}464:1161--1164.

\bibitem{showman2009}
{Showman}~AP, {Fortney}~JJ, {Lian}~Y, et~al. {Atmospheric Circulation of Hot
  Jupiters: Coupled Radiative-Dynamical General Circulation Model Simulations
  of HD 189733b and HD 209458b}. \apj. 2009 Jul;\hspace{0pt}699:564--584.

\bibitem{laughlin2009}
{Laughlin}~G, {Deming}~D, {Langton}~J, et~al. {Rapid heating of the atmosphere
  of an extrasolar planet}. \nat. 2009 Jan;\hspace{0pt}457:562--564.

\bibitem{lewis2017}
{Lewis}~NK, {Parmentier}~V, {Kataria}~T, et~al. {Atmospheric Circulation and
  Cloud Evolution on the Highly Eccentric Extrasolar Planet HD 80606b}. ArXiv
  e-prints. 2017 Jun;\hspace{0pt}.

\bibitem{clampin2009}
{Clampin}~M. {Comparative Planetology: Transiting Exoplanet Science with JWST}.
  In: astro2010: The Astronomy and Astrophysics Decadal Survey; (ArXiv
  Astrophysics e-prints; Vol. 2010); 2009.

\bibitem{barstow2016}
{Barstow}~JK, {Irwin}~PGJ. {Habitable worlds with JWST: transit spectroscopy of
  the TRAPPIST-1 system?} \mnras. 2016 Sep;\hspace{0pt}461:L92--L96.

\bibitem{goyal2017}
{Goyal}~JM, {Mayne}~N, {Sing}~DK, et~al. {A library of ATMO forward model
  transmission spectra for hot Jupiter exoplanets}. ArXiv e-prints. 2017
  Oct;\hspace{0pt}.

\bibitem{robinson2011}
{Robinson}~TD, {Meadows}~VS, {Crisp}~D, et~al. {Earth as an Extrasolar Planet:
  Earth Model Validation Using EPOXI Earth Observations}. Astrobiology. 2011
  Jun;\hspace{0pt}11:393--408.

\bibitem{morley2015}
{Morley}~CV, {Fortney}~JJ, {Marley}~MS, et~al. {Thermal Emission and Reflected
  Light Spectra of Super Earths with Flat Transmission Spectra}. \apj. 2015
  Dec;\hspace{0pt}815:110--131.

\bibitem{biller2017}
Biller~BA, Bonnefoy~M. Exoplanet atmosphere measurements from direct imaging.
  Cham: Springer International Publishing; 2017. p. 1--28.
  \urlprefix\url{https://doi.org/10.1007/978-3-319-30648-3_101-1}.

\bibitem{lajoie2016}
{Lajoie}~CP, {Soummer}~R, {Pueyo}~L, et~al. {Small-grid dithers for the JWST
  coronagraphs}. In: Space Telescopes and Instrumentation 2016: Optical,
  Infrared, and Millimeter Wave; (\procspie; Vol. 9904); Jul.; 2016. p. 99045K.

\bibitem{marois2008}
{Marois}~C, {Macintosh}~B, {Barman}~T, et~al. {Direct Imaging of Multiple
  Planets Orbiting the Star HR 8799}. Science. 2008 Nov;\hspace{0pt}322:1348.

\bibitem{marois2010}
{Marois}~C, {Zuckerman}~B, {Konopacky}~QM, et~al. {Images of a fourth planet
  orbiting HR 8799}. \nat. 2010 Dec;\hspace{0pt}468:1080--1083.

\bibitem{galicher2011}
{Galicher}~R, {Marois}~C, {Macintosh}~B, et~al. {M-band Imaging of the HR 8799
  Planetary System Using an Innovative LOCI-based Background Subtraction
  Technique}. \apjl. 2011 Oct;\hspace{0pt}739:L41.

\bibitem{marley2012}
{Marley}~MS, {Saumon}~D, {Cushing}~M, et~al. {Masses, Radii, and Cloud
  Properties of the HR 8799 Planets}. \apj. 2012 Aug;\hspace{0pt}754:135--151.

\bibitem{marley2007}
{Marley}~MS, {Fortney}~JJ, {Hubickyj}~O, et~al. {On the Luminosity of Young
  Jupiters}. \apj. 2007 Jan;\hspace{0pt}655:541--549.

\bibitem{fortney2008}
{Fortney}~JJ, {Lodders}~K, {Marley}~MS, et~al. {A Unified Theory for the
  Atmospheres of the Hot and Very Hot Jupiters: Two Classes of Irradiated
  Atmospheres}. \apj. 2008 May;\hspace{0pt}678:1419--1435.

\bibitem{mordasini2009}
{Mordasini}~C, {Alibert}~Y, {Benz}~W. {Extrasolar planet population synthesis.
  I. Method, formation tracks, and mass-distance distribution}. \aap. 2009
  Jul;\hspace{0pt}501:1139--1160.

\bibitem{oberg2011}
{{\"O}berg}~KI, {Murray-Clay}~R, {Bergin}~EA. {The Effects of Snowlines on C/O
  in Planetary Atmospheres}. \apjl. 2011 Dec;\hspace{0pt}743:L16.

\bibitem{heap2000}
{Heap}~SR, {Lindler}~DJ, {Lanz}~TM, et~al. {Space Telescope Imaging
  Spectrograph Coronagraphic Observations of {$\beta$} Pictoris}. \apj. 2000
  Aug;\hspace{0pt}539:435--444.

\bibitem{kalas2005}
{Kalas}~P, {Graham}~JR, {Clampin}~M. {A planetary system as the origin of
  structure in Fomalhaut's dust belt}. \nat. 2005
  Jun;\hspace{0pt}435:1067--1070.

\bibitem{smith2010}
{Smith}~R, {Wyatt}~MC. {Warm dusty discs: exploring the A star 24 {$\mu$}m
  debris population}. \aap. 2010 Jun;\hspace{0pt}515:A95.

\bibitem{bruzual2003}
{Bruzual}~G, {Charlot}~S. {Stellar population synthesis at the resolution of
  2003}. \mnras. 2003 Oct;\hspace{0pt}344:1000--1028.

\bibitem{sarajedini2007}
{Sarajedini}~A, {Bedin}~LR, {Chaboyer}~B, et~al. {The ACS Survey of Galactic
  Globular Clusters. I. Overview and Clusters without Previous Hubble Space
  Telescope Photometry}. \aj. 2007 Apr;\hspace{0pt}133:1658--1672.

\bibitem{dalcanton2009}
{Dalcanton}~JJ, {Williams}~BF, {Seth}~AC, et~al. {The ACS Nearby Galaxy Survey
  Treasury}. \apjs. 2009 Jul;\hspace{0pt}183:67--108.

\bibitem{dalcanton2012}
{Dalcanton}~JJ, {Williams}~BF, {Lang}~D, et~al. {The Panchromatic Hubble
  Andromeda Treasury}. \apjs. 2012 Jun;\hspace{0pt}200:18--54.

\bibitem{dunham2014}
{Dunham}~MM, {Stutz}~AM, {Allen}~LE, et~al. {The Evolution of Protostars:
  Insights from Ten Years of Infrared Surveys with Spitzer and Herschel}.
  Protostars and Planets VI. 2014;\hspace{0pt}:195--218.

\bibitem{carr2008}
{Carr}~JS, {Najita}~JR. {Organic Molecules and Water in the Planet Formation
  Region of Young Circumstellar Disks}. Science. 2008 Mar;\hspace{0pt}319:1504.

\bibitem{salyk2008}
{Salyk}~C, {Pontoppidan}~KM, {Blake}~GA, et~al. {H$_{2}$O and OH Gas in the
  Terrestrial Planet-forming Zones of Protoplanetary Disks}. \apjl. 2008
  Mar;\hspace{0pt}676:L49.

\bibitem{pontoppidan2010}
{Pontoppidan}~KM, {Salyk}~C, {Blake}~GA, et~al. {A Spitzer Survey of
  Mid-infrared Molecular Emission from Protoplanetary Disks. I. Detection
  Rates}. \apj. 2010 Sep;\hspace{0pt}720:887--903.

\bibitem{pontoppidan2009}
{Pontoppidan}~KM, {Meijerink}~R, {Dullemond}~CP, et~al. {A New Raytracer for
  Modeling AU-Scale Imaging of Lines from Protoplanetary Disks}. \apj. 2009
  Oct;\hspace{0pt}704:1482--1494.

\bibitem{sabbi2016}
{Sabbi}~E, {Lennon}~DJ, {Anderson}~J, et~al. {Hubble Tarantula Treasury
  Project. III. Photometric Catalog and Resulting Constraints on the
  Progression of Star Formation in the 30 Doradus Region}. \apjs. 2016
  Jan;\hspace{0pt}222:11--35.

\bibitem{bastian2010}
{Bastian}~N, {Covey}~KR, {Meyer}~MR. {A Universal Stellar Initial Mass
  Function? A Critical Look at Variations}. \araa. 2010
  Sep;\hspace{0pt}48:339--389.

\bibitem{kalirai2013}
{Kalirai}~JS, {Anderson}~J, {Dotter}~A, et~al. {Ultra-Deep Hubble Space
  Telescope Imaging of the Small Magellanic Cloud: The Initial Mass Function of
  Stars with M $<$ 1$M_\odot$}. \apj. 2013 Feb;\hspace{0pt}763:110--117.

\bibitem{geha2013}
{Geha}~M, {Brown}~TM, {Tumlinson}~J, et~al. {The Stellar Initial Mass Function
  of Ultra-faint Dwarf Galaxies: Evidence for IMF Variations with Galactic
  Environment}. \apj. 2013 Jul;\hspace{0pt}771:29--37.

\bibitem{caiazzo2017}
{Caiazzo}~I, {Heyl}~JS, {Richer}~H, et~al. {Globular cluster absolute ages from
  cooling brown dwarfs}. ArXiv e-prints. 2017 Jan;\hspace{0pt}.

\bibitem{baraffe2015}
{Baraffe}~I, {Homeier}~D, {Allard}~F, et~al. {New evolutionary models for
  pre-main sequence and main sequence low-mass stars down to the
  hydrogen-burning limit}. \aap. 2015 May;\hspace{0pt}577:A42.

\bibitem{fontaine2001}
{Fontaine}~G, {Brassard}~P, {Bergeron}~P. {The Potential of White Dwarf
  Cosmochronology}. \pasp. 2001 Apr;\hspace{0pt}113:409--435.

\bibitem{cioni2011}
{Cioni}~MRL, {Clementini}~G, {Girardi}~L, et~al. {The VMC survey. I. Strategy
  and first data}. \aap. 2011 Mar;\hspace{0pt}527:A116.

\bibitem{jofre2011}
{Jofr{\'e}}~P, {Weiss}~A. {The age of the Milky Way halo stars from the Sloan
  Digital Sky Survey}. \aap. 2011 Sep;\hspace{0pt}533:A59.

\bibitem{dotter2010}
{Dotter}~A, {Sarajedini}~A, {Anderson}~J, et~al. {The ACS Survey of Galactic
  Globular Clusters. IX. Horizontal Branch Morphology and the Second Parameter
  Phenomenon}. \apj. 2010 Jan;\hspace{0pt}708:698--716.

\bibitem{brown2006}
{Brown}~TM, {Smith}~E, {Ferguson}~HC, et~al. {The Detailed Star Formation
  History in the Spheroid, Outer Disk, and Tidal Stream of the Andromeda
  Galaxy}. \apj. 2006 Nov;\hspace{0pt}652:323--353.

\bibitem{brown2014}
{Brown}~TM, {Tumlinson}~J, {Geha}~M, et~al. {The Quenching of the Ultra-faint
  Dwarf Galaxies in the Reionization Era}. \apj. 2014
  Dec;\hspace{0pt}796:91--103.

\bibitem{weisz2014}
{Weisz}~DR, {Dolphin}~AE, {Skillman}~ED, et~al. {The Star Formation Histories
  of Local Group Dwarf Galaxies. I. Hubble Space Telescope/Wide Field Planetary
  Camera 2 Observations}. \apj. 2014 Jul;\hspace{0pt}789:147--169.

\bibitem{chaboyer2008}
{Chaboyer}~B. {Distances and ages of globular clusters}. In: {Jin}~WJ,
  {Platais}~I, {Perryman}~MAC, editors. A Giant Step: from Milli- to
  Micro-arcsecond Astrometry; (IAU Symposium; Vol. 248); Jul.; 2008. p.
  440--442.

\bibitem{correnti2016}
{Correnti}~M, {Gennaro}~M, {Kalirai}~JS, et~al. {Constraining Globular Cluster
  Age Uncertainties using the IR Color-Magnitude Diagram}. \apj. 2016
  May;\hspace{0pt}823:18--34.

\bibitem{burrows1997}
{Burrows}~A, {Marley}~M, {Hubbard}~WB, et~al. {A Nongray Theory of Extrasolar
  Giant Planets and Brown Dwarfs}. \apj. 1997 Dec;\hspace{0pt}491:856--875.

\bibitem{hansen2013}
{Hansen}~BMS, {Kalirai}~JS, {Anderson}~J, et~al. {An age difference of two
  billion years between a metal-rich and a metal-poor globular cluster}. \nat.
  2013 Aug;\hspace{0pt}500:51--53.

\bibitem{boylan-kolchin2015}
{Boylan-Kolchin}~M, {Weisz}~DR, {Johnson}~BD, et~al. {The Local Group as a time
  machine: studying the high-redshift Universe with nearby galaxies}. \mnras.
  2015 Oct;\hspace{0pt}453:1503--1512.

\bibitem{bellini2014}
{Bellini}~A, {Anderson}~J, {van der Marel}~RP, et~al. {Hubble Space Telescope
  Proper Motion (HSTPROMO) Catalogs of Galactic Globular Clusters. I. Sample
  Selection, Data Reduction, and NGC 7078 Results}. \apj. 2014
  Dec;\hspace{0pt}797:115--147.

\bibitem{watkins2015a}
{Watkins}~LL, {van der Marel}~RP, {Bellini}~A, et~al. {Hubble Space Telescope
  Proper Motion (HSTPROMO) Catalogs of Galactic Globular Cluster. II. Kinematic
  Profiles and Maps}. \apj. 2015 Apr;\hspace{0pt}803:29--50.

\bibitem{vandermarel2012}
{van der Marel}~RP, {Fardal}~M, {Besla}~G, et~al. {The M31 Velocity Vector. II.
  Radial Orbit toward the Milky Way and Implied Local Group Mass}. \apj. 2012
  Jul;\hspace{0pt}753:8--21.

\bibitem{deason2013}
{Deason}~AJ, {Van der Marel}~RP, {Guhathakurta}~P, et~al. {The Velocity
  Anisotropy of Distant Milky Way Halo Stars from Hubble Space Telescope Proper
  Motions}. \apj. 2013 Mar;\hspace{0pt}766:24--34.

\bibitem{boylan-kolchin2013}
{Boylan-Kolchin}~M, {Bullock}~JS, {Sohn}~ST, et~al. {The Space Motion of Leo I:
  The Mass of the Milky Way's Dark Matter Halo}. \apj. 2013
  May;\hspace{0pt}768:140--150.

\bibitem{kallivayalil2015}
{Kallivayalil}~N, {Wetzel}~AR, {Simon}~JD, et~al. {A Hubble Astrometry
  Initiative: Laying the Foundation for the Next-Generation Proper-Motion
  Survey of the Local Group}. ArXiv e-prints. 2015 Mar;\hspace{0pt}.

\bibitem{ferguson2002}
{Ferguson}~AMN, {Irwin}~MJ, {Ibata}~RA, et~al. {Evidence for Stellar
  Substructure in the Halo and Outer Disk of M31}. \aj. 2002
  Sep;\hspace{0pt}124:1452--1463.

\bibitem{bullock2005}
{Bullock}~JS, {Johnston}~KV. {Tracing Galaxy Formation with Stellar Halos. I.
  Methods}. \apj. 2005 Dec;\hspace{0pt}635:931--949.

\bibitem{matsuura2009}
{Matsuura}~M, {Barlow}~MJ, {Zijlstra}~AA, et~al. {The global gas and dust
  budget of the Large Magellanic Cloud: AGB stars and supernovae, and the
  impact on the ISM evolution}. \mnras. 2009 Jun;\hspace{0pt}396:918--934.

\bibitem{boyer2012}
{Boyer}~ML, {Srinivasan}~S, {Riebel}~D, et~al. {The Dust Budget of the Small
  Magellanic Cloud: Are Asymptotic Giant Branch Stars the Primary Dust Source
  at Low Metallicity?} \apj. 2012 Mar;\hspace{0pt}748:40--49.

\bibitem{maraston2005}
{Maraston}~C. {Evolutionary population synthesis: models, analysis of the
  ingredients and application to high-z galaxies}. \mnras. 2005
  Sep;\hspace{0pt}362:799--825.

\bibitem{meixner2006}
{Meixner}~M, {Gordon}~KD, {Indebetouw}~R, et~al. {Spitzer Survey of the Large
  Magellanic Cloud: Surveying the Agents of a Galaxy's Evolution (SAGE). I.
  Overview and Initial Results}. \aj. 2006 Dec;\hspace{0pt}132:2268--2288.

\bibitem{gordon2011}
{Gordon}~KD, {Meixner}~M, {Meade}~MR, et~al. {Surveying the Agents of Galaxy
  Evolution in the Tidally Stripped, Low Metallicity Small Magellanic Cloud
  (SAGE-SMC). I. Overview}. \aj. 2011 Oct;\hspace{0pt}142:102--116.

\bibitem{gordon2016}
{Gordon}~KD, {Fouesneau}~M, {Arab}~H, et~al. {The Panchromatic Hubble Andromeda
  Treasury. XV. The BEAST: Bayesian Extinction and Stellar Tool}. \apj. 2016
  Aug;\hspace{0pt}826:104--123.

\bibitem{roman-duval2014}
{Roman-Duval}~J, {Gordon}~KD, {Meixner}~M, et~al. {Dust and Gas in the
  Magellanic Clouds from the HERITAGE Herschel Key Project. II. Gas-to-dust
  Ratio Variations across Interstellar Medium Phases}. \apj. 2014
  Dec;\hspace{0pt}797:86--109.

\bibitem{draine2007}
{Draine}~BT, {Dale}~DA, {Bendo}~G, et~al. {Dust Masses, PAH Abundances, and
  Starlight Intensities in the SINGS Galaxy Sample}. \apj. 2007
  Jul;\hspace{0pt}663:866--894.

\bibitem{gordon2014}
{Gordon}~KD, {Roman-Duval}~J, {Bot}~C, et~al. {Dust and Gas in the Magellanic
  Clouds from the HERITAGE Herschel Key Project. I. Dust Properties and
  Insights into the Origin of the Submillimeter Excess Emission}. \apj. 2014
  Dec;\hspace{0pt}797:85--103.

\bibitem{searle1978}
{Searle}~L, {Zinn}~R. {Compositions of halo clusters and the formation of the
  galactic halo}. \apj. 1978 Oct;\hspace{0pt}225:357--379.

\bibitem{blumenthal1984}
{Blumenthal}~GR, {Faber}~SM, {Primack}~JR, et~al. {Formation of galaxies and
  large-scale structure with cold dark matter}. \nat. 1984
  Oct;\hspace{0pt}311:517--525.

\bibitem{conselice2014}
{Conselice}~CJ. {The Evolution of Galaxy Structure Over Cosmic Time}. \araa.
  2014 Aug;\hspace{0pt}52:291--337.

\bibitem{madau2014}
{Madau}~P, {Dickinson}~M. {Cosmic Star-Formation History}. \araa. 2014
  Aug;\hspace{0pt}52:415--486.

\bibitem{shapley2011}
{Shapley}~AE. {Physical Properties of Galaxies from z = 2-4}. \araa. 2011
  Sep;\hspace{0pt}49:525--580.

\bibitem{noeske2007}
{Noeske}~KG, {Weiner}~BJ, {Faber}~SM, et~al. {Star Formation in AEGIS Field
  Galaxies since z=1.1: The Dominance of Gradually Declining Star Formation,
  and the Main Sequence of Star-forming Galaxies}. \apjl. 2007
  May;\hspace{0pt}660:L43--L46.

\bibitem{salmon2015}
{Salmon}~B, {Papovich}~C, {Finkelstein}~SL, et~al. {The Relation between Star
  Formation Rate and Stellar Mass for Galaxies at 3.5 $<$= z $<$= 6.5 in
  CANDELS}. \apj. 2015 Feb;\hspace{0pt}799:183--209.

\bibitem{genzel2010}
{Genzel}~R, {Tacconi}~LJ, {Gracia-Carpio}~J, et~al. {A study of the gas-star
  formation relation over cosmic time}. \mnras. 2010
  Oct;\hspace{0pt}407:2091--2108.

\bibitem{faber2007}
{Faber}~SM, {Willmer}~CNA, {Wolf}~C, et~al. {Galaxy Luminosity Functions to
  z\~{}1 from DEEP2 and COMBO-17: Implications for Red Galaxy Formation}. \apj.
  2007 Aug;\hspace{0pt}665:265--294.

\bibitem{lotz2006}
{Lotz}~JM, {Madau}~P, {Giavalisco}~M, et~al. {The Rest-Frame Far-Ultraviolet
  Morphologies of Star-forming Galaxies at z \~{} 1.5 and 4}. \apj. 2006
  Jan;\hspace{0pt}636:592--609.

\bibitem{vanderwel2014}
{van der Wel}~A, {Franx}~M, {van Dokkum}~PG, et~al. {3D-HST+CANDELS: The
  Evolution of the Galaxy Size-Mass Distribution since z = 3}. \apj. 2014
  Jun;\hspace{0pt}788:28--46.

\bibitem{behroozi2015}
{Behroozi}~PS, {Silk}~J. {A Simple Technique for Predicting High-redshift
  Galaxy Evolution}. \apj. 2015 Jan;\hspace{0pt}799:32.

\bibitem{mason2015}
{Mason}~CA, {Trenti}~M, {Treu}~T. {The Galaxy UV Luminosity Function before the
  Epoch of Reionization}. \apj. 2015 Nov;\hspace{0pt}813:21.

\bibitem{finkelstein2016}
{Finkelstein}~SL. {Observational Searches for Star-Forming Galaxies at z $>$
  6}. PASA. 2016 Aug;\hspace{0pt}33:e037.

\bibitem{blanton2009}
{Blanton}~MR, {Moustakas}~J. {Physical Properties and Environments of Nearby
  Galaxies}. \araa. 2009 Sep;\hspace{0pt}47:159--210.

\bibitem{silk2012}
{Silk}~J, {Mamon}~GA. {The current status of galaxy formation}. Research in
  Astronomy and Astrophysics. 2012 Aug;\hspace{0pt}12:917--946.

\bibitem{smoot1992}
{Smoot}~GF, {Bennett}~CL, {Kogut}~A, et~al. {Structure in the COBE differential
  microwave radiometer first-year maps}. \apjl. 1992
  Sep;\hspace{0pt}396:L1--L5.

\bibitem{spergel2003}
{Spergel}~DN, {Verde}~L, {Peiris}~HV, et~al. {First-Year Wilkinson Microwave
  Anisotropy Probe (WMAP) Observations: Determination of Cosmological
  Parameters}. \apjs. 2003 Sep;\hspace{0pt}148:175--194.

\bibitem{planckcollaboration2014}
{Planck Collaboration}, {Ade}~PAR, {Aghanim}~N, et~al. {Planck 2013 results. I.
  Overview of products and scientific results}. \aap. 2014
  Nov;\hspace{0pt}571:A1.

\bibitem{bromm2011}
{Bromm}~V, {Yoshida}~N. {The First Galaxies}. \araa. 2011
  Sep;\hspace{0pt}49:373--407.

\bibitem{couchman1986}
{Couchman}~HMP, {Rees}~MJ. {Pregalactic evolution in cosmologies with cold dark
  matter}. \mnras. 1986 Jul;\hspace{0pt}221:53--62.

\bibitem{tegmark1997}
{Tegmark}~M, {Silk}~J, {Rees}~MJ, et~al. {How Small Were the First Cosmological
  Objects?} \apj. 1997 Jan;\hspace{0pt}474:1--12.

\bibitem{bromm2004}
{Bromm}~V, {Larson}~RB. {The First Stars}. \araa. 2004
  Sep;\hspace{0pt}42:79--118.

\bibitem{williams2000}
{Williams}~RE, {Baum}~S, {Bergeron}~LE, et~al. {The Hubble Deep Field South:
  Formulation of the Observing Campaign}. \aj. 2000
  Dec;\hspace{0pt}120:2735--2746.

\bibitem{giavalisco2004}
{Giavalisco}~M, {Ferguson}~HC, {Koekemoer}~AM, et~al. {The Great Observatories
  Origins Deep Survey: Initial Results from Optical and Near-Infrared Imaging}.
  \apjl. 2004 Jan;\hspace{0pt}600:L93--L98.

\bibitem{scoville2007}
{Scoville}~N, {Aussel}~H, {Brusa}~M, et~al. {The Cosmic Evolution Survey
  (COSMOS): Overview}. \apjs. 2007 Sep;\hspace{0pt}172:1--8.

\bibitem{davis2007}
{Davis}~M, {Guhathakurta}~P, {Konidaris}~NP, et~al. {The All-Wavelength
  Extended Groth Strip International Survey (AEGIS) Data Sets}. \apjl. 2007
  May;\hspace{0pt}660:L1--L6.

\bibitem{grogin2011}
{Grogin}~NA, {Kocevski}~DD, {Faber}~SM, et~al. {CANDELS: The Cosmic Assembly
  Near-infrared Deep Extragalactic Legacy Survey}. \apjs. 2011
  Dec;\hspace{0pt}197:35--73.

\bibitem{koekemoer2011}
{Koekemoer}~AM, {Faber}~SM, {Ferguson}~HC, et~al. {CANDELS: The Cosmic Assembly
  Near-infrared Deep Extragalactic Legacy Survey -- The Hubble Space Telescope
  Observations, Imaging Data Products, and Mosaics}. \apjs. 2011
  Dec;\hspace{0pt}197:36--71.

\bibitem{trenti2011}
{Trenti}~M, {Bradley}~LD, {Stiavelli}~M, et~al. {The Brightest of Reionizing
  Galaxies Survey: Design and Preliminary Results}. \apjl. 2011
  Feb;\hspace{0pt}727:L39.

\bibitem{postman2012}
{Postman}~M, {Coe}~D, {Ben{\'{\i}}tez}~N, et~al. {The Cluster Lensing and
  Supernova Survey with Hubble: An Overview}. \apjs. 2012
  Apr;\hspace{0pt}199:25--47.

\bibitem{lotz2017}
{Lotz}~JM, {Koekemoer}~A, {Coe}~D, et~al. {The Frontier Fields: Survey Design
  and Initial Results}. \apj. 2017 Mar;\hspace{0pt}837:97.

\bibitem{ouchi2009}
{Ouchi}~M, {Mobasher}~B, {Shimasaku}~K, et~al. {Large Area Survey for z = 7
  Galaxies in SDF and GOODS-N: Implications for Galaxy Formation and Cosmic
  Reionization}. \apj. 2009 Dec;\hspace{0pt}706:1136--1151.

\bibitem{tilvi2013}
{Tilvi}~V, {Papovich}~C, {Tran}~KVH, et~al. {Discovery of Lyman Break Galaxies
  at z \~{} 7 from the zFourGE Survey}. \apj. 2013 May;\hspace{0pt}768:56--72.

\bibitem{bouwens2015}
{Bouwens}~RJ, {Illingworth}~GD, {Oesch}~PA, et~al. {UV Luminosity Functions at
  Redshifts z $~$ 4 to z $~$ 10: 10,000 Galaxies from HST Legacy Fields}. \apj.
  2015 Apr;\hspace{0pt}803:34--82.

\bibitem{ellis2013}
{Ellis}~RS, {McLure}~RJ, {Dunlop}~JS, et~al. {The Abundance of Star-forming
  Galaxies in the Redshift Range 8.5-12: New Results from the 2012 Hubble Ultra
  Deep Field Campaign}. \apjl. 2013 Jan;\hspace{0pt}763:L7.

\bibitem{coe2013}
{Coe}~D, {Zitrin}~A, {Carrasco}~M, et~al. {CLASH: Three Strongly Lensed Images
  of a Candidate z {\ap} 11 Galaxy}. \apj. 2013 Jan;\hspace{0pt}762:32--52.

\bibitem{oesch2013}
{Oesch}~PA, {Bouwens}~RJ, {Illingworth}~GD, et~al. {Probing the Dawn of
  Galaxies at z \~{} 9-12: New Constraints from HUDF12/XDF and CANDELS data}.
  \apj. 2013 Aug;\hspace{0pt}773:75--93.

\bibitem{oesch2014}
{Oesch}~PA, {Bouwens}~RJ, {Illingworth}~GD, et~al. {The Most Luminous z \~{}
  9-10 Galaxy Candidates Yet Found: The Luminosity Function, Cosmic
  Star-formation Rate, and the First Mass Density Estimate at 500 Myr}. \apj.
  2014 May;\hspace{0pt}786:108--126.

\bibitem{bouwens2016}
{Bouwens}~RJ, {Oesch}~PA, {Labb{\'e}}~I, et~al. {The Bright End of the z $~$ 9
  and z $~$ 10 UV Luminosity Functions Using All Five CANDELS Fields*}. \apj.
  2016 Oct;\hspace{0pt}830:67--88.

\bibitem{mcleod2015}
{McLeod}~DJ, {McLure}~RJ, {Dunlop}~JS, et~al. {New redshift z $~$ 9 galaxies in
  the Hubble Frontier Fields: implications for early evolution of the UV
  luminosity density}. \mnras. 2015 Jul;\hspace{0pt}450:3032--3044.

\bibitem{gonzalez2010}
{Gonz{\'a}lez}~V, {Labb{\'e}}~I, {Bouwens}~RJ, et~al. {The Stellar Mass Density
  and Specific Star Formation Rate of the Universe at z \~{} 7}. \apj. 2010
  Apr;\hspace{0pt}713:115--130.

\bibitem{labbe2010}
{Labb{\'e}}~I, {Gonz{\'a}lez}~V, {Bouwens}~RJ, et~al. {Star Formation Rates and
  Stellar Masses of z = 7-8 Galaxies from IRAC Observations of the WFC3/IR
  Early Release Science and the HUDF Fields}. \apjl. 2010
  Jun;\hspace{0pt}716:L103--L108.

\bibitem{dunlop2013}
{Dunlop}~JS, {Rogers}~AB, {McLure}~RJ, et~al. {The UV continua and inferred
  stellar populations of galaxies at z $~$ 7-9 revealed by the Hubble
  Ultra-Deep Field 2012 campaign}. \mnras. 2013 Jul;\hspace{0pt}432:3520--3533.

\bibitem{robertson2013}
{Robertson}~BE, {Furlanetto}~SR, {Schneider}~E, et~al. {New Constraints on
  Cosmic Reionization from the 2012 Hubble Ultra Deep Field Campaign}. \apj.
  2013 May;\hspace{0pt}768:71--87.

\bibitem{windhorst2006}
{Windhorst}~RA, {Cohen}~SH, {Jansen}~RA, et~al. {How JWST can measure first
  light, reionization and galaxy assembly}. \nar. 2006
  Mar;\hspace{0pt}50:113--120.

\bibitem{zackrisson2012}
{Zackrisson}~E, {Zitrin}~A, {Trenti}~M, et~al. {Detecting gravitationally
  lensed Population III galaxies with the Hubble Space Telescope and the James
  Webb Space Telescope}. \mnras. 2012 Dec;\hspace{0pt}427:2212--2223.

\bibitem{stiavelli2009}
{Stiavelli}~M. {From First Light to Reionization: The End of the Dark Ages}.
  Darmstadt, Germany; 2009.

\bibitem{wyithe2006}
{Wyithe}~JSB, {Loeb}~A. {Suppression of dwarf galaxy formation by cosmic
  reionization}. \nat. 2006 May;\hspace{0pt}441:322--324.

\bibitem{livermore2017}
{Livermore}~RC, {Finkelstein}~SL, {Lotz}~JM. {Directly Observing the Galaxies
  Likely Responsible for Reionization}. \apj. 2017 Feb;\hspace{0pt}835:113.

\bibitem{vogelsberger2014}
{Vogelsberger}~M, {Genel}~S, {Springel}~V, et~al. {Introducing the Illustris
  Project: simulating the coevolution of dark and visible matter in the
  Universe}. \mnras. 2014 Oct;\hspace{0pt}444:1518--1547.

\bibitem{snyder2017}
{Snyder}~GF, {Lotz}~JM, {Rodriguez-Gomez}~V, et~al. {Massive close pairs
  measure rapid galaxy assembly in mergers at high redshift}. \mnras. 2017
  Jun;\hspace{0pt}468:207--216.

\bibitem{paardekooper2013}
{Paardekooper}~JP, {Khochfar}~S, {Dalla Vecchia}~C. {The First Billion Years
  project: proto-galaxies reionizing the Universe}. \mnras. 2013
  Feb;\hspace{0pt}429:L94--L98.

\bibitem{finkelstein2012}
{Finkelstein}~SL, {Papovich}~C, {Ryan}~RE, et~al. {CANDELS: The Contribution of
  the Observed Galaxy Population to Cosmic Reionization}. \apj. 2012
  Oct;\hspace{0pt}758:93--109.

\bibitem{whalen2013}
{Whalen}~DJ, {Fryer}~CL, {Holz}~DE, et~al. {Seeing the First Supernovae at the
  Edge of the Universe with JWST}. \apjl. 2013 Jan;\hspace{0pt}762:L6.

\bibitem{hartwig2017}
{Hartwig}~T, {Bromm}~V, {Loeb}~A. {Survey strategies for the first supernovae
  with JWST}. ArXiv e-prints. 2017 Nov;\hspace{0pt}.

\bibitem{haiman2009}
{Haiman}~Z. {Observing the First Stars and Black Holes}. Astrophysics and Space
  Science Proceedings. 2009;\hspace{0pt}10:385--418.

\bibitem{windhorst2018}
{Windhorst}~RA, {Timmes}~FX, {Wyithe}~JSB, et~al. {On the Observability of
  Individual Population III Stars and Their Stellar-mass Black Hole Accretion
  Disks through Cluster Caustic Transits}. The Astrophysical Journal Supplement
  Series. 2018 Feb;\hspace{0pt}234:41--80.

\bibitem{tumlinson2017}
{Tumlinson}~J, {Peeples}~MS, {Werk}~JK. {The Circumgalactic Medium}. \araa.
  2017 Aug;\hspace{0pt}55:389--432.

\bibitem{hopkins2012}
{Hopkins}~PF, {Quataert}~E, {Murray}~N. {Stellar feedback in galaxies and the
  origin of galaxy-scale winds}. \mnras. 2012 Apr;\hspace{0pt}421:3522--3537.

\bibitem{somerville2015}
{Somerville}~RS, {Dav{\'e}}~R. {Physical Models of Galaxy Formation in a
  Cosmological Framework}. \araa. 2015 Aug;\hspace{0pt}53:51--113.

\bibitem{kirkpatrick2017}
{Kirkpatrick}~A, {Alberts}~S, {Pope}~A, et~al. {The AGN-Star Formation
  Connection: Future Prospects with JWST}. \apj. 2017 Nov;\hspace{0pt}849:111.

\bibitem{ferrarese2000}
{Ferrarese}~L, {Merritt}~D. {A Fundamental Relation between Supermassive Black
  Holes and Their Host Galaxies}. \apjl. 2000 Aug;\hspace{0pt}539:L9--L12.

\bibitem{kennicutt2003}
{Kennicutt}~RC~Jr, {Armus}~L, {Bendo}~G, et~al. {SINGS: The SIRTF Nearby
  Galaxies Survey}. \pasp. 2003 Aug;\hspace{0pt}115:928--952.

\bibitem{calzetti2007}
{Calzetti}~D, {Kennicutt}~RC, {Engelbracht}~CW, et~al. {The Calibration of
  Mid-Infrared Star Formation Rate Indicators}. \apj. 2007
  Sep;\hspace{0pt}666:870--895.

\bibitem{calzetti2010}
{Calzetti}~D, {Wu}~SY, {Hong}~S, et~al. {The Calibration of Monochromatic
  Far-Infrared Star Formation Rate Indicators}. \apj. 2010
  May;\hspace{0pt}714:1256--1279.

\bibitem{caputi2011}
{Caputi}~KI, {Cirasuolo}~M, {Dunlop}~JS, et~al. {The stellar mass function of
  the most-massive galaxies at 3 $<=$ z $<$ 5 in the UKIDSS Ultra Deep Survey}.
  \mnras. 2011 May;\hspace{0pt}413:162--176.

\bibitem{atek2010}
{Atek}~H, {Malkan}~M, {McCarthy}~P, et~al. {The WFC3 Infrared Spectroscopic
  Parallel (WISP) Survey}. \apj. 2010 Nov;\hspace{0pt}723:104--115.

\bibitem{springel2005}
{Springel}~V, {White}~SDM, {Jenkins}~A, et~al. {Simulations of the formation,
  evolution and clustering of galaxies and quasars}. \nat. 2005
  Jun;\hspace{0pt}435:629--636.

\bibitem{wuyts2010}
{Wuyts}~S, {Cox}~TJ, {Hayward}~CC, et~al. {On Sizes, Kinematics, M/L Gradients,
  and Light Profiles of Massive Compact Galaxies at z \~{} 2}. \apj. 2010
  Oct;\hspace{0pt}722:1666--1684.

\bibitem{lotz2011}
{Lotz}~JM, {Jonsson}~P, {Cox}~TJ, et~al. {The Major and Minor Galaxy Merger
  Rates at z $<$ 1.5}. \apj. 2011 Dec;\hspace{0pt}742:103--124.

\bibitem{vandokkum2010}
{van Dokkum}~PG, {Whitaker}~KE, {Brammer}~G, et~al. {The Growth of Massive
  Galaxies Since z = 2}. \apj. 2010 Feb;\hspace{0pt}709:1018--1041.

\bibitem{kriek2010}
{Kriek}~M, {Labb{\'e}}~I, {Conroy}~C, et~al. {The Spectral Energy Distribution
  of Post-starburst Galaxies in the NEWFIRM Medium-band Survey: A Low
  Contribution from TP-AGB Stars}. \apjl. 2010 Oct;\hspace{0pt}722:L64--L69.

\bibitem{riess1998}
{Riess}~AG, {Filippenko}~AV, {Challis}~P, et~al. {Observational Evidence from
  Supernovae for an Accelerating Universe and a Cosmological Constant}. \aj.
  1998 Sep;\hspace{0pt}116:1009--1038.

\bibitem{perlmutter1999}
{Perlmutter}~S, {Aldering}~G, {Goldhaber}~G, et~al. {Measurements of {$\Omega$}
  and {$\Lambda$} from 42 High-Redshift Supernovae}. \apj. 1999
  Jun;\hspace{0pt}517:565--586.

\bibitem{riess2006}
{Riess}~AG, {Livio}~M. {The First Type Ia Supernovae: An Empirical Approach to
  Taming Evolutionary Effects in Dark Energy Surveys from SNe Ia at z$>$2}.
  \apj. 2006 Sep;\hspace{0pt}648:884--889.

\bibitem{treu2015}
{Treu}~T, {Ellis}~RS. {Gravitational Lensing: Einstein's unfinished symphony}.
  Contemporary Physics. 2015 Jan;\hspace{0pt}56:17--34.

\bibitem{gardner2012}
{Gardner}~JP. {The James Webb Space Telescope: extending the science}. In:
  Space Telescopes and Instrumentation 2012: Optical, Infrared, and Millimeter
  Wave; (\procspie; Vol. 8442); Sep.; 2012. p. 844228.

\bibitem{levan2017}
{Levan}~AJ, {Lyman}~JD, {Tanvir}~NR, et~al. {The Environment of the Binary
  Neutron Star Merger GW170817}. \apjl. 2017 Oct;\hspace{0pt}848:L28.

\end{thebibliography}


\end{document}